\documentclass[11pt,aps,showpacs,nofootinbib,prd,aps,epsf,floats,amsmath,amssymb,amsfonts]{revtex4}
\usepackage{amsmath, amssymb}
\newcommand{\mathsym}[1]{{}}

\usepackage{graphicx}
\usepackage{amsmath}
\usepackage{amssymb}
\usepackage{bm}
\newcommand{\be}{\begin{equation}}
\newcommand{\ee}{\end{equation}}
\newcommand{\bea}{\begin{eqnarray}}
\newcommand{\eea}{\end{eqnarray}}

\newcommand{\rem}[1]{}
\newsavebox{\PSLASH}
 \sbox{\PSLASH}{$p$\hspace{-1.8mm}/}
 
\renewcommand{\theequation}{\thesection.\arabic{equation}}
\newcounter{saveeqn}
\newcommand{\add}{\addtocounter{equation}{1}}
\newcommand{\alpheqn}{\setcounter{saveeqn}{\value{equation}}%
\setcounter{equation}{0}%
\renewcommand{\theequation}{\mbox{\thesection.\arabic{saveeqn}{\alph{equation}}}}}
\newcommand{\reseteqn}{\setcounter{equation}{\value{saveeqn}}%
\renewcommand{\theequation}{\thesection.\arabic{equation}}}

 \newsavebox{\notrightarrow}
 \sbox{\notrightarrow}{$\to$\hspace{-4mm}/}
 
 \newsavebox{\PARTIALSLASH}
 \sbox{\PARTIALSLASH}{$\partial$\hspace{-1.6mm}/}
 
 \newsavebox{\ASLASH}
 \sbox{\ASLASH}{$A$\hspace{-2.1mm}/}
 
 \newsavebox{\KSLASH}
 \sbox{\KSLASH}{$k$\hspace{-1.8mm}/}
 
 \newsavebox{\LSLASH}
 \sbox{\LSLASH}{$\ell$\hspace{-1.8mm}/}
 
 \newsavebox{\QSLASH}
 \sbox{\QSLASH}{$q$\hspace{-1.8mm}/}
 
 \newsavebox{\DSLASH}
 \sbox{\DSLASH}{$D$\hspace{-2.2mm}/}
 
 \newsavebox{\DbfSLASH}
 \sbox{\DbfSLASH}{${\mathbf D}$\hspace{-2.8mm}/}
 
 \newsavebox{\DELVECRIGHT}
 \sbox{\DELVECRIGHT}{$\stackrel{\rightarrow}{\partial}$}
 
 \newcommand{\blue}{\IfColor{\textCadetBlue}{}}
\newcommand{\black}{\IfColor{\textBlack}{}}
\newcommand{\red}{\IfColor{\textRed}{}}
\newcommand{\green}{\IfColor{\textOliveGreen}{}}
\newcommand{\lila}{\IfColor{\textRedViolet}{}}

\newcommand{\parspace}{\par\vspace{0.3cm}\par}

\begin{document}

\title{Phase diagram of hot magnetized two-flavor color superconducting quark matter}
\author{Sh. Fayazbakhsh}\email{fayyazbaksh@physics.sharif.ir}
\author{N. Sadooghi}\email{sadooghi@physics.sharif.ir}
\affiliation{Department of Physics, Sharif University of Technology,
P.O. Box 11155-9161, Tehran-Iran}
\begin{abstract}
A two-flavor color superconducting (2SC) Nambu--Jona-Lasinio (NJL)
model is introduced at finite temperature $T$, chemical potential
$\mu$ and in the presence of a constant magnetic field $\tilde{e}B$.
The effect of $(T,\mu,\tilde{e}B)$ on the formation of chiral and
color symmetry breaking condensates is studied. The complete phase
portrait of the model in $T-\mu$, $\mu-\tilde{e}B$, and
$T-\tilde{e}B$ phase spaces for various fixed $\tilde{e}B$, $T$, and
$\mu$ is explored. A threshold magnetic field $\tilde{e}B_{t}\simeq
0.5$ GeV$^{2}$ is found above which the dynamics of the system is
solely dominated by the lowest Landau level (LLL) and the effects of
$T$ and $\mu$ are partly compensated by $\tilde{e}B$.
\end{abstract}
 \maketitle
\section{Introduction}\label{introduction}
\noindent Recently, the study of the properties of quark matter in
the presence of strong uniform magnetic fields has attracted much
attention. Possible effects caused by strong magnetic fields include
magnetic catalysis \cite{miransky1995, catalysis}, modification of
the nature of electroweak phase transition \cite{ayala2008},
spontaneous creation of axial currents \cite{son2007}, formation of
$\pi_{0}$ domain walls \cite{stephanov2007} and chiral density waves
\cite{klimenko2010}, chiral magnetic effect \cite{kharzeev2008,
fukushima2009}, and last but not least the influence on possible
color-superconducting phases
\cite{berges1998,gorbar2000,manuel2006,ferrer2006, warringa2007,
shovkovy2007,mandal2009, fayaz2010}. In this paper, we will focus on
the magnetic catalysis and its possible effects on the phase diagram
of a magnetized two-flavor color superconducting NJL model at finite
temperature and chemical potential.\footnote{Magnetic catalysis has
various application in cosmology \cite{cosmology}, condensed matter
physics \cite{condensed}, and particle physics \cite{particle,
inagaki2003, sato1997}.} In particular, the dependence of the
included meson and diquark masses on thermodynamic parameters, and
possible interplay between these parameters on the formation of
meson and diquark condensates and on the nature of phase transition
will be scrutinized.
\par
At zero temperature, it is known that strong magnetic fields enhance
the production of chiral and diquark condensates, albeit through
different mechanisms, as it is described in \cite{manuel2006}.
Whereas magnetic catalysis of chiral symmetry breaking
\cite{miransky1995} is mainly responsible for dynamical mass
generation and enhances the production of chiral condensates by
increasing the particle-antiparticle interaction strength,  a
certain modification in the density of states of charged quarks near
the Fermi surface, depending on the external magnetic field,
reinforces the pairing of charged quarks and is made responsible for
the enhancement of diquark production by a penetrating strong
magnetic field \cite{manuel2006}. In other words, in contrast to the
effect of magnetic catalysis of chiral symmetry breaking, which is
essentially based on a dimensional reduction of the dynamics of
fermions from $D=3+1$ to $D=1+1$ dimensions to the regime of LLL
dominance \cite{miransky1995}, the pairing mechanism by color
superconductivity, involving the charged quarks near the Fermi
surface, is already $D=1+1$ dimensional, and therefore the external
magnetic field does not lead to any further dimensional reduction
\cite{manuel2006}. It is the goal of the present paper to explore
possible effects of finite $T$ and $\mu$ on the above mechanisms of
chiral and diquark production in the presence of strong magnetic
fields. We will show that in our setup a certain threshold magnetic
field exists, above which the dynamics of the system is solely
dominated by the lowest Landau level and the effects of $T$ and
$\mu$ are partly compensated by very large $\tilde{e}B$. The largest
observed magnetic field in nature is $\sim 10^{12}-10^{13}$
Gau\ss~in pulsars and up to $\sim 10^{14}-10^{15}$ Gau\ss~on the
surface of some magnetars, where the inner field is estimated to be
of order $\sim 10^{18}-10^{20}$ Gau\ss~\cite{ferrer2010}. In the
early universe, magnetic fields of order $\sim 10^{47}$ Gau\ss~may
be produced at the beginning of inflation \cite{rubinstein2001}.
Superconductive cosmic strings, if they exist, may have magnetic
fields up to $\sim 10^{47}-10^{48}$ Gau\ss~in their vicinities
\cite{witten1985}. In \cite{shabad2006}, it is shown that ``the
maximum value of magnetic field that delimits the range of values
admitted without revising QED'' is of order $\sim 10^{42}$
Gau\ss.~There are also evidences for strong magnetic field creation
in non-central heavy ion experiments \cite{kharzeev51-STAR}. The
early estimate of the magnetic field for the RHIC energy was made in
\cite{mclerran2007}, where it was shown that the magnitude of the
magnetic field for an earlier stage of noncentral Au-Au collision at
energy $\sqrt{s_{NN}}=200$ GeV and the impact parameter $\sim 4$ fm
is about $eB\simeq 1.3~m_{\pi}^{2}\simeq 0.025$ GeV$^{2}$, that
corresponds to $B\simeq 4.3\times 10^{18}$ Gau\ss.\footnote{Here,
$m_{\pi}=140$ MeV. Moreover, $eB=1$ GeV$^{2}$ corresponds to
$B=1.69\times 10^{20}$ Gau\ss.} Using a microscopic transport model,
the authors in \cite{skokov2009} estimate the lowest bound of the
maximal magnetic field strength at the LHC energy
$\sqrt{s_{NN}}=4.5$ TeV with the same impact parameter, to be
$eB\simeq 15~m_{\pi}^{2}\simeq 0.3$ GeV$^{2}$, which is equivalent
to $B\simeq 5\times 10^{19}$ Gau\ss. Numerically, it is not
\textit{a priori} clear whether these amounts of magnetic fields are
large enough to justify the LLL approximation, as it is done in
\cite{miransky1995} to demonstrate the magnetic catalysis of chiral
symmetry breaking and in \cite{manuel2006} to demonstrate the effect
of strong magnetic field on color superconductivity.
\par
We have tried to answer to this question, among others, in
\cite{fayaz2010}, where a color neutral and dense 2SC-NJL model has
been introduced in a (rotated) constant magnetic field at zero
temperature. To do this, first the dependence of chiral and color
symmetry breaking condensates on the chemical potential $\mu$ and
the rotated magnetic field $\tilde{e}B$ is determined analytically
in a lowest Landau level (LLL) approximation. Then, the meson and
diquark masses are computed numerically for arbitrary magnetic
fields. Comparing these analytical and numerical data, we have found
a certain threshold magnetic field $\tilde{e}B_{t}\simeq 0.45-0.5$
GeV$^{2}$, corresponding to $B\simeq 8-8.5\times 10^{19}$
Gau\ss,~above which the system turns out to be dominated by LLL.
Below $\tilde{e}B_{t}$, the chiral and diquark mass gaps oscillate
with the external magnetic field. These oscillations are the results
of the well-known van Alfven-de Haas effect \cite{alfven}, that
occurs whenever Landau levels pass the quark Fermi surface. They are
also observed in \cite{inagaki2003, ebert1999}, where the dependence
of chiral symmetry breaking mass gaps on constant magnetic fields is
explored. Similar oscillations are also perceived in
\cite{shovkovy2007}, where diquark mass gap and the magnetization
corresponding to the superconducting magnetized color-flavor locked
(CFL) phase are determined as a function of external magnetic
fields. For $\tilde{e}B>\tilde{e}B_{t}$, the system enters a
``linear regime'', where the mass gaps and the magnetization depend
linearly on the external magnetic field. In this regime, where the
system is believed to be solely affected by the dynamics of the
fermions in the LLL, the numerical data coincide with the analytical
results (see \cite{fayaz2010} for more detail). What concerns the
phase diagram of the model introduced in \cite{fayaz2010}, it is
shown that a first order phase transition occurs between the chiral
symmetry breaking ($\chi$SB) and the color symmetry breaking (CSC)
phases at $T=0$ and $\mu\sim 350-450$ MeV. This transition is then
followed by a second order phase transition between the CSC phase
into the normal quark matter.\footnote{As it is known from
\cite{rajagopal2007}, in the regime of low temperature and large
chemical potential, the 2SC phase goes over into the three-flavor
CFL phase. In a two-flavor model, however, where no CFL phase can be
built, only a simple transition from the 2SC to the normal phase is
assumed to exist. Same assumption is also made in \cite{fayaz2010}.
As we will note in Sec. III, our numerical results for low
temperature and large chemical potential will be only of theoretical
nature.}
\par
In the present paper, the same magnetized 2SC model will be
considered at finite temperature. We are in particular interested on
the additional effects of finite temperature, and, will focus on
possible interplay between $T,\mu$ and $\tilde{e}B$ on the formation
of chiral and color symmetry breaking condensates and on the nature
of phase transitions. Our results may be relevant for the physics of
heavy ion collisions, where, recently the question of accessibility
of color superconducting quark matter phases is pointed out
\cite{blaschke2010}. The authors in \cite{blaschke2010} use a
Polyakov-NJL (PNJL) model at finite temperature and density and
present the corresponding QCD phase diagram including, among others,
mixed phase regions of first order transition of 2SC-CFL quark
matter and second order 2SC-normal phase transition. From this phase
diagram, they conclude that color superconductor phase is already
accessible at the present nuclotron-M energies $4<E<8$ AGeV, and,
that possible transition from 2SC  to normal quark matter becomes
attainable in the planned FAIR-CBM and NICA-MPD experiments at
$2<E<40$ AGeV. In this paper, the additional effect of constant
magnetic field, that is believed to be created in non-central heavy
ion collisions will be explored from purely theoretical point of
view. A complete phenomenological answer to the question of
accessibility of color superconducting phases in heavy ion
collisions is out of the scope of this paper, and will be postponed
to future publications.
\par
The organization of the paper is as follows: In Sec. II, we
introduce our model for the hot magnetized two-flavor
superconducting quark matter and determine the corresponding
thermodynamic potential. In Sec. III, our numerical results on the
dependence of the meson and diquark gaps on chemical potential,
magnetic field, and temperature will be presented in Secs. III.A.1
-- III.A.3. In Sec. III.B, the complete $T-\mu$, $T-\tilde{e}B$ and
$\mu-\tilde{e}B$ phase portraits of the system at various fixed
$\tilde{e}B,\mu$ and $T$ will be illustrated. The above mentioned
threshold magnetic field, $\tilde{e}B_{t}$, will be determined by
comparing the analytical and the numerical results, corresponding to
the second order critical lines of the transition from the $\chi$SB
and CSC phases to the normal phase, in a $T-\mu$ plane for
$\tilde{e}B=0.5,0.7$ GeV$^{2}$ (Sec. III.B.1) and in a
$T-\tilde{e}B$ plane for $\mu=0$ MeV (Sec. III.B.2). The details of
the analytical computations, that lead to second order critical
surfaces of these transitions in a $(T,\mu,\tilde{e}B)$ phase space,
will be presented in App. A. Section IV is devoted to concluding
remarks.
\section{Magnetized 2SC quark matter at finite $T$ and $\mu$}
\noindent In Sec. II.A, we briefly review our results from
\cite{fayaz2010} and introduce a two-flavor NJL model including the
meson and diquark condensates at finite temperature and density and
in the presence of a constant and uniform (rotated) magnetic field.
In Sec. II.B, the corresponding one-loop thermodynamic potential
will be determined in a mean field approximation.
\subsection{The Model}
\noindent The Lagrangian density of a two-flavor gauged NJL model is
given by
\begin{eqnarray}\label{Fb1}
\lefteqn{{\cal{L}}_{f}=\overline{\psi}(x)[i\gamma^{\mu}(\partial_{\mu}-ieQ
A_{\mu}-igT^{8}G^{8}_{\mu})+\mu\gamma^{0}]\psi(x)}\nonumber\\
&&+G_{S}[(\overline{\psi}(x)\psi(x))^2+(\overline{\psi}(x)i\gamma_{5}\vec{\tau}\psi(x))^2]+G_{D}[(i\overline{\psi}^C(x)\varepsilon_{f}\epsilon_{c}^{3}\gamma_{5}\psi(x))(i\overline{\psi}(x)
\varepsilon_{f}\epsilon_{c}^{3}\gamma_{5}\psi^C(x))].
\end{eqnarray}
Here, $\psi^C=C\overline{\psi}^T$ and $\overline{\psi}^C=\psi^TC$
are charge-conjugate spinors, and $C=i\gamma^2\gamma^0$ is
charge-conjugation matrix, $\vec{\tau}=(\tau_{1},\tau_{2},\tau_{3})$
are Pauli matrices. Moreover, $(\varepsilon_{f})_{ij}$ and
$(\epsilon^{3}_c)^{ab}\equiv(\epsilon_c)^{ab3}$ are antisymmetric
matrices in color and flavor spaces, respectively. For a theory with
two quark flavors and three color degrees of freedom, the flavor
indices $i,j=(1,2)=(u,d)$, and the color indices
$a,b=(1,2,3)=(r,g,b)$. The quarks are taken to be massless $m_u=m_d=
0$. The quark chemical potential which is responsible for nonzero
baryonic density of quark matter is denoted by $\mu$. Here,
$T^{8}=\frac{\lambda_{8}}{2}$, where
$\lambda_{8}=\frac{1}{\sqrt3}~\mbox{diag}(1,1,-2)$ the
$8^{\mbox{\small{th}}}$ Gell-Mann $\lambda$-matrix. The scalar and
diquark couplings are denoted by $G_S$ and $G_D$, respectively. The
charge matrix $Q\equiv Q_{f}\otimes {\mathbf{1}}_{c}$, where
$Q_{f}\equiv\mbox{diag}\left(2/3,-1/3\right)$ is the fermionic
charge matrix coupled to $U(1)$ gauge field $A_{\mu}$. The same
setup with an additional color chemical potential $\mu_{8}$,
imposing the color neutrality of the theory, is also used in
\cite{fayaz2010} to study the effect of magnetic field on quark
matter under extreme conditions. Following the same steps as in
\cite{fayaz2010} to determine the Lagrangian density containing the
chiral and diquark condensates in an appropriate Nambu-Gorkov form,
we define first the mesonic fields
\begin{eqnarray}\label{Fb2}
\sigma=-2 G_{S}(\overline{\psi}\psi),
\qquad\mbox{and}\qquad\vec{\pi}=-2
G_{S}(\overline{\psi}i\gamma^5\vec{\tau}\psi),
\end{eqnarray}
as well as the diquarks fields
\begin{eqnarray}\label{Fb3}
\hspace{-0.5cm}\Delta=-2
G_{D}(i\overline{\psi}^C\varepsilon_{f}\epsilon_{c}^{3}\gamma_{5}\psi),\qquad\mbox{and}\qquad\Delta^{*}=-2
G_{D}(i\overline{\psi}\varepsilon_{f}\epsilon_{c}^{3}\gamma_{5}\psi^C).
\end{eqnarray}
Combining then the gauge fields $A_{\mu}$ and $G_{\mu}^{8}$, using
the ``rotated'' charge operator $\tilde{Q}=Q_{f}\otimes
1_{c}-1_{f}\otimes \left(\frac{\lambda_{8}}{2\sqrt{3}}\right)_{c}$,
the rotated massless $U_{em}(1)$ field,
$\tilde{A}_{\mu}=A_{\mu}\cos\theta-G_{\mu}^{8}\sin\theta$, as well
as the massive in-medium $8^{\mbox{\tiny{th}}}$ gluon field,
$\tilde{G}_{\mu}^{8}=A_{\mu}\sin\theta+G_{\mu}^{8}\cos\theta$ can be
derived (see \cite{fayaz2010} for more details). Replacing
$\tilde{A}_{\mu}$ with an external gauge field
$\tilde{A}_{\mu}^{ext}=(0,0,Bx,0)$ in the Landau gauge, a constant
rotated background $U(1)$ magnetic field directed in the third
direction $\widetilde{\mathbf{B}}=B{\mathbf{e}}_{3}$ is induced.
Neglecting then the massive gauge boson $\tilde{G}_{\mu}^{8}$, the
total modified bosonized Lagrangian density, $\tilde{\cal{L}}=
\tilde{\cal{L}}_{k}+\tilde{\cal{L}}_{f}$, in the presence of a
uniform magnetic field arises. It consists of a kinetic term
\begin{eqnarray}\label{Fb4}
\tilde{\cal{L}}_{k}\equiv-\left(
\frac{\sigma^{2}}{4G_{S}}+\frac{|\Delta|^2}{4G_{D}}+\frac{B^{2}}{2}\right),
\end{eqnarray}
and an interaction term
\begin{eqnarray}\label{Fb5}
\hspace{-0.1cm}\tilde{\cal{L}}_{f}=
\overline{\psi}(x)[i\gamma^{\mu}(\partial_{\mu}-i\tilde{e}\tilde{Q}\tilde{A}_{\mu}^{ext})+\mu\gamma^{0}-\sigma]\psi(x)-\frac{1}{2}\big[\Delta^{*}(i\overline{\psi}^C(x)\varepsilon_f\epsilon^{3}_c\gamma_{5}\psi(x))+\Delta(i\overline{\psi}(x)
\varepsilon_f\epsilon^{3}_c\gamma_{5}\psi^C(x))\big].\nonumber\\
\end{eqnarray}
Assuming that the vacuum of the system is characterized by
$\langle\sigma\rangle\neq 0$ and $\langle\vec{\pi}\rangle=0$, we
have neglected the $\vec{\pi}$ mesons. Moreover, using the
definition of the rotated charge operator $\tilde{Q}$ in a
6-dimensional flavor-color representation
$\left(u_{r},u_{g},u_{b},d_{r},d_{g},d_{b}\right)$, the rotated
$\tilde{q}$ charges of different quarks, in units of
$\tilde{e}=e\cos\theta$, are given by
\par
\begin{center}
\begin{tabular}{c|c c c c c c}
    \hline\hline
    quarks&$u_{r}$ & $u_{g}$ & $u_{b}$ & $d_{r}$ & $d_{g}$ & $d_{b}$ \\
    \hline
    $\tilde{q}$&$+\frac{1}{2}$ & $+\frac{1}{2}$ & 1 & $-\frac{1}{2}$ & $-\frac{1}{2}$ & 0 \\
    \hline\hline
\end{tabular}
\end{center}
\par\noindent
To bring the above Lagrangian density $\tilde{\cal{L}}_{f}$ in a
more appropriate Nambu-Gorkov form, we introduce at this stage the
rotated charge projectors $\Omega_{\tilde{q}}$
\begin{eqnarray}\label{Fb6}
\Omega_{0}= \mbox{diag}(0,0,0,0,0,1),&\qquad&
\Omega_{1}=\mbox{diag}(0,0,1,0,0,0),\nonumber\\
\Omega_{+\frac{1}{2}}= \mbox{diag}(1,1,0,0,0,0),&\qquad&
\Omega_{-\frac{1}{2}}=\mbox{diag}(0,0,0,1,1,0),
\end{eqnarray}
that satisfy
$\tilde{Q}\Omega_{\tilde{q}}=\tilde{q}\Omega_{\tilde{q}}$. The
Nambu-Gorkov bispinors are then defined by
$$\Psi_{\tilde{q}}=\left(\begin{array}{c}
\psi_{\tilde{q}}\\
\psi^{C}_{-\tilde{q}}
\end{array}\right),$$ where
$\psi_{\tilde{q}}(x)\equiv\Omega_{\tilde{q}}\psi(x)$. In terms of
$\bar{\Psi}_{\tilde{q}}$ and $\Psi_{\tilde{q}}$, the Lagrangian
density $\tilde{\cal{L}}_{f}$ from (\ref{Fb5}) in the Nambu-Gorkov
form reads (for more details see \cite{fayaz2010})
\begin{eqnarray}\label{Fb7}
\tilde{\cal{L}}_{f}=\frac{1}{2}\sum_{\tilde
q\in\{0,1,\pm\frac{1}{2}\}}\overline{\Psi}_{\tilde
q}(x){\cal{S}}_{\tilde q}{\Psi}_{\tilde q}(x).
\end{eqnarray}
For $\tilde{q}\in\{0,1\}$, ${\cal{S}}_{\tilde{q}}$  is given by
\begin{eqnarray}\label{Fb8}
{\cal{S}}_{\tilde{q}\in\{0,1\}}\equiv\left(
\begin{array}{cc}
[G^{+}_{(\tilde{q})}]^{-1}   & 0 \\ 0 & [G^{-}_{(\tilde{q})}]^{-1} \\
\end{array}
\right),
\end{eqnarray}
and for $\tilde{q}\in\{-\frac{1}{2},+\frac{1}{2}\}$, it reads
\begin{eqnarray}\label{Fb9}
{\cal{S}}_{\tilde{q}\in\{-\frac{1}{2},+\frac{1}{2}\}}\equiv\left(
\begin{array}{cc}
[G^{+}_{(\tilde{q})}]^{-1} & -\kappa\Omega_{-\tilde{q}}\;\;  \\ -\kappa^{\prime}\Omega_{\tilde{q}} & [G^{-}_{(\tilde{q})}]^{-1} \\
\end{array}
\right).
\end{eqnarray}
Here,
\begin{eqnarray}\label{Fb10}
[G^{\pm}_{(\tilde{q})}]^{-1}\equiv
\gamma^{\mu}(i\partial_{\mu}+\tilde{e}\tilde{q}\tilde{A}_{\mu}-\sigma\pm\mu\delta_{\mu0}),
\end{eqnarray}
and $\kappa^{ij,ab}_{\alpha\beta}\equiv
i\Delta\tau_{2}^{ij}\lambda_{2}^{ab}\gamma^{5}_{\alpha\beta}$ as
well as
$\kappa^{\prime}\equiv\gamma_{0}\kappa^{\dagger}\gamma_{0}=i\Delta^{*}\tau_{2}\lambda_{2}\gamma^{5}$.
In the next section, the Lagrangian density $\tilde{\cal{L}}$ with
$\tilde{\cal{L}}_{k}$ in (\ref{Fb4}) and $\tilde{\cal{L}}_{f}$ in
(\ref{Fb7})-(\ref{Fb10}) will be used to determine the thermodynamic
potential of this model in the mean field approximation.
\subsection{Thermodynamic potential}
\noindent The quantum effective action of the theory,
$\Gamma_{\mbox{\tiny{eff}}}$, is defined by integrating out the
fermionic degrees of freedom using the path integral
\begin{eqnarray}\label{Fb11}
e^{i\Gamma_{\mbox{\tiny{eff}}}[\sigma, \Delta, \Delta^{*}]}=\int
{\cal{D}}\psi{\cal{D}}\bar{\psi}\exp\left(i\int d^{4}x~
\tilde{\cal{L}}\right).
\end{eqnarray}
At one loop level, it consists of two parts: the tree-level and the
one-loop effective action, $\Gamma_{\mbox{\tiny{eff}}}^{(0)}$ and
$\Gamma_{\mbox{\tiny{eff}}}^{(1)}$. In the mean field approximation,
where the order parameter $\sigma\equiv \langle\sigma(x)\rangle$,
$\Delta\equiv \langle\Delta(x)\rangle$ and $\Delta^{*}\equiv
\langle\Delta^{*}(x)\rangle$ are constant, the tree-level part of
$\Gamma_{\mbox{\tiny{eff}}}$ is given by
\begin{eqnarray}\label{Fb12}
\Gamma_{\mbox{\tiny{eff}}}^{(0)}[\sigma,\Delta,\Delta^{*};B]=-{\cal{V}}\left(\frac{\sigma^{2}}{4G_{S}}+\frac{|\Delta|^{2}}{4G_{D}}+\frac{B^{2}}{2}\right).
\end{eqnarray}
Here, ${\cal{V}}$ is a four-dimensional space-time volume and
$|\Delta|^{2}=\Delta\Delta^{*}$. The one-loop effective action is
given by
\begin{eqnarray}\label{Fb13}
\Gamma_{\mbox{\tiny{eff}}}^{(1)}[\sigma,\Delta,\Delta^{*};B]=-\frac{i}{2}\sum_{\tilde
q}{\mbox{Tr}}_{\mbox{\tiny{NG}}cfsx}~\ln[{{\cal{S}}_{\tilde
q}}^{-1}],
\end{eqnarray}
where ${\cal{S}}_{\tilde{q}}$ is defined in
(\ref{Fb8})-(\ref{Fb10}). In (\ref{Fb13}) the trace operation (Tr)
includes apart from a two-dimensional trace in the Nambu-Gorkov
space (NG), a trace over color ($c$), flavor ($f$), and spinor ($s$)
degrees of freedom as well as a trace over a four-dimensional
space-time coordinate ($x$). After performing the trace operation
over the NG,$c,f,$ and $s$ using the method described in
\cite{fayaz2010}, we arrive at
\begin{eqnarray}\label{Fb14}
\tilde{\Gamma}^{(1)}_{\mbox{\tiny{eff}}}(\bar{p})=\sum\limits_{\kappa\in\{r,g,b\}}\tilde{\Gamma}^{(1)/c}_{\mbox{\tiny{eff}}}(\bar{p}),
\end{eqnarray}
that includes the contribution of the blue ($b$), red ($r$), and
green ($g$) quarks
\begin{eqnarray}\label{Fb15}
\tilde{\Gamma}^{(1)/\mbox{\tiny{b}}}_{\mbox{\tiny{eff}}}(\bar{p})&=&-i\sum\limits_{\tilde{q}\in\{0,1\}}\ln{\det}_{x}[\{({E}_{\tilde{q}}
+\mu)^2-p_{0}^2\}\{({E}_{\tilde{q}}-\mu)^2-p_{0}^2\}],
\nonumber\\
\sum\limits_{c\in\{r,g\}}\tilde{\Gamma}^{(1)/c}_{\mbox{\tiny{eff}}}(\bar{p})&=&-2i\sum\limits_{\tilde{q}
\in\{+\frac{1}{2},-\frac{1}{2}\}}
\ln{\det}_{x}[({{E}^{(+1)}_{\tilde{q}}}^{2}-{p}_{0}^2)(
{{E}^{(-1)}_{\tilde{q}}}^{2}-{p}_{0}^2)].
\end{eqnarray}
Here, $\bar{p}$ is a modified four-momentum defined by
\cite{fukushima2009}
\begin{eqnarray}\label{Fb16}
\begin{array}{rclccrcl}
{\bar{p}}_{\tilde q\neq 0}^{\mu}&=&(p_{0},0,\frac{\tilde q}{|\tilde
q|}\sqrt{2|\tilde q\tilde{e} B| n},p_{3}),&\qquad&\mbox{for}&{\tilde{q}}&=&1,\pm\frac{1}{2},\\
{\bar{p}}_{\tilde{q}=0}^{\mu}&=&(p_{0},{\mathbf{p}}),&\qquad&\mbox{for}&\tilde{q}&=&0,
\end{array}
\end{eqnarray}
with ${\mathbf{p}}\equiv (p_{1},p_{2},p_{3})$. In (\ref{Fb15}),
$E_{\tilde{q}}$ are given by the dispersion relations corresponding
to the neutral and charged particles \cite{fayaz2010}
\begin{eqnarray}\label{Fb17}
\begin{array}{rclccrcl}
E_{\tilde{q}}&=&\sqrt{2
|\tilde{q}\tilde{e}B|n+p_{3}^{2}+\sigma^{2}},&\qquad&\mbox{for}&{\tilde{q}}&=&1,\pm\frac{1}{2},\\
E_{0}&=&\sqrt{{\mathbf{p}}^{2}+\sigma^{2}},&\qquad&\mbox{for}&{\tilde{q}}&=&0.\\
\end{array}
\end{eqnarray}
Moreover, we have $E_{\tilde{q}}^{(\pm
1)}=\sqrt{(E_{\tilde{q}}\pm\mu)+|\Delta|^{2}}$. Performing the
remaining determinant in the coordinate space, a space-time volume
factor ${\cal{V}}$ arises. Combining then the resulting expression
with the tree-level part of the effective action from (\ref{Fb12}),
the effective action of the theory can be given in terms of the
effective thermodynamic (mean field) potential
$\Omega_{\mbox{\tiny{eff}}}$ as
$\Gamma_{\mbox{\tiny{eff}}}=-{\cal{V}}\Omega_{\mbox{\tiny{eff}}}$.
Introducing now discrete Matsubara frequencies by replacing $p_{0}$
with $p_{0}=i\omega_{\ell}$, where
$\omega_{\ell}=\frac{\pi}{\beta}(2\ell+1)$ and $\beta\equiv T^{-1}$,
the one-loop effective potential
$\Omega_{\mbox{\tiny{eff}}}^{(1)}\equiv
-\frac{1}{V\beta}\tilde{\Gamma}_{\mbox{\tiny{eff}}}^{(1)}$ is first
given by\footnote{In imaginary time formulation, the
four-dimensional space-time volume ${\cal{V}}$ is replaced by
${\cal{V}}\to V\beta$, where $V$ is the three-dimensional space
volume and $\beta=1/T$ the compactification radius of imaginary time
coordinate.}
\begin{eqnarray}\label{Fb18}
\Omega^{(1)}_{\mbox{\tiny{eff}}}(\tilde{e}B,T,\mu)=-\frac{1}{V\beta}
\sum\limits_{\mathbf{p}}\sum\limits_{\ell=-\infty}^{\infty}
\sum\limits_{\kappa=\pm{1}} \bigg\{\sum\limits_{\tilde{q}\in\{0,1\}}
\ln{[\omega_{\ell}^{2}+({E}_{\tilde{q}}+\kappa\mu)^2]}
+2\sum\limits_{\tilde{q} \in\{+\frac{1}{2},-\frac{1}{2}\}}
\ln{[\omega_{\ell}^2+{E}^{(\kappa)2}_{\tilde{q}}]}\bigg\}.\nonumber\\
\end{eqnarray}
Converting then the logarithms into proper-time integrals over the
dimensionful variable $s$, and using the Poisson resummation formula
\begin{eqnarray}\label{Fb19}
\sum\limits_{\ell=-\infty}^{\infty}e^{-s(\frac{2\pi\ell}{\beta}+x)^{2}}=\frac{\beta}{2\sqrt{\pi
s}}[1+2\sum\limits_{\ell=1}^{\infty}}\cos{(x\beta\ell)~
e^{-\frac{\beta^{2}\ell^{2}}{4s}}],
\end{eqnarray}
to separate the resulting expression into a temperature dependent
and a temperature independent part, we arrive at the one-loop
effective potential of our model
\begin{eqnarray}\label{Fb20}
\Omega_{\mbox{\tiny{eff}}}^{(1)}(\tilde{e}B,T,\mu)&=&\frac{1}{2\sqrt{\pi}V}\sum\limits_{\mathbf{p}}\sum\limits_{\kappa=\pm{1}}\bigg\{\sum\limits_{\tilde{q}\in\{0,1\}}\int_{0}^{\infty}\frac{ds}{s^{\frac{3}{2}}}
e^{-s({E}_{\tilde{q}}+\kappa\mu)^2}+2\sum\limits_{\tilde{q}
\in\{+\frac{1}{2},-\frac{1}{2}\}}\int_{0}^{\infty}\frac{ds}{s^{\frac{3}{2}}}e^{-s{E}^{(\kappa)2}_{\tilde{q}}}\bigg\}\nonumber\\
&&\times
[1+2\sum\limits_{\ell=1}^{\infty}(-1)^{\ell}~e^{-\frac{\beta^{2}\ell^{2}}{4s}}],
\end{eqnarray}
where $E_{\tilde{q}}$ are defined in (\ref{Fb17}). Replacing at this
stage the discrete sum over momenta with continuous integrations
over momenta by making use of
\begin{eqnarray}\label{Fb21}
\frac{1}{V}\sum_{\mathbf{p}}f(\bar{\mathbf{p}}_{\tilde{q}=0})\to\int
\frac{d^{3}p}{(2\pi)^{3}}~f(\mathbf{p}),
\end{eqnarray}
for neutral, and
\begin{eqnarray}\label{Fb22}
\frac{1}{V}\sum_{\mathbf{p}}f(\bar{\mathbf{p}}_{\tilde{q}\neq 0})\to
|\tilde{q}\tilde{e}B|\sum\limits_{n=0}^{+\infty}\alpha_{n}\int
_{-\infty}^{+\infty} \frac{dp_{3}}{8\pi^{2}}~f(n,p_{3}),
\end{eqnarray}
for charged particles,\footnote{Only charged particles interact with
the external magnetic field.} and eventually adding the tree level
part of the effective potential to the resulting expression, the
full mean field effective potential at one-loop level is given
by\footnote{In \cite{fayaz2010}, the same effective potential
(\ref{Fb23}) was determined using a different method (see
(3.19)-(3.22) in \cite{fayaz2010}).}
\begin{eqnarray}\label{Fb23}
\lefteqn{\hspace{-1cm}
\Omega_{\mbox{\tiny{eff}}}(\tilde{e}B,T,\mu)=\frac{\sigma^{2}}{4G_{S}}+\frac{|\Delta|^{2}}{4G_{D}}+\frac{B^{2}}{2}+\frac{1}{2\sqrt{\pi}}\sum\limits_{\kappa=\pm{1}}\bigg\{\int\frac{d^{3}p}{(2\pi)^{3}}\int_{0}^{\infty}
\frac{ds}{s^{\frac{3}{2}}}e^{-s({E}_{0}+\kappa\mu)^2}
}\nonumber\\
&&+\tilde{e}B\sum\limits_{n=0}^{\infty}\alpha_{n}\int_{0}^{\infty}\frac{ds}{s^{\frac{3}{2}}}
\int_{0}^{\infty}\frac{dp_{3}}{4\pi^{2}}(e^{-s({E}_{+1}+\kappa\mu)^2}
+2e^{-s{E}^{(\kappa)2}_{|\pm
1/2|}})\bigg\}[1+2\sum\limits_{\ell=1}^{\infty}(-1)^{\ell}~
e^{-\frac{\beta^{2}\ell^{2}}{4s}}].
\end{eqnarray}
Note that in (\ref{Fb22}) as well as (\ref{Fb23}), $n$ denotes the
discrete Landau levels and $\alpha_{n}=2-\delta_{n0}$ is introduced
to consider the fact that Landau levels with $n>0$ are doubly
degenerate \cite{ferrer2006, shovkovy2007}. Moreover, we have used
$E_{+1/2}=E_{-1/2}$. In the next section, we will use (\ref{Fb23})
to determine numerically the chiral and diquark gaps and to present
the complete phase structure of the magnetized two-flavor
superconducting NJL model at finite $T$ and $\mu$.
\section{Numerical results}
\setcounter{equation}{0} \noindent In the previous section, we have
determined the effective potential (\ref{Fb23}) of two-flavor NJL
model including meson and diquark condensates at finite temperature,
chemical potential and in the presence of constant magnetic fields
in the mean field approximation at one-loop level. It is the purpose
of this paper to have a complete understanding on the effect of
these external parameters on the quark matter in the 2SC phase. This
will complete our analysis in \cite{fayaz2010}, where only the
effect of $\mu,\tilde{e}B$ was considered at $T=0$. We start this
section with presenting the numerical results on the $\mu,T$ and
$\tilde{e}B$ dependence of the chiral and diquark condensates. We
then continue with exploring the $T-\mu$, $T-\tilde{e}B$, and
$\mu-\tilde{e}B$ phase diagrams for fixed values of $\tilde{e}B$,
$\mu$ and $T$, respectively. Before presenting our results, we will
fix, in the subsequent paragraphs, our notations and describe our
numerical method.
\par
To determine the chiral and diquark gaps, the thermodynamic
potential $\Omega_{\mbox{\tiny{eff}}}$ from (\ref{Fb23}) is to be
minimized. To solve the corresponding gap equations
\begin{eqnarray}\label{D1}
\frac{\partial\Omega_{\mbox{\tiny{eff}}}(\sigma,\Delta;T,\mu,\tilde{e}B)}{\partial\sigma}\bigg|_{\sigma_{B},\Delta_{B}}=0,\qquad
\mbox{and}\qquad
\frac{\partial\Omega_{\mbox{\tiny{eff}}}(\sigma,\Delta;T,\mu,\tilde{e}B)}{\partial\Delta}\bigg|_{\sigma_{B},\Delta_{B}}=0,
\end{eqnarray}
numerically, we have to fix the free parameters of the model. Our
specific choice of parameters is \cite{huang2002,fayaz2010}
\begin{eqnarray}\label{D2}
\Lambda=0.6533~\mbox{GeV},~G_{S}=5.0163~\mbox{GeV}^{-2},\qquad\mbox{and}\qquad
G_{D}=\frac{3}{4}G_{S},
\end{eqnarray}
where $\Lambda$ is the momentum cutoff and $G_{S}$ as well as
$G_{D}$ are the chiral and diquark couplings. Using this special set
of parameters, we have shown in \cite{fayaz2010}, that at $T=0$ no
mixed phase characterized, with $(\sigma_{B}\neq 0,\Delta_{B}\neq
0)$, will appear. The same feature persists at finite $T$. Moreover,
for vanishing magnetic field and at zero temperature, the parameters
in (\ref{D2}) yield the meson mass $\sigma_{0}\simeq 323.8$ MeV at
$\mu=250$ MeV, and the diquark mass $\Delta_{0}\simeq 126$ MeV at
$\mu=460$ MeV \cite{fayaz2010}.\footnote{Although our free
parameters $\Lambda, G_{D}$, and $G_{S}$ coincide with the
parameters used in \cite{huang2002}, the numerical value of
$\sigma_{0}$ is different from what is reported in \cite{huang2002}.
The reason for this difference is apparently in the choice of the
cutoff function. Whereas in \cite{huang2002} a sharp momentum cutoff
is used, we have used smooth cutoff function (\ref{D3}) to perform
the momentum integrations numerically.} To perform the momentum
integration over ${\mathbf{p}}$ and $p_{3}$ in (\ref{Fb23})
numerically, we have introduced, as in \cite{fayaz2010}, smooth
cutoff functions (form factors)
\begin{eqnarray}\label{D3}
f_{\Lambda}=\frac{1}{1+\exp\left(\frac{|{\mathbf{p}}|-\Lambda}{A}\right)},\qquad\mbox{and}\qquad
f_{\Lambda,B}^{n}=\frac{1}{1+\exp\left(\frac{\sqrt{p_{3}^{2}+2|\tilde{q}\tilde{e}B|n}-\Lambda}{A}\right)},
\end{eqnarray}
that correspond to neutral and charged particles,
respectively.\footnote{In (\ref{Fb23}), the integrals proportional
to $\tilde{e}B$  and including a summation over Landau levels $n$
arise from charged quarks with charges $\tilde{q}=\pm
\frac{1}{2},+1$.} In (\ref{D3}), $A$ is a free parameter and is
chosen to be $A=0.05\Lambda$. Similar smooth cutoff function (form
factor) is also used in \cite{warringa2007}. Here, as in
\cite{warringa2007}, the free parameter $A$ determines the sharpness
of the cutoff scheme. At this stage, let us notice that the
solutions of (\ref{D1}) are in general ``local'' minima of the
theory. Keeping $(\sigma,\Delta)\neq (0,0)$ and looking for global
minima of the system described by complete
$\Omega_{\mbox{\tiny{eff}}}(\sigma,\Delta;T,\mu,\tilde{e}B)$ from
(\ref{Fb23}), it turns out that in the regime $\mu\in[0,800]$ MeV,
$T\in [0,250]$ MeV and $\tilde{e}B\in [0,0.8]$ GeV$^{2}$, the system
exhibits two ``global '' minima. They are given by $(\sigma_{B}\neq
0, \Delta_{B}=0)$ and $(\sigma_{B}=0,\Delta_{B}\neq 0)$. We will
denote the regime characterized by these two global minima by the
$\chi$SB and the CSC phases, respectively. According to the above
descriptions, in order to determine the chiral and diquark
condensates, we will use instead of the gap equations (\ref{D1}),
\begin{eqnarray}\label{D4}
\frac{\partial\Omega_{\mbox{\tiny{eff}}}(\sigma,\Delta_{B}=0;T,\mu,\tilde{e}B)}{\partial\sigma}\bigg|_{\sigma_{B}}=0,\qquad
\end{eqnarray}
in the $\chi$SB phase and
\begin{eqnarray}\label{D5}
\frac{\partial\Omega_{\mbox{\tiny{eff}}}(\sigma_{B}=0,\Delta;T,\mu,\tilde{e}B)}{\partial\Delta}\bigg|_{\Delta_{B}}=0,
\end{eqnarray}
in the CSC phase to simplify our computations. As it turns out,
apart from the $\chi$SB and the CSC phases, there is also a normal
phase characterized by $(\sigma_{B}=0,\Delta_{B}=0)$. Following the
method presented in \cite{sato1997}, we will determine the critical
lines of first order transition between the $\chi$SB phase and the
normal phase using
\begin{eqnarray}\label{D6}
\frac{\partial
\Omega_{\mbox{\tiny{eff}}}(\sigma,\Delta_{B}=0)}{\partial
\sigma}\bigg|_{\sigma_{B}}=0,
\qquad\mbox{and}\qquad\Omega_{\mbox{\tiny{eff}}}(\sigma_{B},\Delta_{B}=0)=\Omega_{\mbox{\tiny{eff}}}(\sigma_{B}=0,\Delta_{B}=0).
\end{eqnarray}
Similarly, the first order phase transition between $\chi$SB and the
CSC phases is determined by solving
\begin{eqnarray}\label{D9}
\frac{\partial
\Omega_{\mbox{\tiny{eff}}}(\sigma,\Delta_{B}=0)}{\partial
\sigma}\Bigg|_{\sigma_{B}}=0,\qquad \frac{\partial
\Omega_{\mbox{\tiny{eff}}}(\sigma_{B}=0,\Delta)}{\partial
\Delta}\Bigg|_{\Delta_{B}}=0,
\end{eqnarray}
and
\begin{eqnarray}
\Omega_{\mbox{\tiny{eff}}}(\sigma_{B}=0,\Delta_{B})=\Omega_{\mbox{\tiny{eff}}}(\sigma_{B},\Delta_{B}=0),
\end{eqnarray}
simultaneously \cite{sato1997}. The second order critical lines
between the $\chi$SB as well as CSC phase and the normal phase will
be also determined by the method described in \cite{sato1997} (see
Eq. (2.35) in \cite{sato1997}): To determine the second line between
the $\chi$SB and the normal phase, we solve
\begin{eqnarray}\label{D9-a}
\lim_{\sigma^{2}\rightarrow 0}\frac{\partial
\Omega_{\mbox{\tiny{eff}}}(\sigma,\Delta=0)}{\partial \sigma^{2}}=0.
\end{eqnarray}
The second line between the CSC and the normal phase is then
determined by solving
\begin{eqnarray}\label{D9-b}
\lim_{\Delta^{2}\rightarrow 0}\frac{\partial
\Omega_{\mbox{\tiny{eff}}}(\sigma=0,\Delta)}{\partial \Delta^{2}}=0.
\end{eqnarray}
The same method is also used in \cite{inagaki2003} to find the
second order phase transition between the $\chi$SB and the normal
phase (see page 9 in \cite{inagaki2003}). To make sure that after
the second order phase transition the global minima of the effective
potential are shifted to $\sigma=0$ in (\ref{D9-a}) and to
$\Delta=0$ in (\ref{D9-b}), and in order to avoid instabilities, an
analysis similar to \cite{berges1998} is also performed. Same method
is also used in \cite{fayaz2010} to determine second order critical
line between the CSC and the normal phase.
\subsection{The $(\mu,T,\tilde{e}B)$ dependence of chiral and diquark
condensates}\label{IIIA}
\subsubsection{The $\mu$-dependence of $\sigma_{B}$ and
$\Delta_{B}$}\label{IIIA1}
\par
\begin{figure}[hbt]
\includegraphics[width=5.3cm,height=4cm]{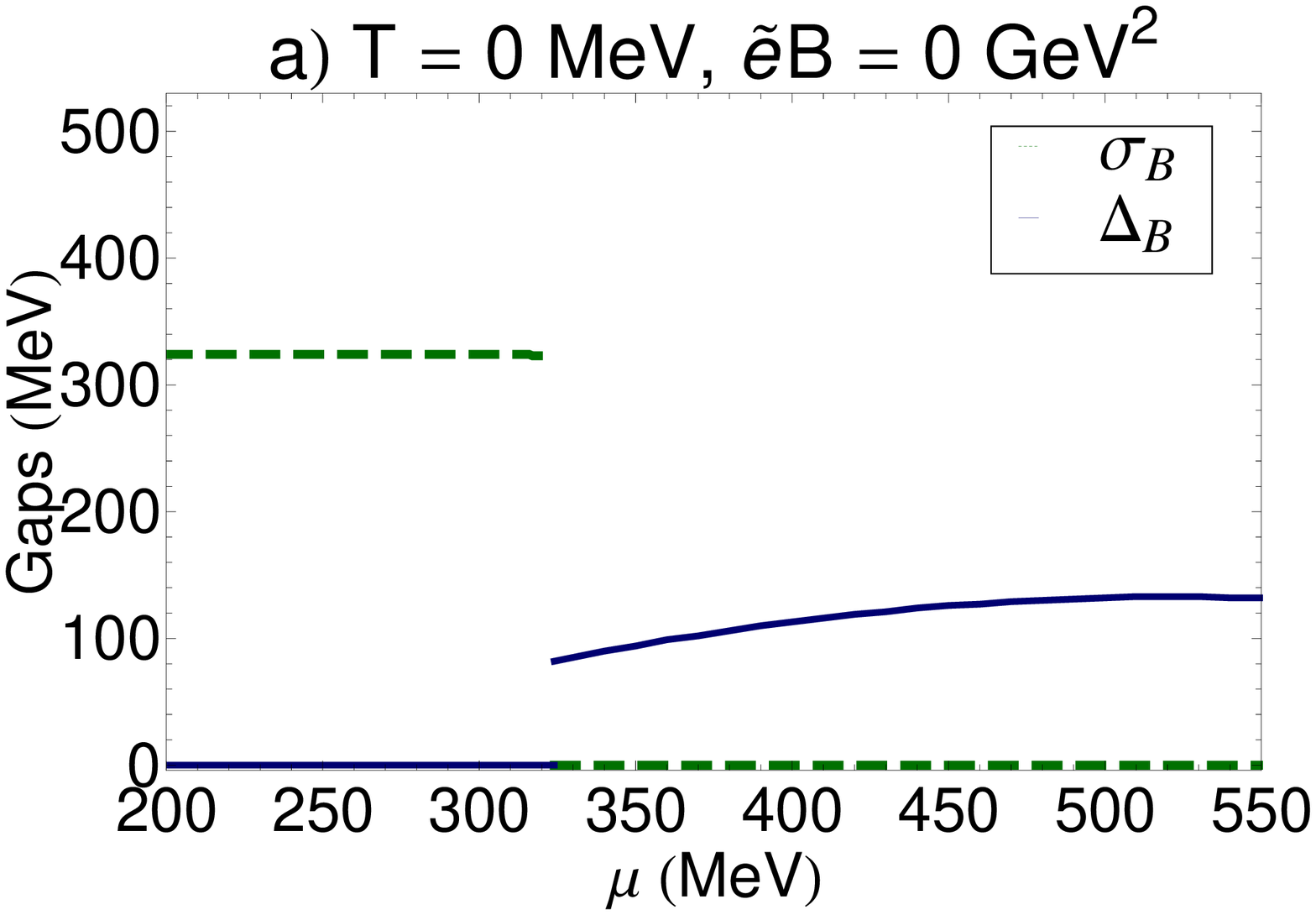}
\hspace{0.2cm}
\includegraphics[width=5.3cm,height=4cm]{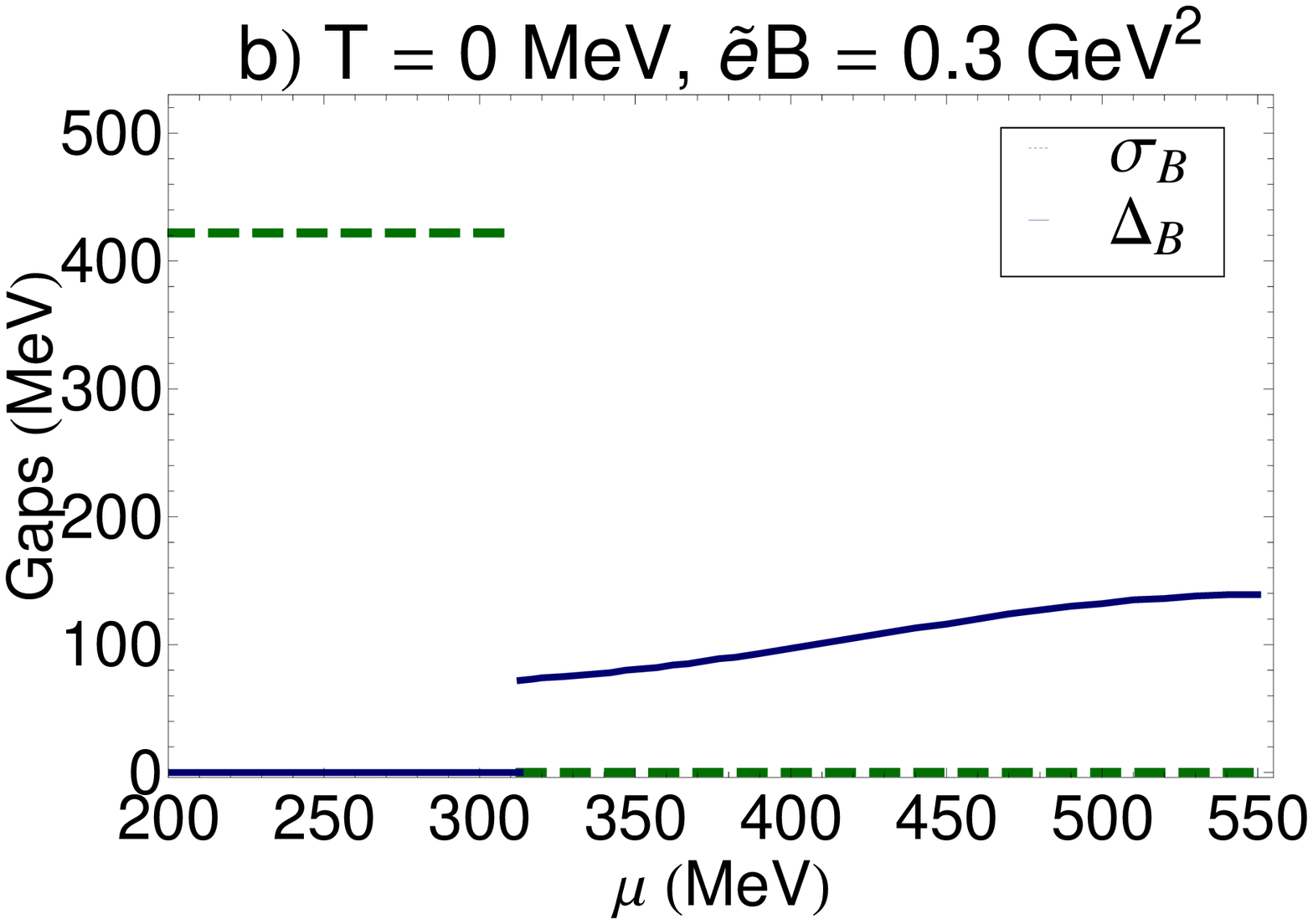}
\hspace{0.2cm}
\includegraphics[width=5.3cm,height=4cm]{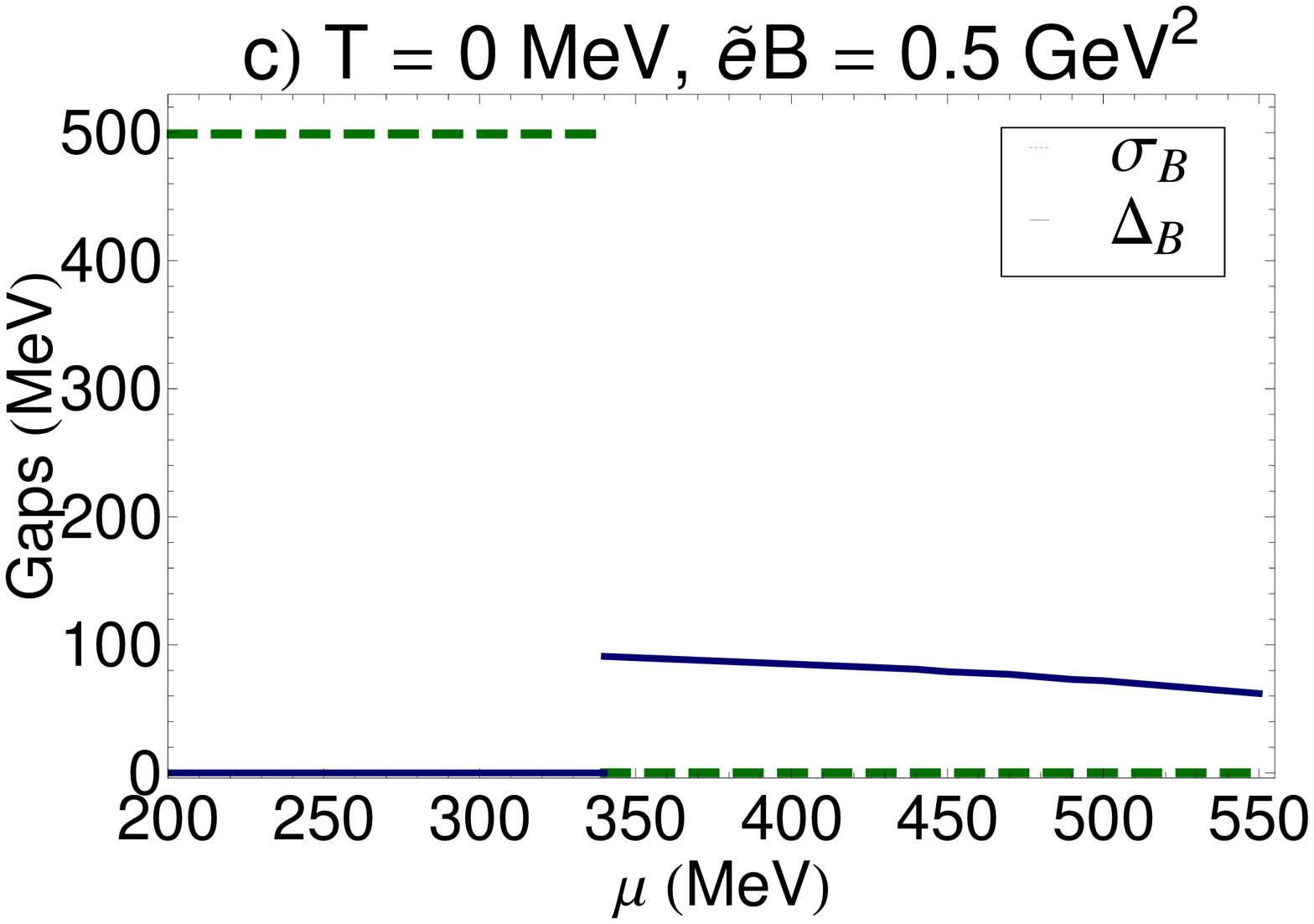}
 \caption{The $\mu$-dependence of $\sigma_{B}$ and $\Delta_{B}$ is demonstrated for $T=0$ MeV and $\tilde{e}B=0,0.3,0.5$ GeV$^{2}$ in (a), (b) and (c), respectively.
 The transitions from $\chi$SB to the CSC phase is of first order,
 the transitions from the CSC to the normal phase is second order.
 }\label{fig0}
\end{figure}

\begin{figure}[hbt]
\includegraphics[width=5.3cm,height=4cm]{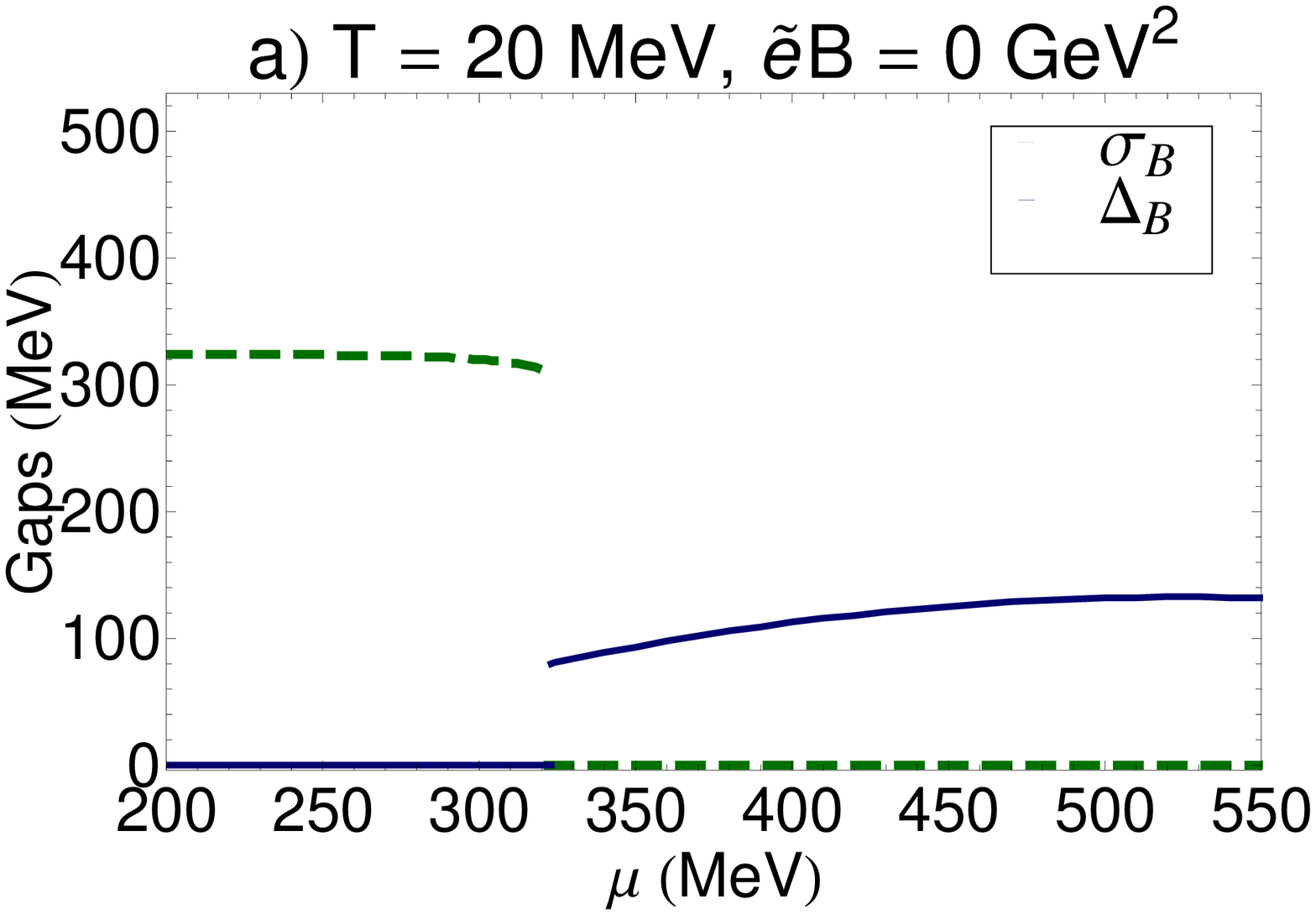}
\hspace{0.2cm}
\includegraphics[width=5.3cm,height=4cm]{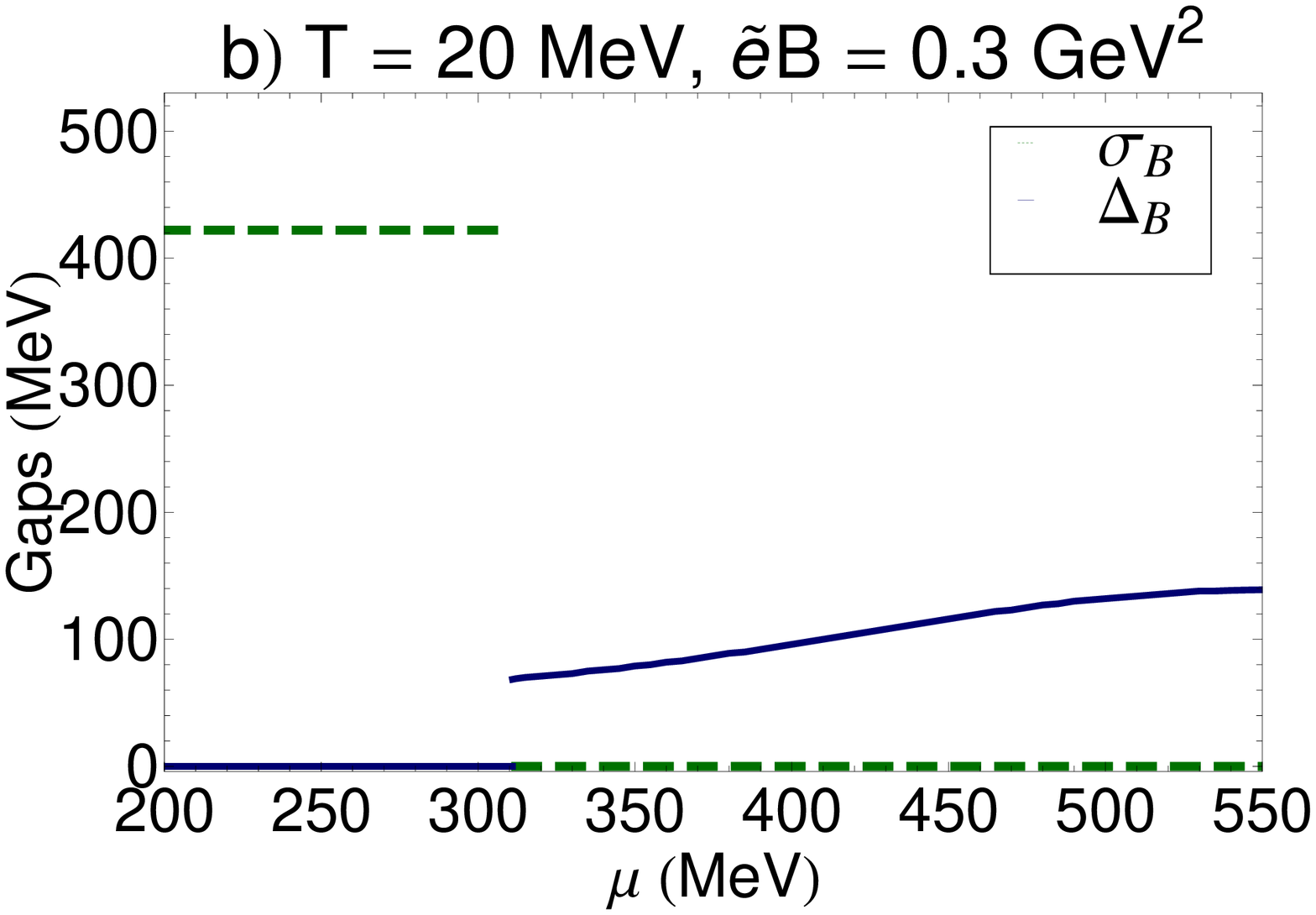}
\hspace{0.2cm}
\includegraphics[width=5.3cm,height=4cm]{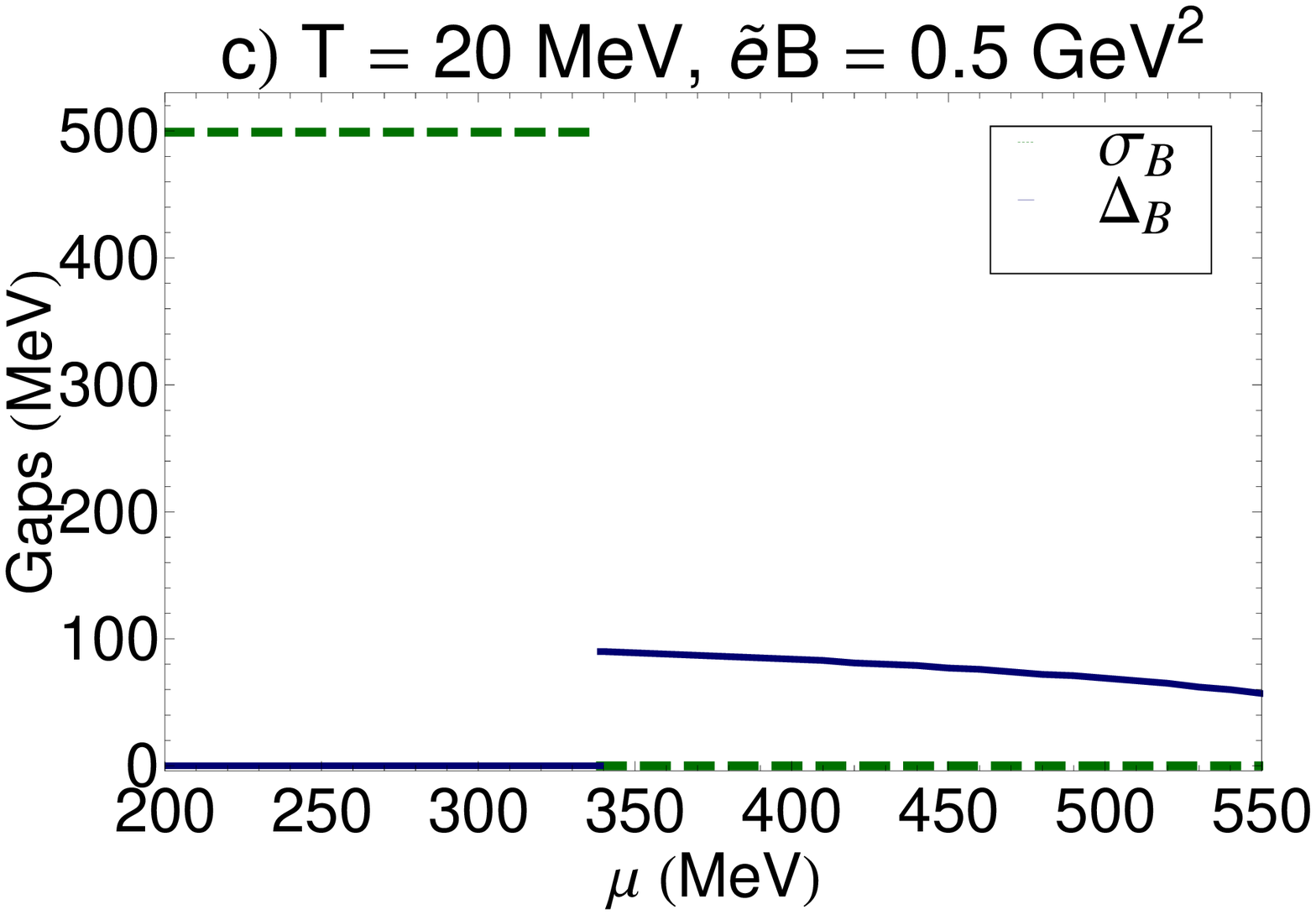}
 \caption{The $\mu$-dependence of $\sigma_{B}$ and $\Delta_{B}$ is demonstrated for $T=20$ MeV and $\tilde{e}B=0,0.3,0.5$ GeV$^{2}$ in (a), (b) and (c), respectively.
 The transitions from $\chi$SB to the CSC phase is of first order,
 the transitions from the CSC to the normal phase is second order.
}\label{fig1}
\end{figure}

\begin{figure}[hbt]
\includegraphics[width=5.3cm,height=4cm]{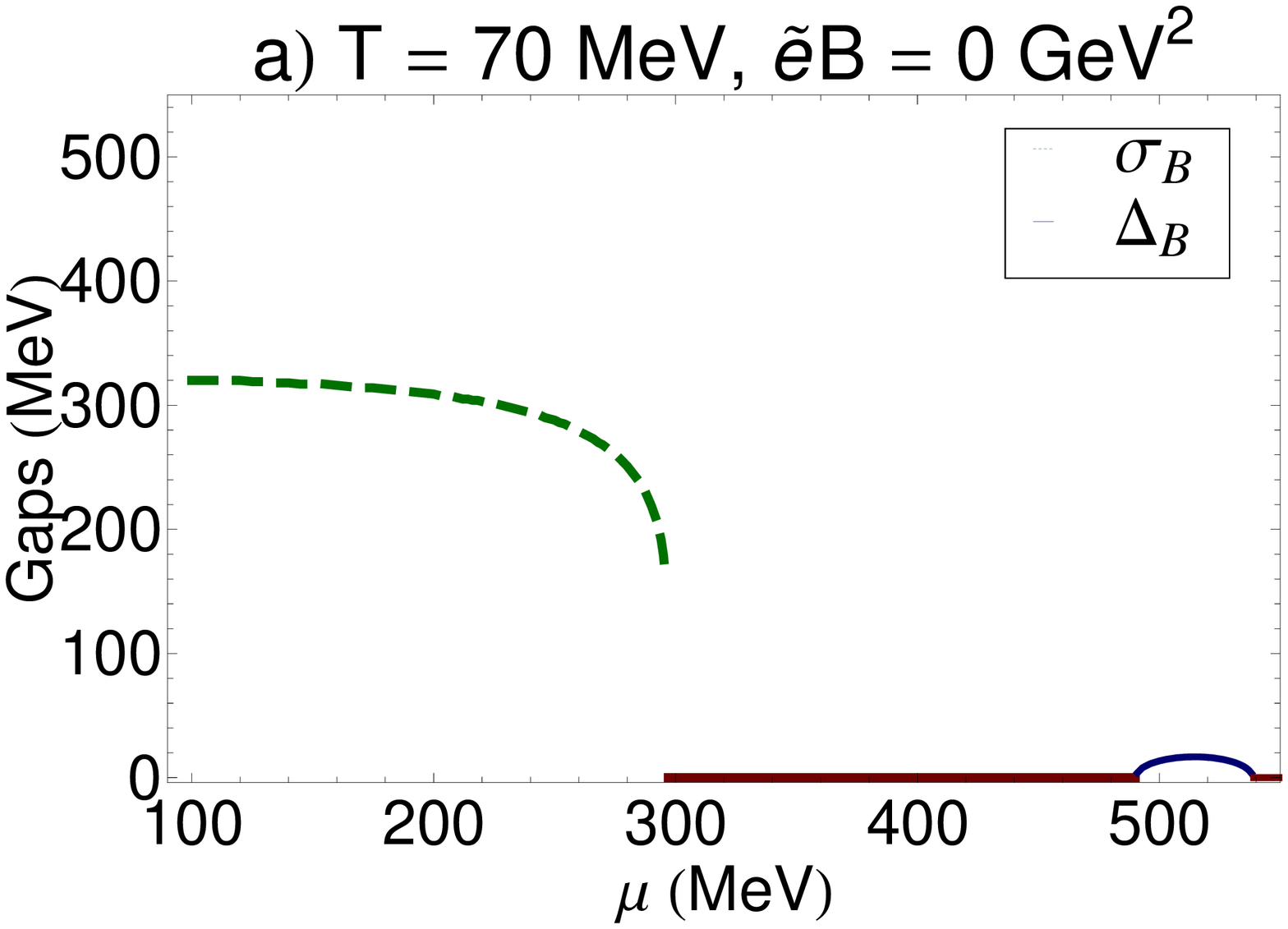}
\hspace{0.2cm}
\includegraphics[width=5.3cm,height=4cm]{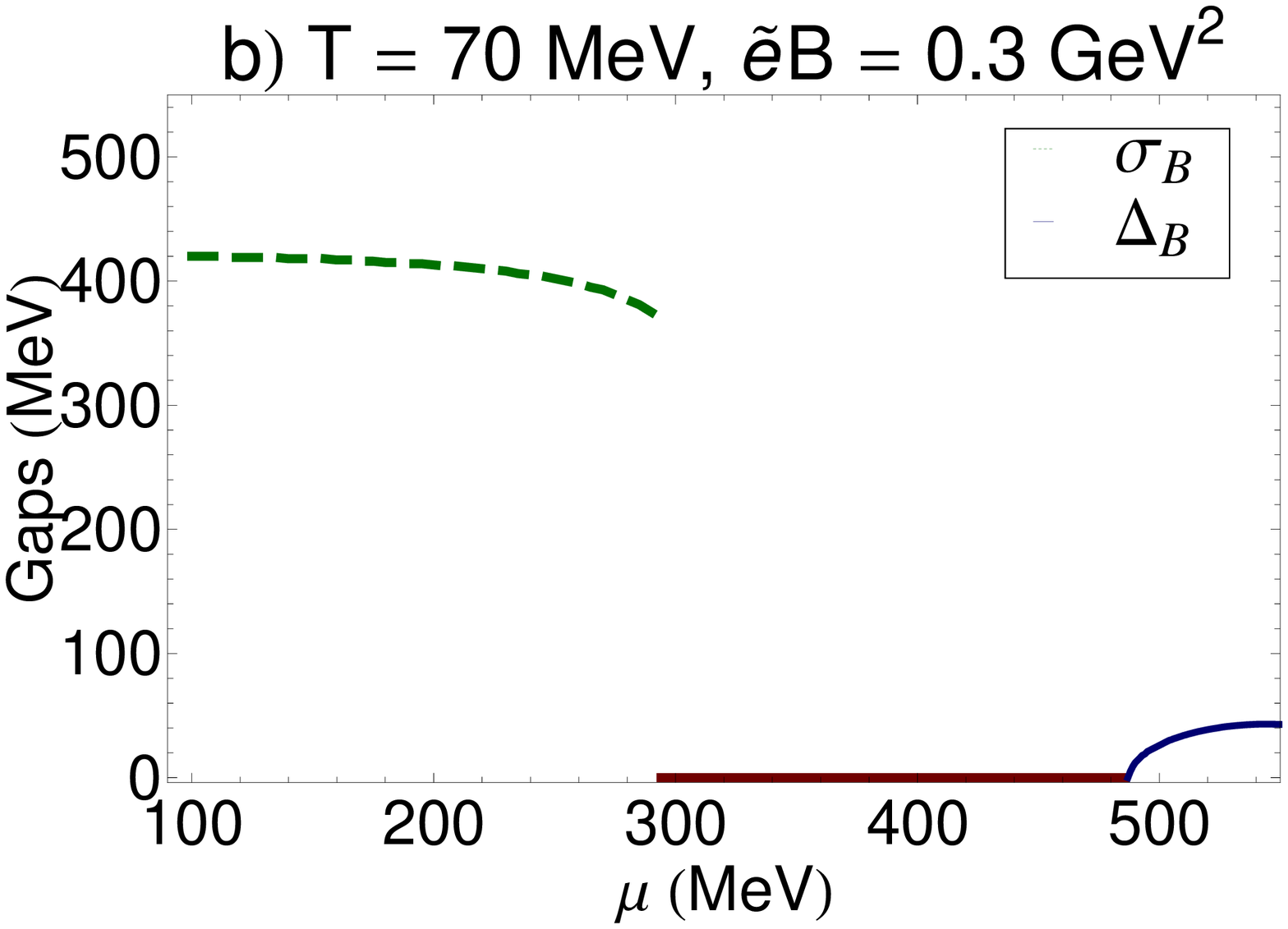}
\hspace{0.2cm}
\includegraphics[width=5.3cm,height=4cm]{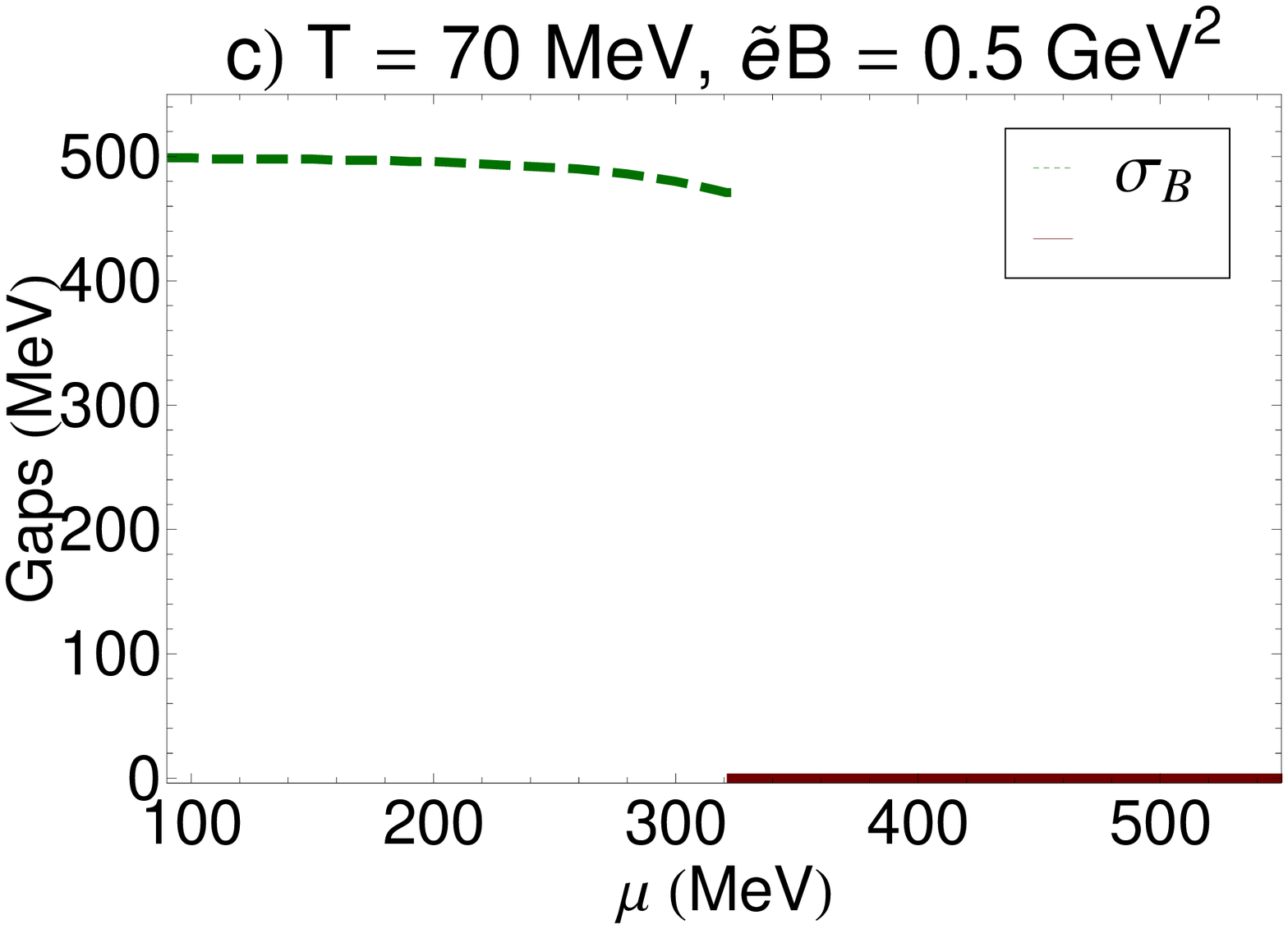}
\caption{The $\mu$-dependence of $\sigma_{B}$ and $\Delta_{B}$ is
demonstrated for $T=70$ MeV and $\tilde{e}B=0,0.3$ GeV$^{2}$ in (a)
and (b), respectively. While the transitions from $\chi$SB to the
normal phase (solid red line) is of first order, the transitions
from the normal to the CSC phase is of second order. c) The
$\mu$-dependence of $\sigma_{B}$ is demonstrated for $T=70$ MeV and
$\tilde{e}B=0.5$ GeV$^{2}$. The transition from the $\chi$SB to the
normal phase is of first order and no CSC phase appears at $\mu>330$
MeV.}\label{fig2}
\end{figure}

\begin{figure}[hbt]
\includegraphics[width=5.3cm,height=4cm]{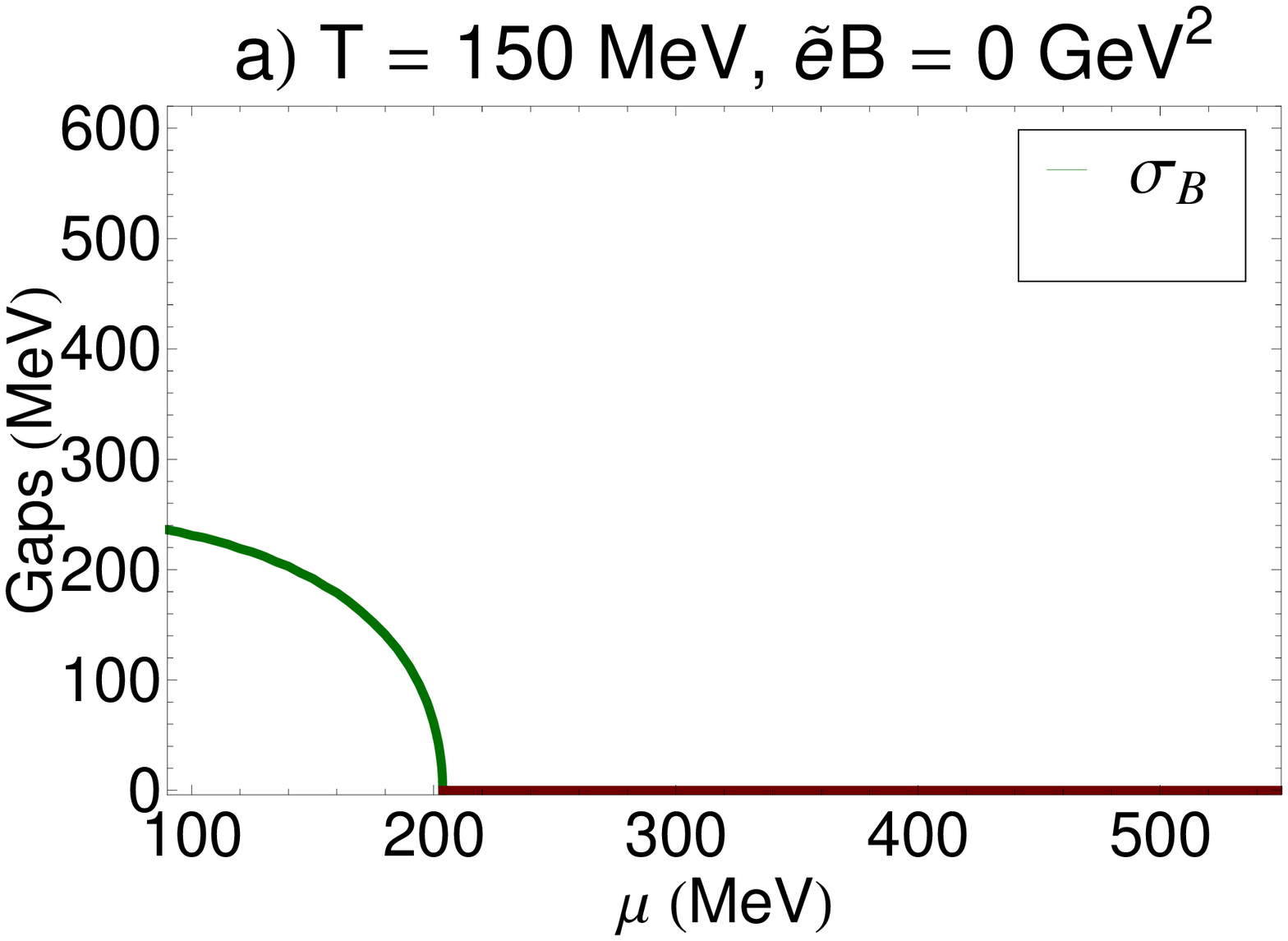}
\hspace{0.2cm}
\includegraphics[width=5.3cm,height=4cm]{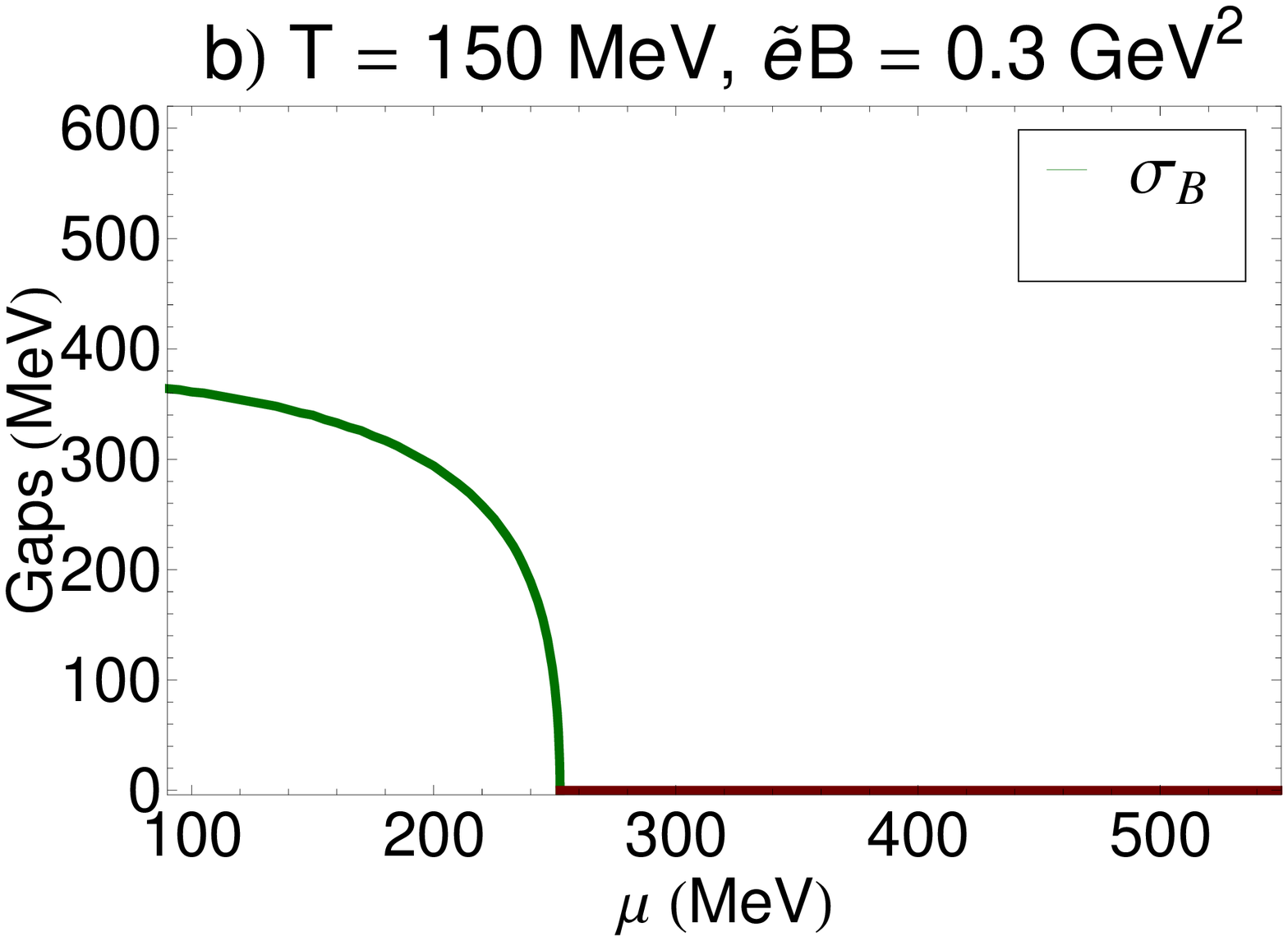}
\hspace{0.2cm}
\includegraphics[width=5.3cm,height=4cm]{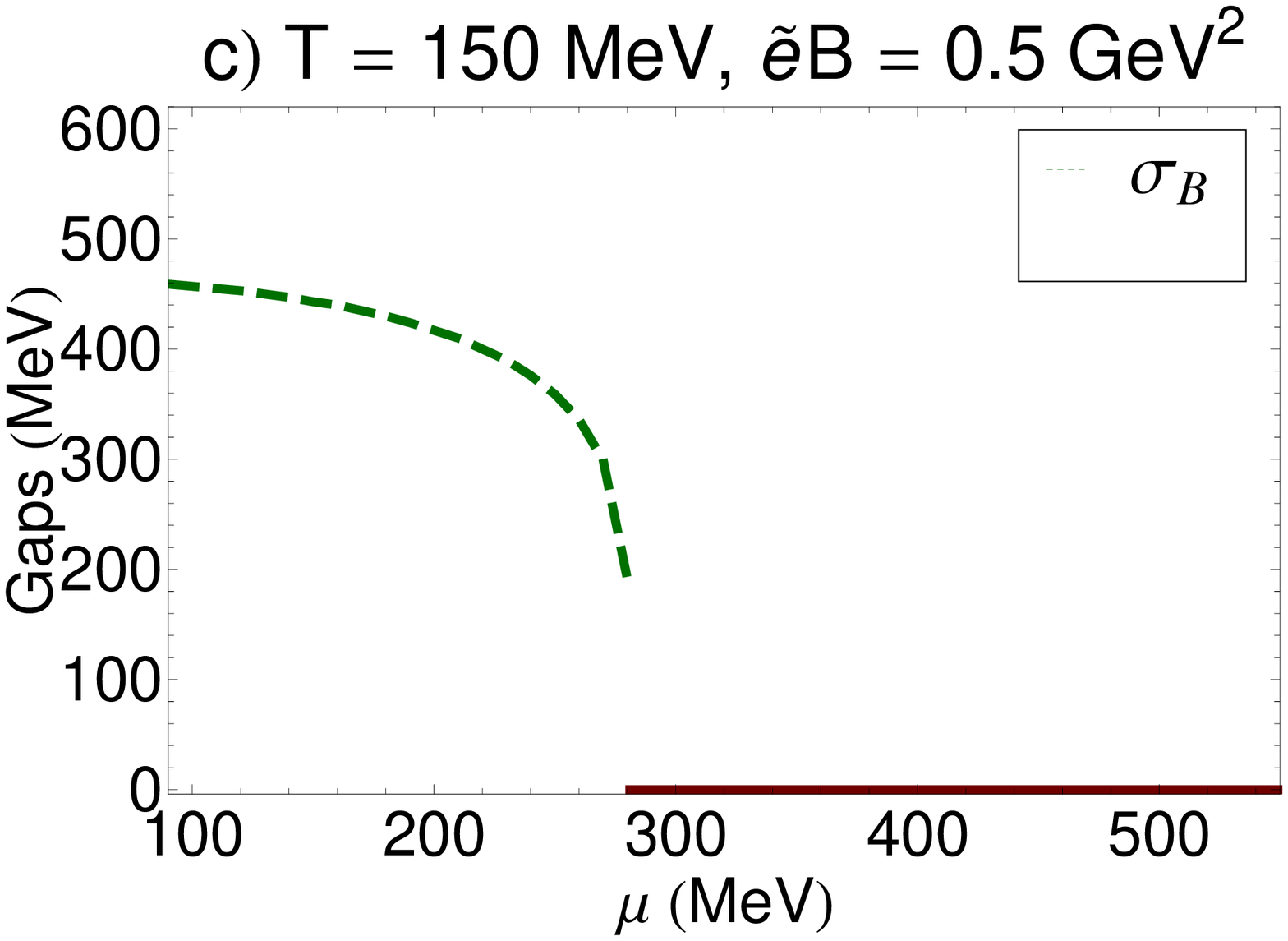}
 \caption{The $\mu$-dependence of $\sigma_{B}$ is demonstrated for $T=150$ MeV and $\tilde{e}B=0,0.3,0.5$ GeV$^{2}$ in (a)-(c).
 While for $\tilde{e}B=0,0.3$ GeV$^{2}$ the transition from $\chi$SB to the normal phase is of second order, for $\tilde{e}B=0.5$ GeV$^{2}$
 this transition is of first order. No CSC phase appears at $T=150$ MeV. The solid red line denotes the normal phase with $(\sigma_{B}=0,\Delta_{B}=0)$. }\label{fig3NN}
\end{figure}

\noindent
The $\mu$-dependence of both gaps at different temperatures,
$T=0,20,70,150$ MeV, and magnetic fields, $\tilde{e}B=0,0.3,0.5$
GeV$^{2}$, are demonstrated in Figs. \ref{fig0}-\ref{fig3NN}, panels
(a)-(c), respectively. The green dashed and solid lines denote the
$\sigma_{B}$ mass gaps. The red solid lines determine the normal
phase with $\sigma_{B}=\Delta_{B}=0$. The diquark gaps $\Delta_{B}$
are demonstrated with blue solid lines. Dashed (solid) lines denote
the first (second) order phase transitions.
\par\noindent
The plots in Figs. \ref{fig0}(a)-\ref{fig0}(c) show that at $T=0$
MeV, the magnetic field enhances the formation of chiral condensate
$\sigma_{B}$. The value of $\sigma_{B}$ is constant in $\mu$, with
$\mu<\mu_{c}$ and $\mu_{c}\simeq 320-350$ MeV, and increases with
$\tilde{e}B$. On the other hand, for small value of $\tilde{e}B<0.5$
GeV$^{2}$, $\Delta_{B}$ increases with $\mu$ in the regime
$\mu_{c}<\mu<600$ MeV [Figs. \ref{fig0}(a) and \ref{fig0}(b)].
Similar observation is also made in \cite{fayaz2010}, where the same
model as in the present paper is studied, and additionally the color
chemical potential  $\mu_{8}$ is assumed to be nonzero. Comparing
the diagrams of Fig. \ref{fig0} with the corresponding diagrams in
Fig. 7 of \cite{fayaz2010}, it turns out that $\mu_{8}$ has no
significant effect on the $\mu$-dependence of $\sigma_{B}$ and
$\Delta_{B}$ at $T=0$ MeV.\footnote{In \cite{shovkovy2003}, in
$\tilde{e}B=0$ case, the color chemical potential $\mu_{8}$ is shown
to be small and its effect is therefore neglected. See also
\cite{2SC}, where two-flavor magnetized color superconducting quark
matter with $\mu_{8}=0$ is studied.}  For $\tilde{e}B=0$ GeV$^{2}$,
the $\mu$-dependence of diquark mass gap in Fig. \ref{fig0}(a) can
be compared with the analytical result
\begin{eqnarray}\label{D9-c1}
\Delta_{0}^{2}={\cal{C}}_{2}(\Lambda^{2}-\mu^{2})\exp\left(-\frac{\Lambda^{2}}{\mu^{2}}
\left(\frac{1}{\hat{g}_{d}}-1\right)\right),
\end{eqnarray}
from \cite{fayaz2010}. In (\ref{D9-c1}),
${\cal{C}}_{2}=4e^{-3}\simeq 0.2$ and $\hat{g}_{d}\equiv
\frac{4G_{D}\Lambda^{2}}{\pi^{2}}$. Plugging the numerical value of
$\Lambda$ and $G_{D}$ from (\ref{D2}) in (\ref{D9-c1}) and plotting
the resulting expression in a $\Delta_{0}$ vs. $\mu$ diagram, the
result is in close agreement with the $\mu$-dependence of diquark
mass gap in Fig. \ref{fig0}(a) in the regime $320<\mu<600$ MeV. In
the same regime of $\mu$, the $\mu$-dependence of $\Delta_{0}$ from
(\ref{D9-c1}) agrees also with the well-known result from
\cite{alford2001}. In contrast, as it can be seen in Fig.
\ref{fig0}(c), for $\tilde{e}B=0.5$ GeV$^{2}$, $\Delta_{B}$
decreases with $\mu\in[350,600[$ MeV. This behavior is expected from
the analytical expression
\begin{eqnarray}\label{D9-c2}
\Delta_{B}^{2}=4(\Lambda_{B}^{2}-\mu^{2})\exp\left(-\frac{\Lambda^{2}}{\Lambda_{B}^{2}}\frac{1}{g_{d}}\right),
\end{eqnarray}
that is also computed in \cite{fayaz2010} using an appropriate LLL
approximation.\footnote{Later we will see that $\tilde{e}B\simeq
0.5$ GeV$^{2}$ is strong enough to justify LLL approximation.} In
(\ref{D9-c2}), $\Lambda_{B}\equiv \sqrt{\tilde{e}B}$ and
$g_{d}\equiv \frac{G_{D}\Lambda^{2}}{\pi^{2}}$. Plugging the
numerical values of $\Lambda$ and $G_{D}$ from (\ref{D2}) in
(\ref{D9-c2}), and plotting the resulting expression in a
$\Delta_{B}$ vs. $\mu$ diagram for $\tilde{e}B=0.5$ GeV$^{2}$, it
turns out that $\Delta_{B}$ decreases with $\mu$ in the regime
$350<\mu<600$ MeV, as it is shown in Fig. \ref{fig0}(c). Similar
expression as (\ref{D9-c2}) is also derived in \cite{manuel2006} for
the diquark mass gap in a three-flavor CFL model in the presence of
strong magnetic field, using the LLL approximation.
\par
In Figs. \ref{fig1}(a)-\ref{fig1}(c) the $\mu$-dependence of
$\sigma_{B}$ and $\Delta_{B}$ is plotted at $T=20$ MeV and for
$\tilde{e}B=0,0.3,0.5$ GeV$^{2}$. Comparing with the results from
Figs. \ref{fig0}(a)-\ref{fig0}(c), it turns out that increasing
temperature up to $T=20$ MeV, has no significant effects on the
results of $T=0$ MeV.
This is in contrast with the situation at $T=70$ MeV. In Figs.
\ref{fig2}(a)-\ref{fig2}(c), the $\mu$-dependence of $\sigma_{B}$
and $\Delta_{B}$ are plotted at $T=70$ MeV and for
$\tilde{e}B=0,0.3,0.5$ GeV$^{2}$. As it turns out, in the regime
$\mu<\mu_{c}$, the chiral condensate $\sigma_{B}$ increases with the
magnetic field. The diquark condensate appears in the regime
$\mu\in[480,600[$ MeV for $\tilde{e}B=0,0.3$ GeV$^{2}$ [Figs.
\ref{fig2}(a) and \ref{fig2}(b)]. For $\tilde{e}B=0.5$ GeV$^{2}$ in
Fig. \ref{fig2}(c), however, no diquark condensate appears in the
relevant regime $\mu< 600$ MeV. Moreover, comparing Figs. \ref{fig1}
and \ref{fig2}, we notice that at $T=70$ MeV, in Fig. \ref{fig2}, in
contrast to the situation at $T=20$ MeV, in Fig. \ref{fig1}, the
first order transition from the $\chi$SB to the normal phase does
not occur over the CSC phase; in Fig. \ref{fig2}, for $\tilde{e}B<
0.5$ GeV$^{2}$, there is a first order phase transition from the
$\chi$SB to the normal phase, then a second order phase transition
occurs from the normal to the CSC phase. For $\tilde{e}B=0.5$
GeV$^{2}$, the phase transition from the $\chi$SB to the normal
phase is of first order but no CSC phase appears in the regime
$\mu<600$ MeV.
\par
\par
At higher temperature, as it is demonstrated for $T=150$ MeV in Fig.
\ref{fig3NN}, no diquark condensate appears at all in the relevant
regime $\mu<600$ MeV [Figs. \ref{fig3NN}(a)-\ref{fig3NN}(c)].
Moreover, whereas for $\tilde{e}B=0,0.3$ GeV$^{2}$, the transitions
from the $\chi$SB to the normal phase are of second order, for
$\tilde{e}B=0.5$ GeV$^{2}$, the $\chi$SB-normal phase transition
turns out to be of first order. The above observations suggest that
while the magnetic field catalyzes the formation of the chiral
condensate $\sigma_{B}$, the diquark condensate $\Delta_{B}$ is
suppressed in the presence of a constant magnetic field. However, as
it turns out, the latter feature depends on the strength of the
magnetic field. This will be shown explicitly in the next section,
where the effect of arbitrary magnetic fields on the formation of
$\sigma_{B}$ and $\Delta_{B}$ for a wide range of $\tilde{e}B\in
[0,0.8]$ GeV$^{2}$ will be explored.
\subsubsection{The $\tilde{e}B$-dependence of $\sigma_{B}$ and
$\Delta_{B}$}\label{IIIB}
\par\noindent The effect of external magnetic field on the formation of chiral
condensates is studied intensively in the literature (see e.g.
\cite{miransky1995, cosmology, condensed, particle}). As it is shown
in \cite{miransky1995}, in the presence of strong magnetic field,
the dynamics of the system is fully described by its dynamics in the
LLL. However, there are, to the best of our knowledge, no evidences
in the literature that fix quantitatively the strength of magnetic
field which is enough to justify a LLL approximation. In
\cite{fayaz2010}, we have answered to this question numerically.
First, we have determined the $\tilde{e}B$ dependence of the gaps in
a LLL approximation, and then, compared the analytical results with
the numerical ones including the effect of all Landau levels. It
turned out that by increasing the magnetic field from $\tilde{e}B=0$
GeV$^{2}$ to $\tilde{e}B=0.8$ GeV$^{2}$, first in the regime
$\tilde{e}B<0.45$ GeV$^{2}$, the mass gaps underly small van
Alfven-de Haas oscillations. Above a certain threshold magnetic
field $\tilde{e}B_{t}\simeq 0.45-0.50$ GeV$^{2}$, the dependence of
the condensates of $\tilde{e}B$ was linear. In this linear regime,
our numerical results were comparable with the analytical results
arising from the solution of the corresponding gap equations in the
LLL approximation.
\begin{figure}[hbt]
\includegraphics[width=8.2cm,height=6cm]{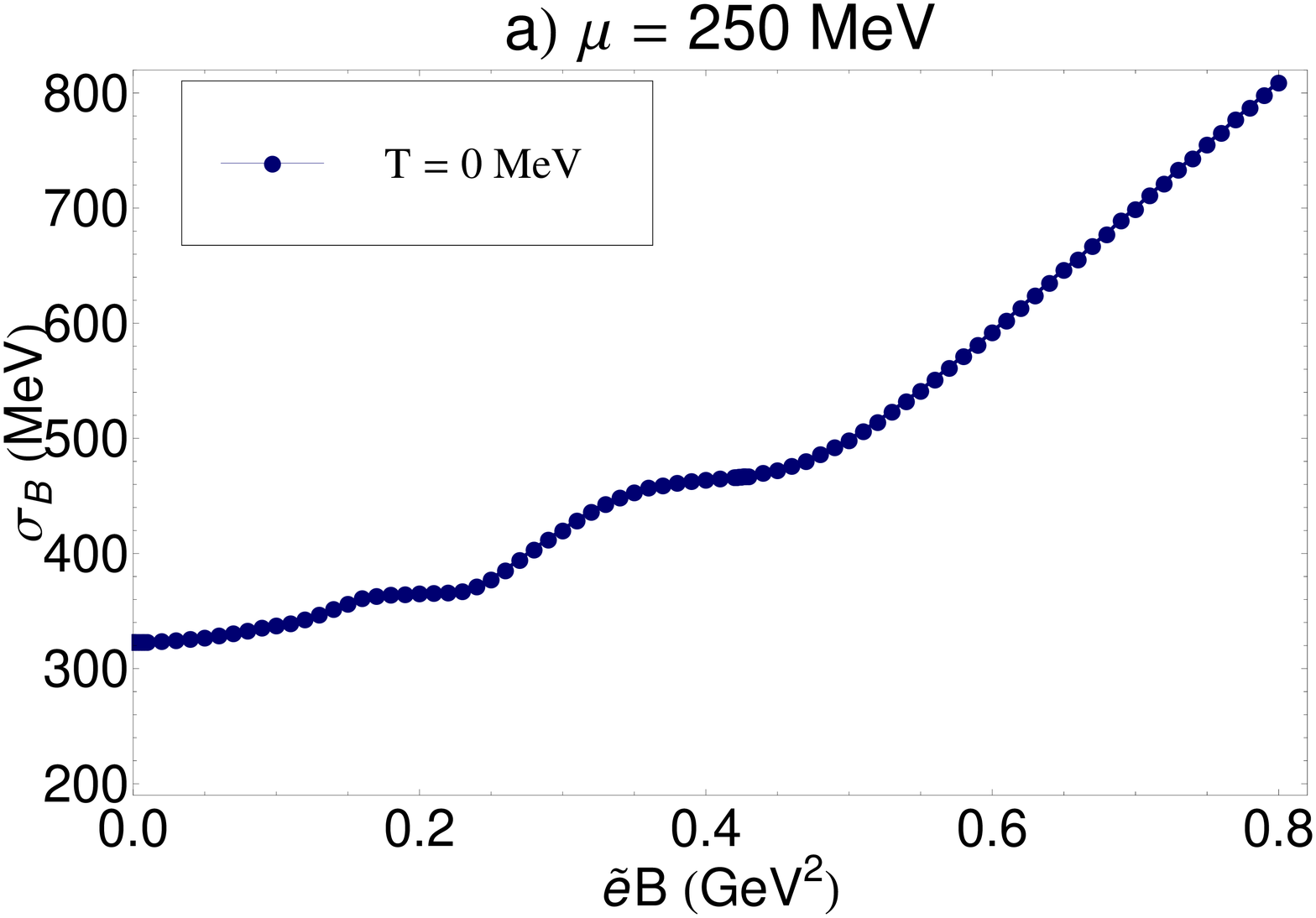}
\hspace{0.2cm}
\includegraphics[width=8.2cm,height=6cm]{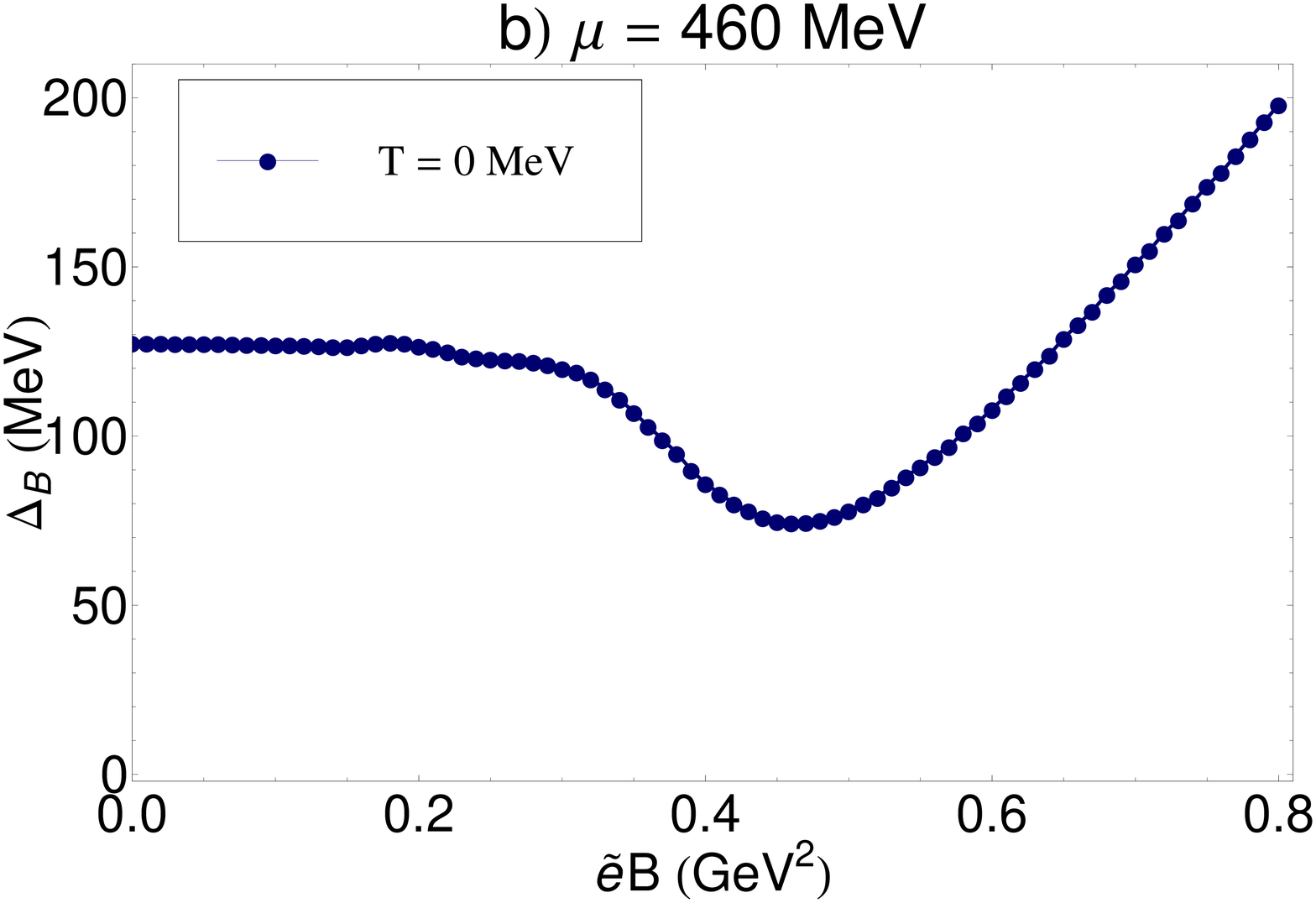}
\caption{(a) The dependence of the chiral gap $\sigma_{B}$ on
$\tilde{e}B$ for fixed $\mu=250$ MeV, at
 $T=0$ MeV. (b) The dependence of diquark gap $\Delta_{B}$ on
$\tilde{e}B$ for fixed $\mu=460$ MeV, at $T=0$ MeV.}\label{T-0}
\end{figure}
\par
In Fig. \ref{T-0}, the $\tilde{e}B$ dependence of meson and diquark
mass gaps, $\sigma_{B}$ and $\Delta_{B}$, are plotted at $T=0$ and
$\mu=250$ MeV [Fig. \ref{T-0}(a)], and $\mu=460$ MeV [Fig.
\ref{T-0}(b)]. A comparison with similar plots from Fig. 2 of
\cite{fayaz2010} shows that color neutrality condition has no
significant effect on the above mentioned threshold magnetic field
$\tilde{e}B_{t}\simeq 0.5$ GeV$^{2}$, and on the behavior of
$\sigma_{B}$ and $\Delta_{B}$ below and above $\tilde{e}B_{t}$,
consisting of van Alfven-de Haas oscillations in
$\tilde{e}B<\tilde{e}B_{t}$ and the linear rise in
$\tilde{e}B>\tilde{e}B_{t}$ regimes.\footnote{See Footnote 9.}
Similar strong van Alphen-de Haas oscillation of 2SC mass gap
$\Delta_{B}$ in the regime $\tilde{e}B\in [0.4,0.6]$ GeV$^{2}$ in
Fig. \ref{T-0}(b) is also observed in \cite{shovkovy2007,
warringa2007} in the three-flavor CFL phase in the presence of
magnetic fields, albeit in another regime of $\tilde{e}B/\mu^{2}$.
\begin{figure}[hbt]
\includegraphics[width=8.2cm,height=6cm]{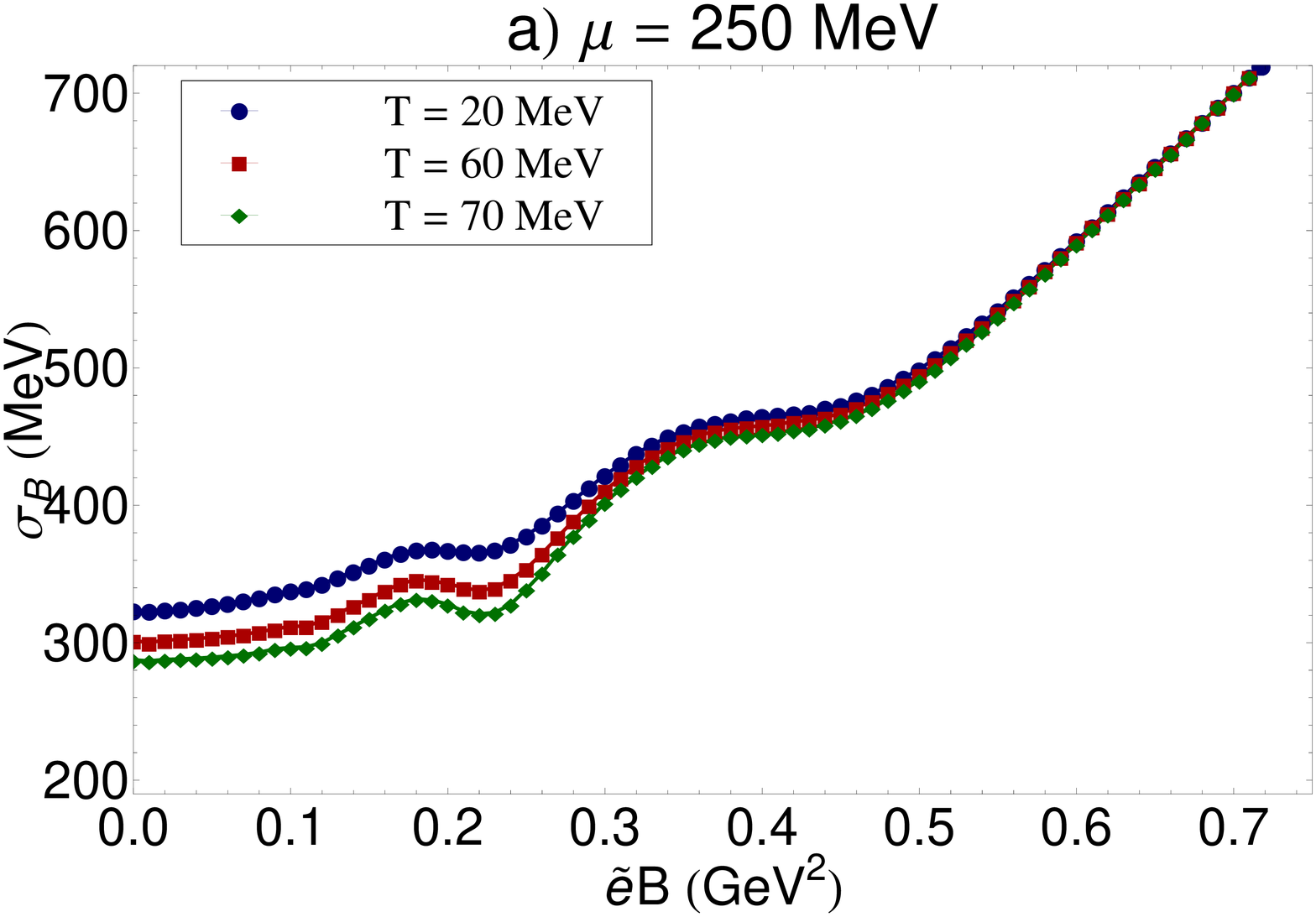}
\hspace{0.2cm}
\includegraphics[width=8.2cm,height=6cm]{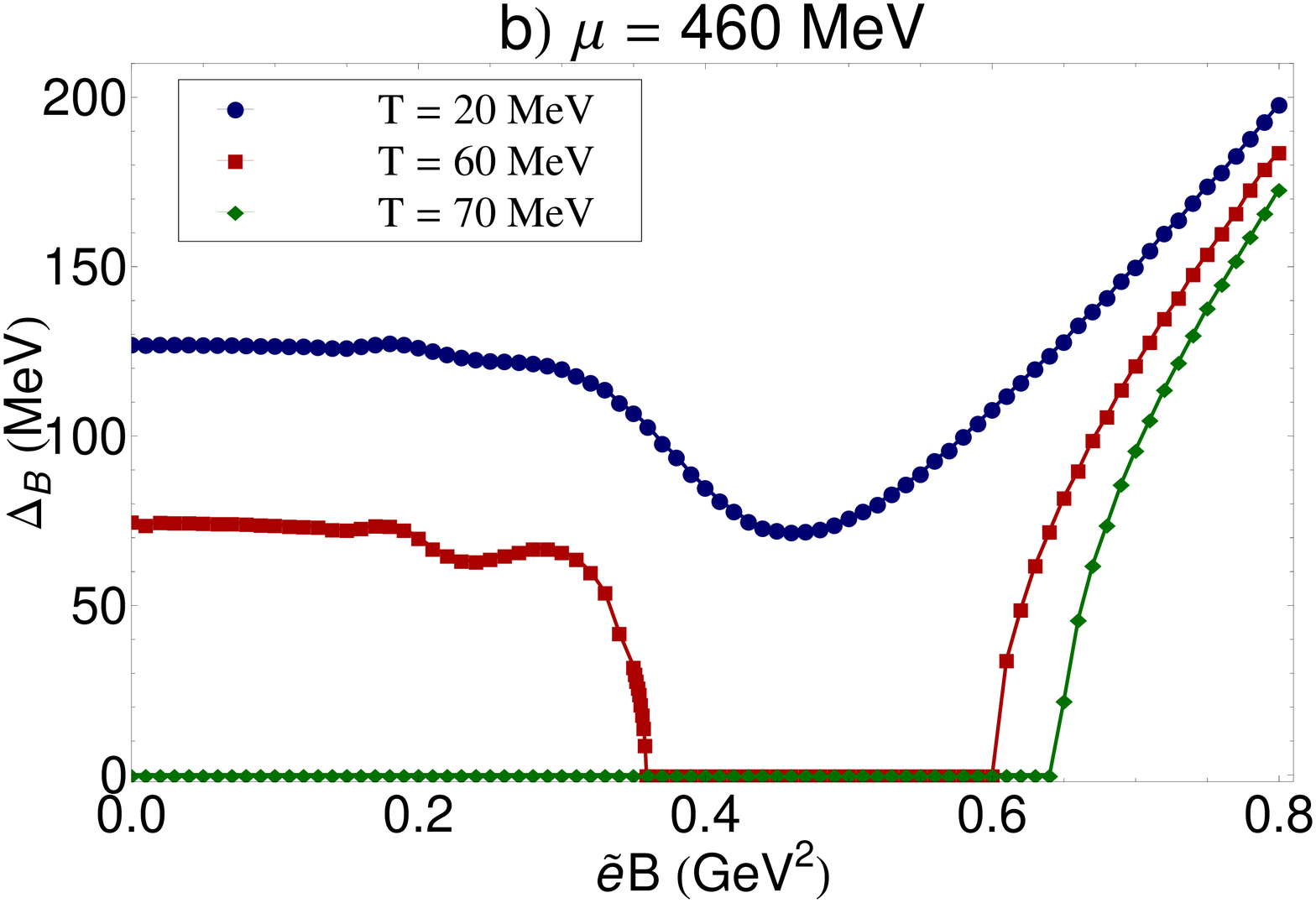}
 \caption{(a) The dependence of the chiral gap $\sigma_{B}$ on $\tilde{e}B$ for fixed $\mu=250$ MeV, at various
 temperatures. (b) The dependence of diquark gap $\Delta_{B}$ on $\tilde{e}B$ for
fixed $\mu=460$ MeV, at various temperatures.}\label{mass-delta-eB}
\end{figure}
\par
In Fig. \ref{mass-delta-eB}(a), the $\tilde{e}B$ dependence of
$\sigma_{B}$ is demonstrated for fixed $\mu=250$ MeV and at various
fixed temperatures $T=20,60,70$ MeV; after small oscillations in the
regime below the threshold magnetic field $\tilde{e}B_{t}$, the
system enters the linear regime, where only the contribution of the
LLL affects the dynamics of the system. For
$\tilde{e}B<\tilde{e}B_{t}$, $\sigma_{B}$ decreases by increasing
the temperature. In the linear regime $\tilde{e}B>\tilde{e}B_{t}$,
however, the effect of temperature is minimized. In Fig.
\ref{mass-delta-eB}(b), the dependence of the diquark gap
$\Delta_{B}$ is demonstrated for fixed $\mu=460$ MeV and at various
temperatures $T=20,60,70$ MeV. At $T=20$ MeV, small oscillations
occur in $\tilde{e}B<\tilde{e}B_{t}$. In contrast, $\Delta_{B}$
monotonically increases with $\tilde{e}B$ in the linear regime
$\tilde{e}B>\tilde{e}B_{t}$. The behavior of $\Delta_{B}$ at $T=20$
MeV is not too much different than its behavior at $T=0$ MeV [see
Fig. \ref{T-0}(b)]. This confirms our previous observation from the
comparison of Figs. \ref{fig0} and \ref{fig1}. By increasing the
temperature to $T=60$ MeV, $\Delta_{B}$ decreases. It vanishes in
the range $\tilde{e}B\in[0.4,0.6]$ GeV$^{2}$. Comparing to the $T=0$
MeV plot from Fig. \ref{T-0}(b), it turns out that the value of
$\Delta_{B}$ at $T=0$ MeV in this regime of $\tilde{e}B\in[0.4,0.6]$
GeV$^{2}$ is in the same order of magnitude as $60$ MeV. This is
related with the well-known BCS ratio of critical temperature to
zero-temperature gap in zero magnetic field case \cite{pisarski1999}
[see the discussion below]. For $\tilde{e}B>0.6$ GeV$^{2}$,
$\Delta_{B}$ increases again with $\tilde{e}B$. The situation
demonstrated in Fig. \ref{mass-delta-eB}(b) at $T=60$ MeV can be
interpreted as a second order phase transition from the CSC to the
normal phase at $T=60$ MeV, $\mu=460$ MeV and $\tilde{e}B\simeq 0.4$
GeV$^{2}$, followed up with a second order phase transition from the
normal to the CSC phase at $T=60$ MeV, $\mu=460$ MeV and
$\tilde{e}B\simeq 0.6$ GeV$^{2}$. We believe that this
CSC-Normal-CSC phase transition are caused by strong van Alfven-de
Haas oscillations in the regime $\tilde{e}B\in[0.4,0.6]$ GeV$^{2}$.
A comparison between the $T=60$ MeV plot of Fig.
\ref{mass-delta-eB}(b) and the $\mu=460$ MeV plot of Fig.
\ref{figTeB}(i) confirms these statements.\footnote{See also the
explanation in the paragraph following Fig. \ref{figTeB}, where we
have compared Fig. \ref{figTeB}(i) with Fig.
\ref{mass-delta-eB}(b).} Let us notice that van Alfven-de Haas
oscillations, which are also observed in metals \cite{alfven}, are
the consequence of oscillatory structure in the density of states of
quarks and occur whenever the Landau level pass the quark Fermi
surface \cite{inagaki2003, shovkovy2007}.
\par
In Fig. \ref{mass-delta-eB}(b), at $T=70$ MeV and in the regime for
$\tilde{e}B\lesssim \tilde{e}B_{t}$, the diquark condensate is not
built at all, i.e. $\Delta_{B}=0$. This result indicates that a
normal phase exists at $T\geq 70$ MeV, $\mu=460$ MeV and in the
regime $\tilde{e}B\lesssim 0.65$ GeV$^{2}$, and that a second order
phase transition from the normal to the CSC phase occurs at
$(T,\mu)\sim(70, 460)$ MeV and critical magnetic field
$\tilde{e}B\simeq 0.65$ GeV$^{2}$. These results coincide with the
observations in the complete $T-\tilde{e}B$ phase diagram from Fig.
\ref{figTeB}(i) for $\mu=460$ MeV. Indeed, our numerical
computations show that for $\mu=460$ MeV, the critical temperatures
in the whole range of $\tilde{e}B\in[0,0.6]$ GeV$^{2}$ are smaller
than $70$ MeV. In $\tilde{e}B=0$ GeV$^{2}$ and $\mu=460$ MeV, it is
$T_{c}\simeq 68.5$ MeV and decreases with $\tilde{e}B\lesssim 0.5$
GeV$^{2}$ [see also Fig. \ref{figTeB}(i)]. Using this value of
$T_{c}$ in zero magnetic field and the numerical value of the
diquark gap at zero temperature and zero magnetic field from Fig.
\ref{T-0}(b), i.e. $\Delta_{B}(T=0)\simeq 127.5$ MeV,  the above
mentioned BCS ratio can be computed. It is given by
$$\frac{T_{c}}{\Delta_{B}(T=0)}\simeq \frac{68.5}{127.5}\simeq 0.54,$$ which is in good
agreement with the BCS ratio
$\frac{T_{c}}{\Delta_{B}(T=0)}=e^{\gamma}\pi^{-1}+{\cal{O}}(g)\simeq
0.567+{\cal{O}}(g)$ of critical temperature to the zero-temperature
2SC mass gap of QCD in zero magnetic field \cite{pisarski1999}.
Here, $\gamma\simeq 0.577$ is the Euler-Mascheroni constant. As for
the regime of strong magnetic fields, $\tilde{e}B>0.6$ GeV$^{2}$ at
fixed $T=70$ MeV, $\Delta_{B}$ increases monotonically with
$\tilde{e}B$, as expected from (\ref{D9-c2}) for $\mu=460$ MeV, and
$\Lambda$ as well as $G_{D}$ from (\ref{D2}).
\begin{figure}[thb]
\includegraphics[width=8cm,height=6cm]{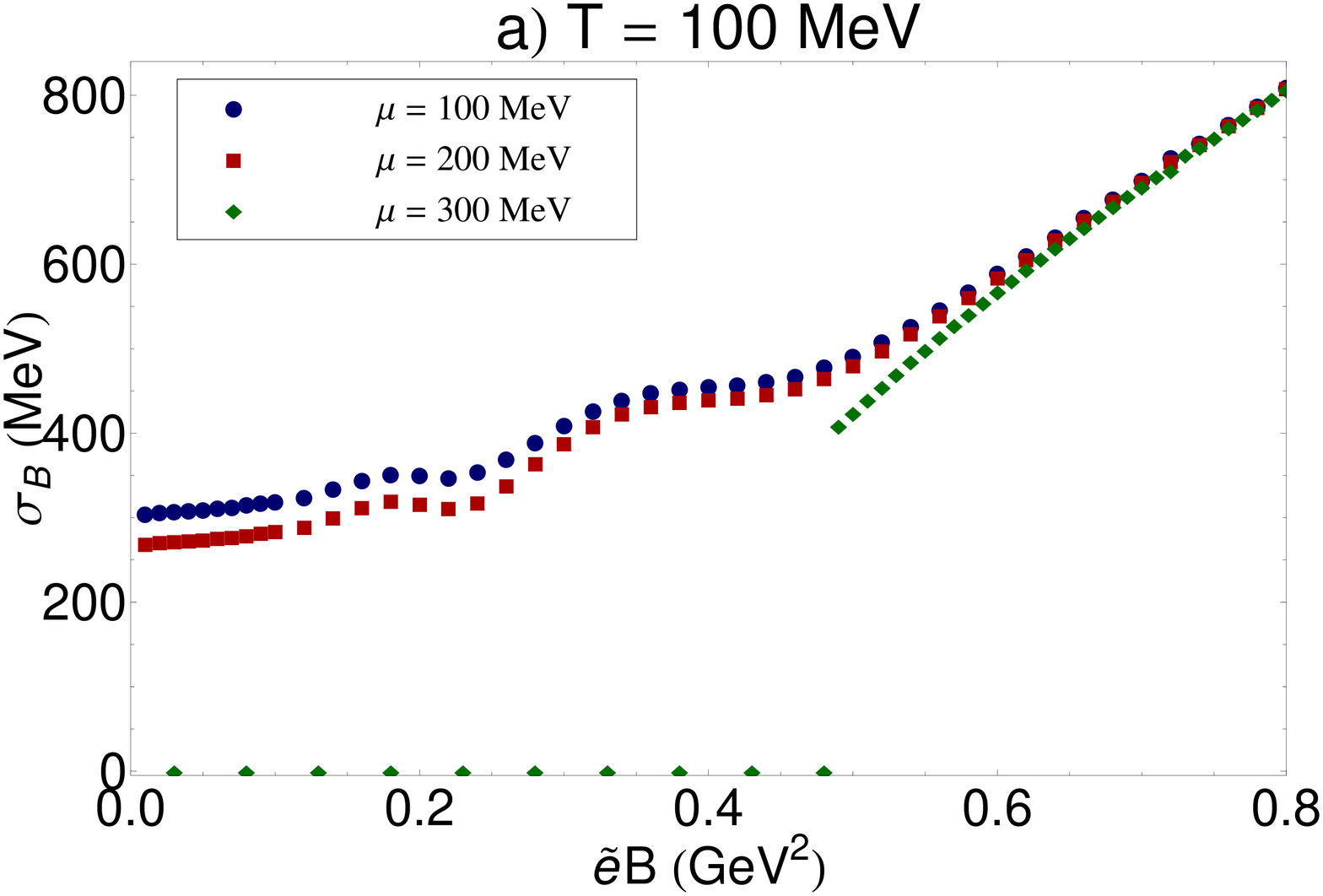}
\hspace{0.3cm}
\includegraphics[width=8cm,height=6cm]{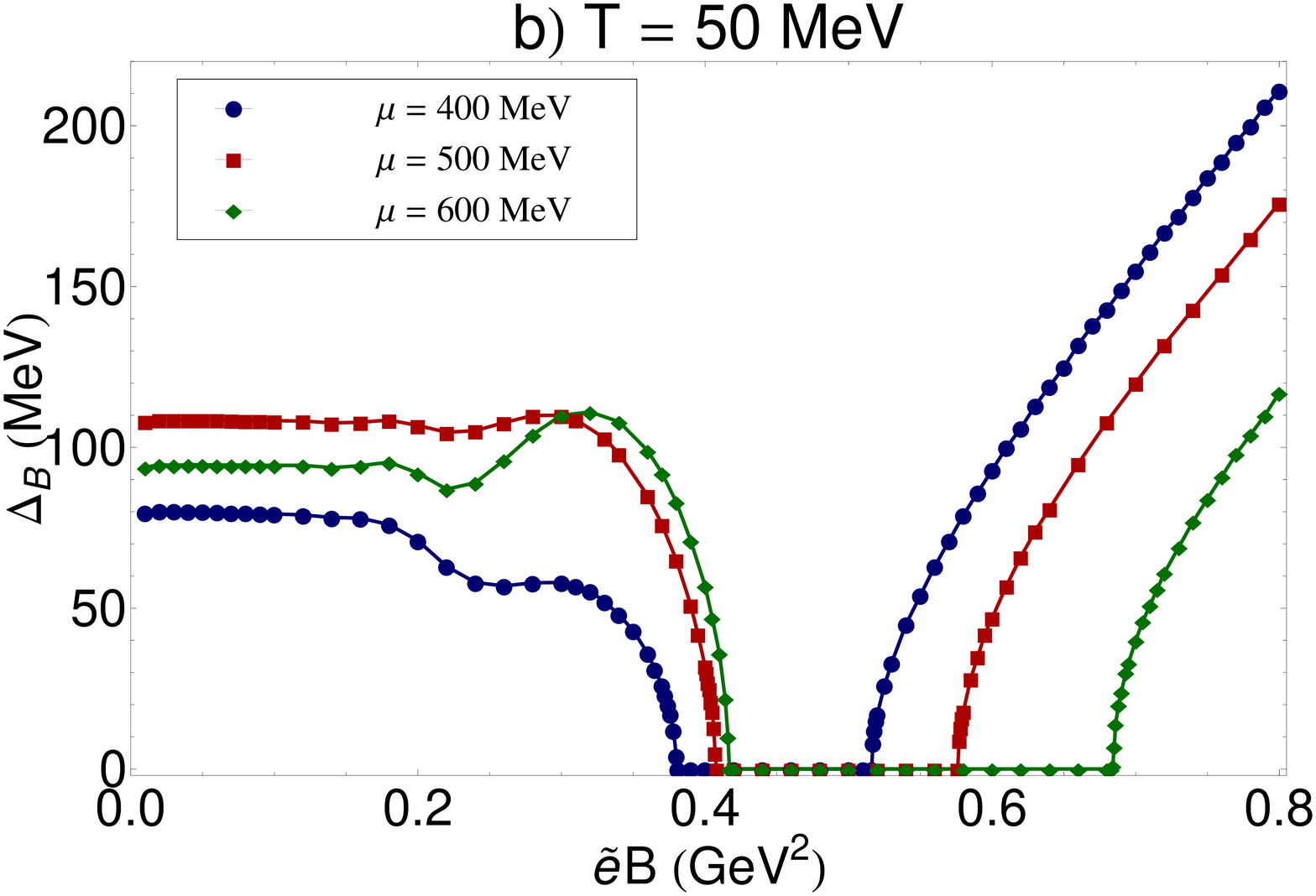}
\caption{(a) The dependence of the chiral gap $\sigma_{B}$ on
$\tilde{e}B$ for fixed $T=100$ MeV, at various chemical potentials
$\mu$. (b) The dependence of diquark gap $\Delta_{B}$ on
$\tilde{e}B$ for fixed $T=50$ MeV, at various chemical
potentials.}\label{sigma-eB-mu}
\end{figure}
\par
In Fig. \ref{sigma-eB-mu}, the dependence of chiral and diquark gaps
on the external magnetic field for constant $T$ are demonstrated for
various $\mu$. As it is shown in Fig. \ref{sigma-eB-mu}(a), the
chiral mass gap $\sigma_{B}$ decreases by increasing $\mu$. Small
van Alfven-de Haas oscillations occur in the regime below the
threshold magnetic field $\tilde{e}B_{t}\simeq 0.5$ GeV$^{2}$ and
die out in the linear regime above $\tilde{e}B_{t}$. In this regime
$\sigma_{B}$ monotonically rises with $\tilde{e}B$, and the effect
of $\mu$ is minimized. In the regime below $\tilde{e}B_{t}$, for
$\mu=300$ MeV, the chiral mass gap $\sigma_{B}$ vanishes. This
indicates a first order phase transition from the normal into the
$\chi$SB phase at fixed $T=100$ MeV, $\mu=300$ MeV and for
$\tilde{e}B\simeq 0.5$ GeV$^{2}$. This result agrees with our
findings from the complete $T-\tilde{e}B$ phase diagram plotted in
Fig. \ref{figTeB}(c) for fixed $\mu=300$ MeV. As it can be checked
in Fig. \ref{figTeB}(c), at fixed $T=100$ MeV, the $\chi$SB phase
appears first for $\tilde{e}B>0.5$ GeV$^{2}$. What concerns the
$\tilde{e}B$-dependence of the diquark mass gap in Fig.
\ref{sigma-eB-mu}(b), it increases with $\mu$ in the regime below
the threshold magnetic field $\tilde{e}B_{t}$. In the regime above
$\tilde{e}B_{t}$, however, $\Delta_{B}$ decreases by increasing
$\mu$. Same phenomenon was also demonstrated in Fig. \ref{fig1}(c)
at low temperature $T=20$ MeV and for $\tilde{e}B=0.5$ near
$\tilde{e}B_{t}$, and was shown to be in full agreement with the
analytical result (\ref{D9-c2}) \cite{fayaz2010}. Note that strong
van Alfven-de Haas oscillations are responsible for vanishing
$\Delta_{B}$ in the regime $\tilde{e}B\in[0.4,0.6]$ GeV$^{2}$. They
induce CSC-Normal-CSC second order phase transitions, as it can be
seen in the same regime of $\tilde{e}B$ in Fig. \ref{figTeB}(i). The
plots in Fig. \ref{sigma-eB-mu}(b) show also that the critical
magnetic fields for the transition from the normal into the CSC
phase increase by increasing the chemical potential [see Fig.
\ref{sigma-eB-mu}(b)].\footnote{Detailed analysis on the dependence
of critical $\tilde{e}B$ on $\mu$ and vice versa will be performed
in Sec. III.B.}
\subsubsection{The $T$-dependence of $\sigma_{B}$ and $\Delta_{B}$}
\par
\begin{figure}[h]
\includegraphics[width=8cm,height=5.6cm]{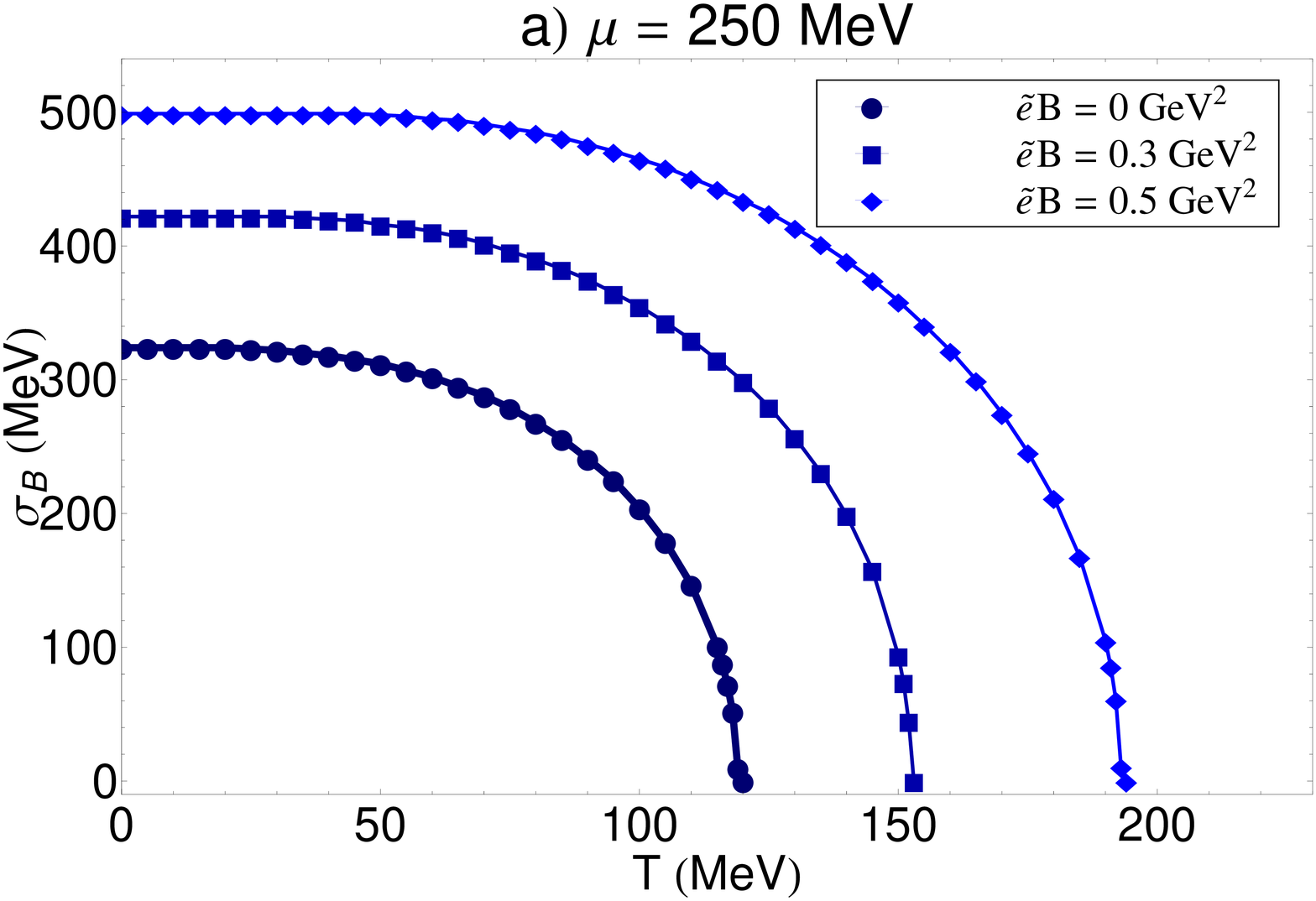}
\hspace{0.2cm}
\includegraphics[width=8cm,height=5.6cm]{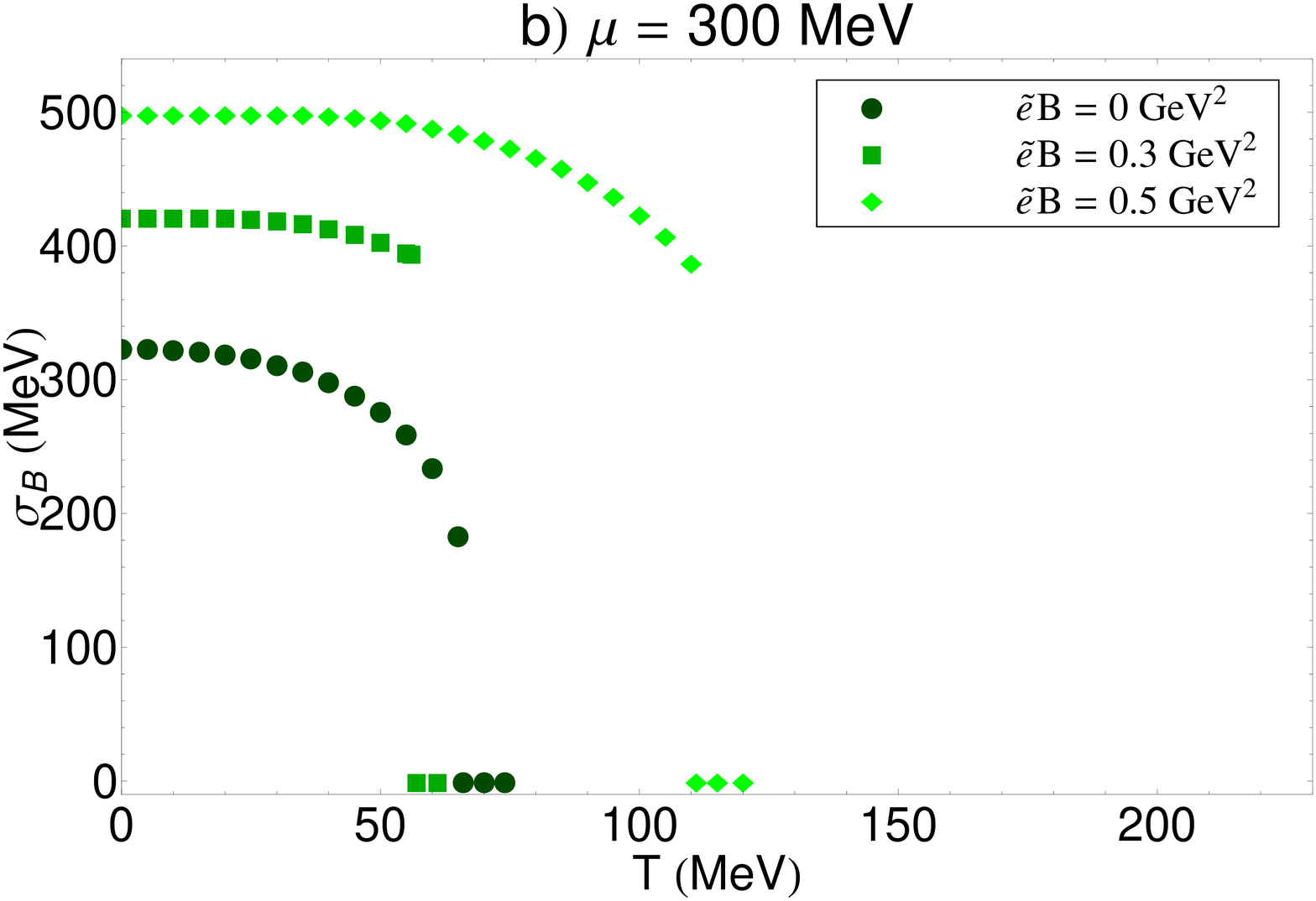}
\caption{The $T$-dependence of $\sigma_{B}$ for fixed $\mu=250$ MeV
(panel a) and  $\mu=300$ MeV (panel b) for various
$\tilde{e}B=0,0.3,0.5$ GeV$^{2}$. Whereas for $\mu=250$ MeV the
transition from the chiral to the normal phase is of second order
(blue curves in panel a), for $\mu=300$ MeV, this transition
 is of first order (green dots in panel b). }\label{sigma-T}
\end{figure}

\noindent
In Fig. \ref{sigma-T}, the temperature dependence of the chiral
condensate is presented for fixed $\mu=250$ MeV (panel a), $\mu=300$
MeV (panel b), and various magnetic fields $\tilde{e}B=0,0.3,0.5$
GeV$^{2}$. Whereas the transition from the $\chi$SB into the normal
phase for $\mu=250$ MeV is of second order [continuous decreasing of
$\sigma_{B}$ to $\sigma_{B}=0$ in Fig. \ref{sigma-T}(a)], it is of
first order for $\mu=300$ MeV [discontinuous transition from
$\sigma_{B}\neq 0$ to $\sigma_{B}=0$ in Fig. \ref{sigma-T}(b)]. As
it is also expected from our previous results, the magnetic field
enhances the production of $\sigma_{B}$, that increases by
increasing the value of the magnetic field.
\par
\begin{figure}[t]
\includegraphics[width=8cm,height=6cm]{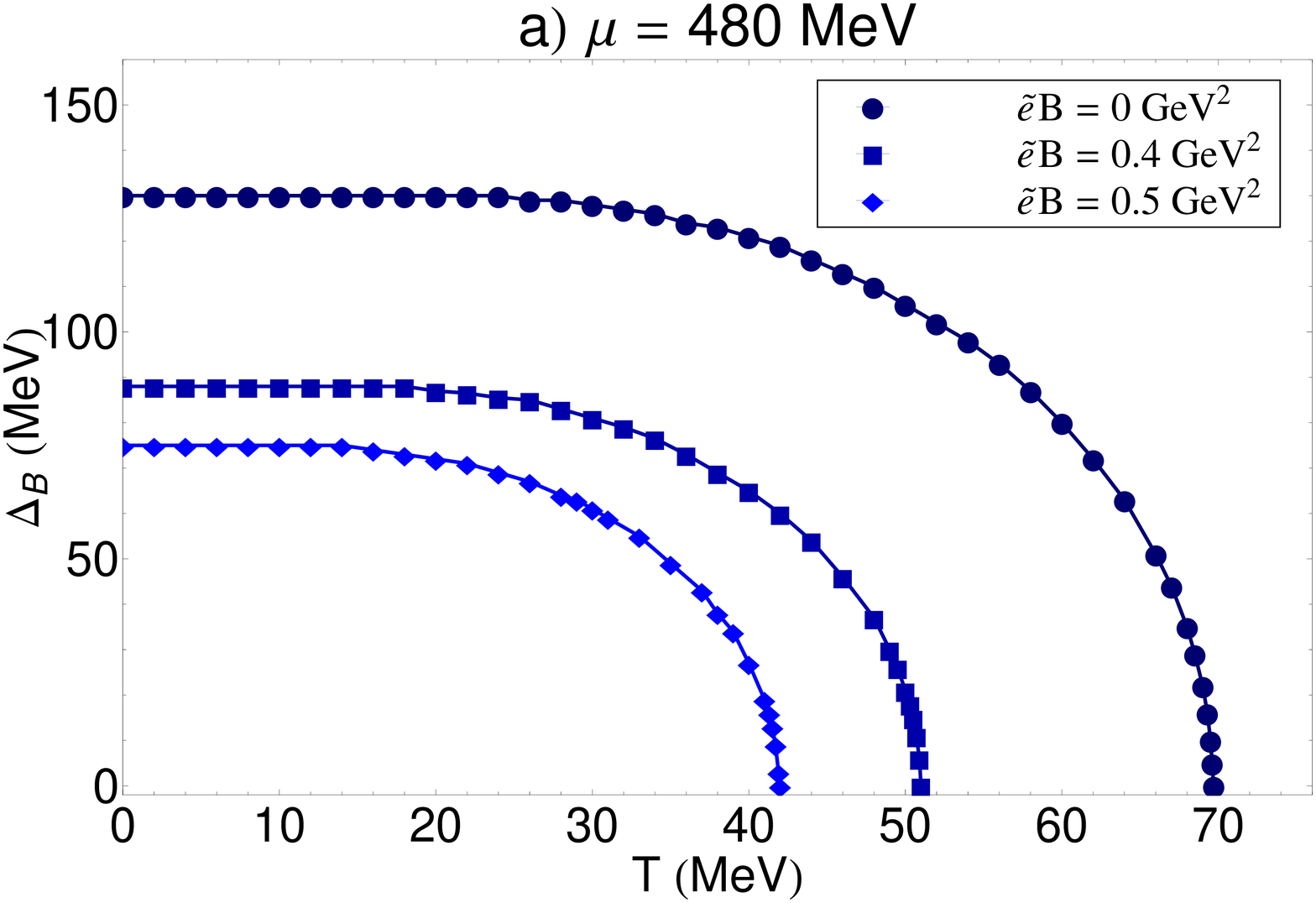}
\hspace{0.2cm}
\includegraphics[width=8cm,height=6cm]{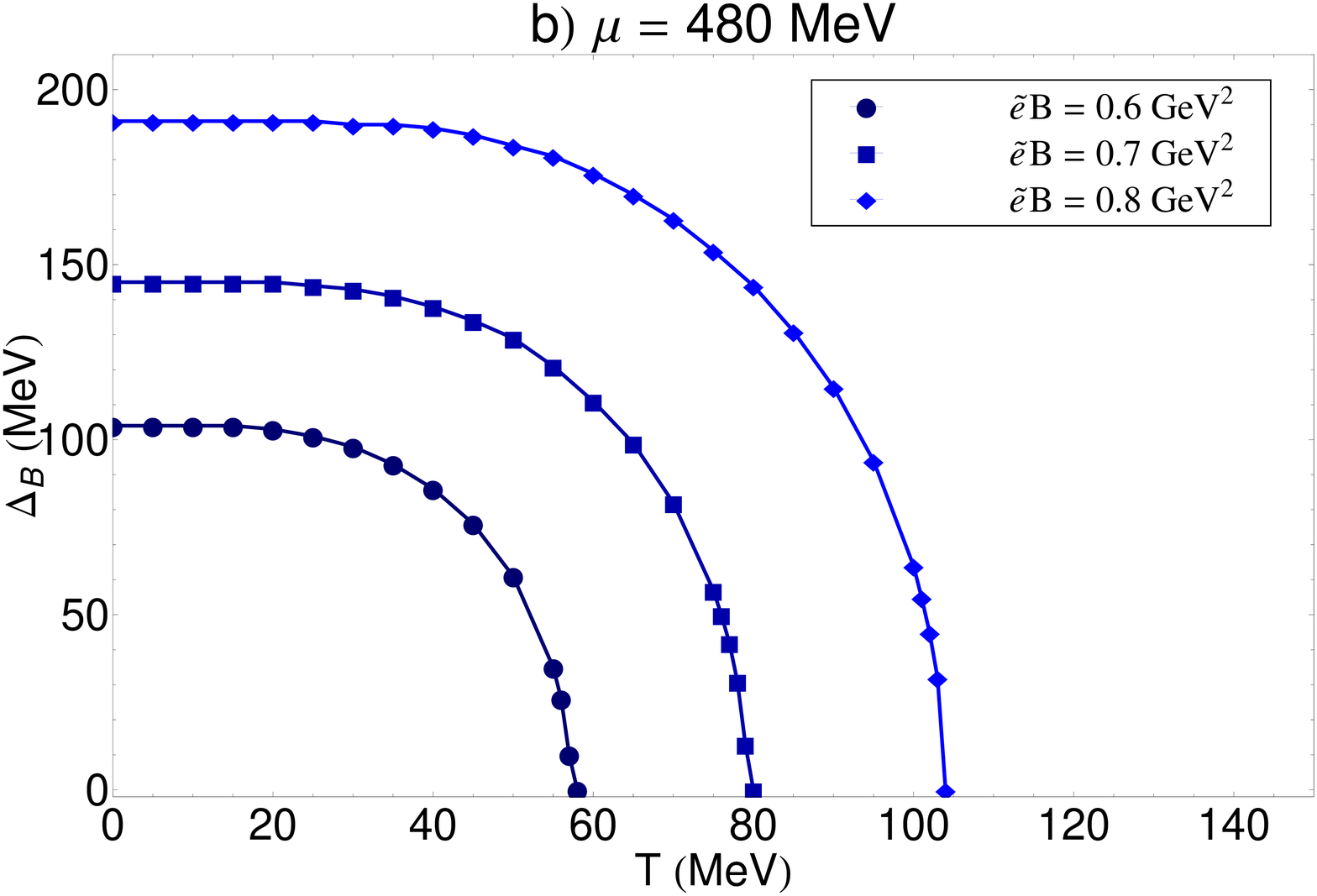}
\caption{The $T$-dependence of $\Delta_{B}$ for fixed $\mu=480$ MeV
for various $\tilde{e}B=0,0.3,0.5$ GeV$^{2}$ below the threshold
magnetic field (panel a) and for $\tilde{e}B=0.6,0.7,0.8$ GeV$^{2}$
above the threshold magnetic field (panel b). Below (above) the
threshold magnetic field the production of diquark is suppressed
(enhanced). }\label{delta-T}
\end{figure}

The temperature dependence of $\Delta_{B}$ is presented in Fig.
\ref{delta-T} for fixed $\mu=480$ MeV and $\tilde{e}B=0,0.4,0.5$
GeV$^{2}$ (panel a), below the threshold magnetic field
$\tilde{e}B_{t}\simeq 0.45-0.50$ GeV$^{2}$, as well as for fixed
$\mu=480$ MeV and $\tilde{e}B=0.6,0.7,0.8$ GeV$^{2}$, above the
threshold magnetic field. As it turns out magnetic fields below the
threshold magnetic field, suppress the production of $\Delta_{B}$,
whereas magnetic fields stronger than the threshold magnetic field
enhances its production [see also Fig. \ref{mass-delta-eB}(b)].
Similarly, for $\tilde{e}B<\tilde{e}B_{t}$, the value of critical
temperature decreases by increasing the magnetic field, whereas for
$\tilde{e}B>\tilde{e}B_{t}$, the critical temperatures increase by
increasing the magnetic field.\footnote{The $\tilde{e}B$ dependence
of critical temperatures will be discussed in detail in III.B.}
\par
\begin{figure}[thb]
\includegraphics[width=5.4cm,height=4cm]{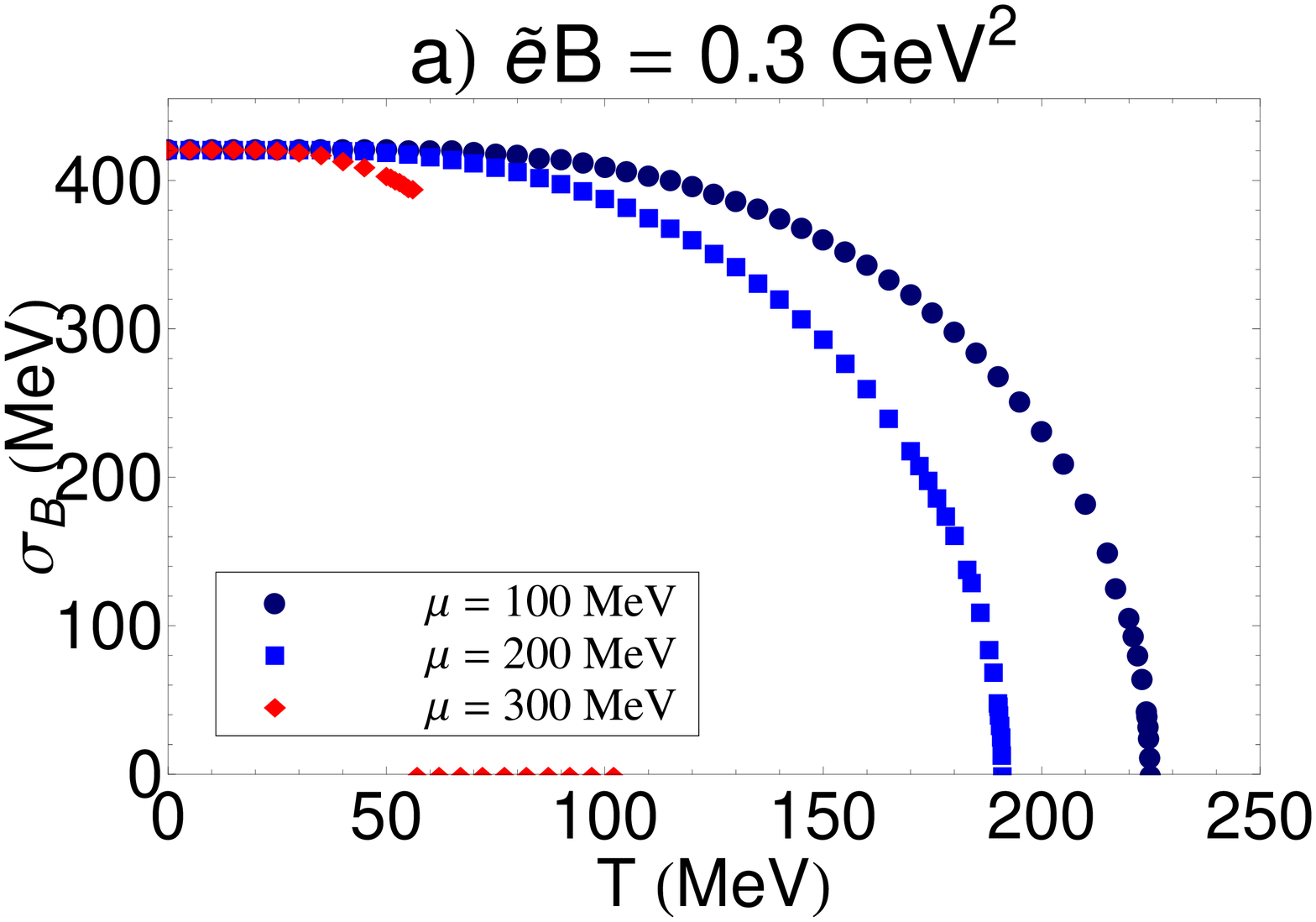}
\hspace{0.1cm}
\includegraphics[width=5.4cm,height=4cm]{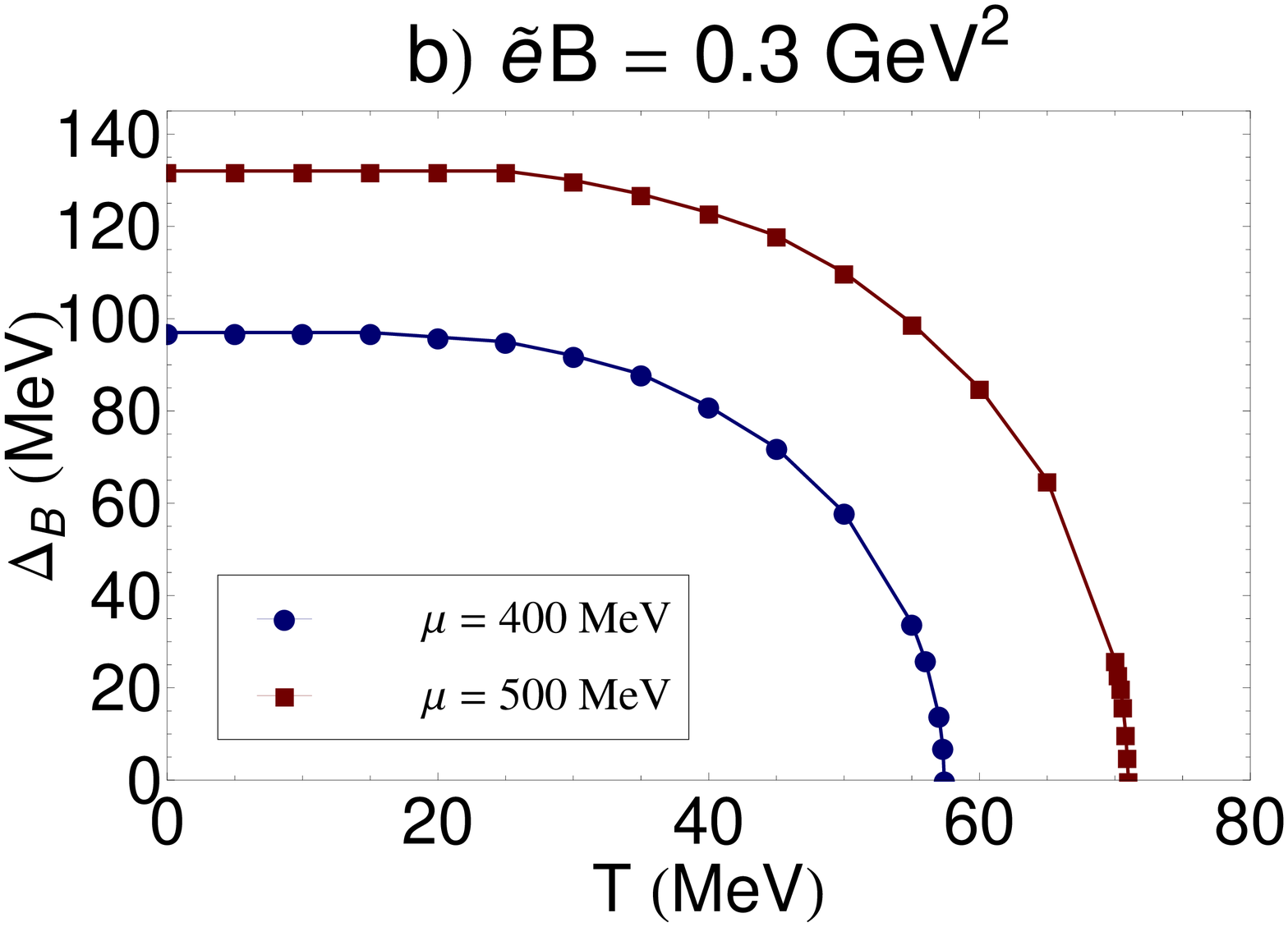}
\hspace{0.1cm}
\includegraphics[width=5.4cm,height=4cm]{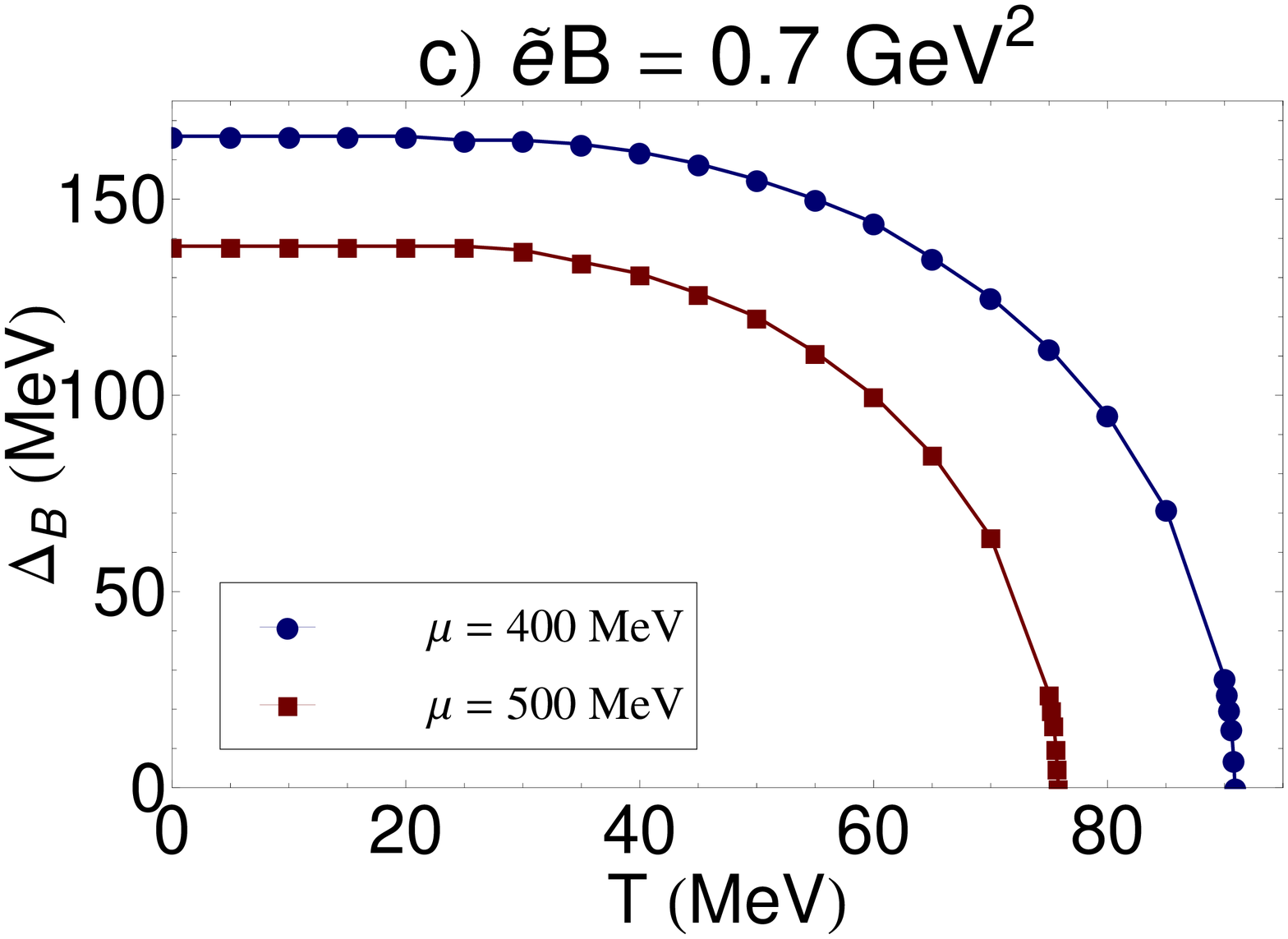}
\caption{T-dependence of $\sigma_{B}$ and $\Delta_{B}$ for different
$\mu$ for fixed $\tilde{e}B$.}\label{sigma-T-mu}
\end{figure}

The same effect is also observed in Fig. \ref{sigma-T-mu}. In Fig.
\ref{sigma-T-mu}(a), the $T$-dependence of $\sigma_{B}$ is plotted
for constant $\tilde{e}B=0.3$ GeV$^{3}$ and various
$\mu=100,200,300$ MeV. At low temperature $T<50$ MeV, $\sigma_{B}$
remains constant, and decreases by increasing the temperature.
Increasing the chemical potential only accelerate this decrease,
i.e. the critical temperature decreases by increasing $\mu$. Whereas
the transition at $\mu=100,~200$ MeV is of second order (blue dots),
a first order phase transition occurs at $\mu=300$ MeV (red dots).
The $T$-dependence of diquark condensate $\Delta_{B}$ is
demonstrated for $\tilde{e}B=0.3$ GeV$^{2}$ below the threshold
magnetic field [Fig. \ref{sigma-T-mu}(b)], and for $\tilde{e}B=0.7$
GeV$^{2}$ above the threshold magnetic field [Fig.
\ref{sigma-T-mu}(c)]. As it turns out, whereas the diquark
condensate increases by increasing $\mu$ for fixed T and
$\tilde{e}B<\tilde{e}B_{t}$ [Fig. \ref{sigma-T-mu}(b)], it decreases
by increasing $\mu$ for $\tilde{e}B>\tilde{e}B_{t}$ [Fig.
\ref{sigma-T-mu}c]. This result coincides with our results from Fig.
\ref{sigma-eB-mu}(b). Moreover, as we can see in Figs.
\ref{sigma-T-mu}(b) and 8(c), the diquark condensates decreases by
increasing the temperature. Similar to what is observed in Fig.
\ref{delta-T}, for $\tilde{e}B=0.3$ GeV$^{2}$, below
$\tilde{e}B_{t}$, the critical temperature increases by increasing
$\mu$ [Fig. \ref{sigma-T-mu}(b)], while for $\tilde{e}B=0.7$
GeV$^{2}$ above $\tilde{e}B_{t}$, the critical temperature decreases
by increasing $\mu$. Later, we will see that this result is also in
full agreement to our result from Fig. \ref{figTeB}(i).
\subsection{Phase diagrams of hot magnetized 2SC quark matter}
\par\noindent
To complete our study on the effect of $(T,\mu,\tilde{e}B)$ on the
quark matter including mesons and diquarks in the presence of
constant magnetic field, we will present in this section the phase
structure of the model in a $T-\mu$ plane for various fixed
$\tilde{e}B$ in Fig. \ref{figTmu}. The $T-\tilde{e}B$ phase diagram
of the model for various fixed $\mu$ is presented in Fig.
\ref{figTeB}, and finally in Figs. \ref{figmueBa} and
\ref{figmueBb}, the $\mu-\tilde{e}B$ phase diagram is explored for
various fixed $T$. As in the previous section, green dashed lines
denote the first order phase transitions and the blue solid lines
the second order transitions.
\subsubsection{$T-\mu$ phase diagram for various fixed $\tilde{e}B$}
\begin{figure}[hbt]
\includegraphics[width=5cm,height=4cm]{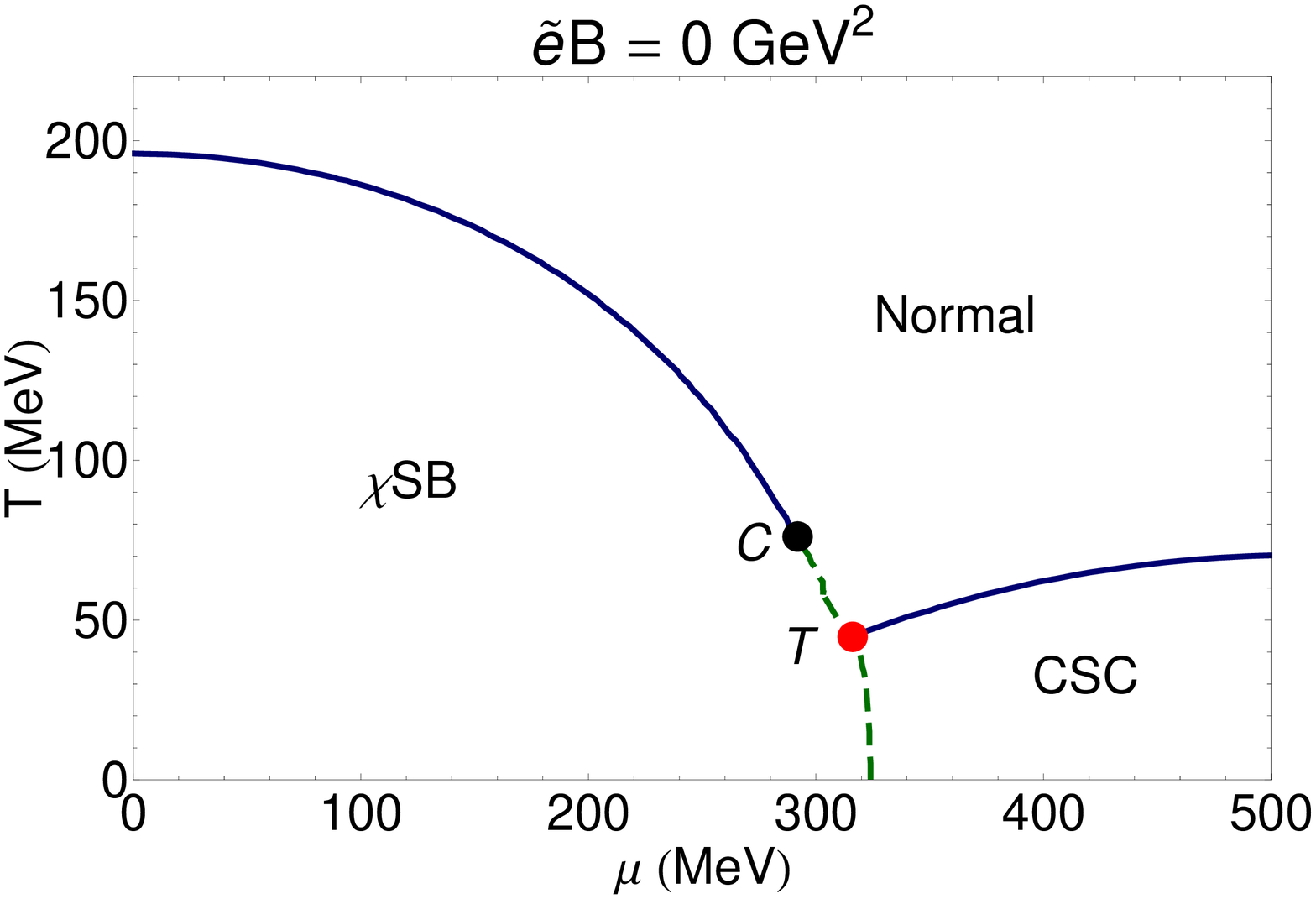}
\hspace{0.3cm}
\includegraphics[width=5cm,height=4cm]{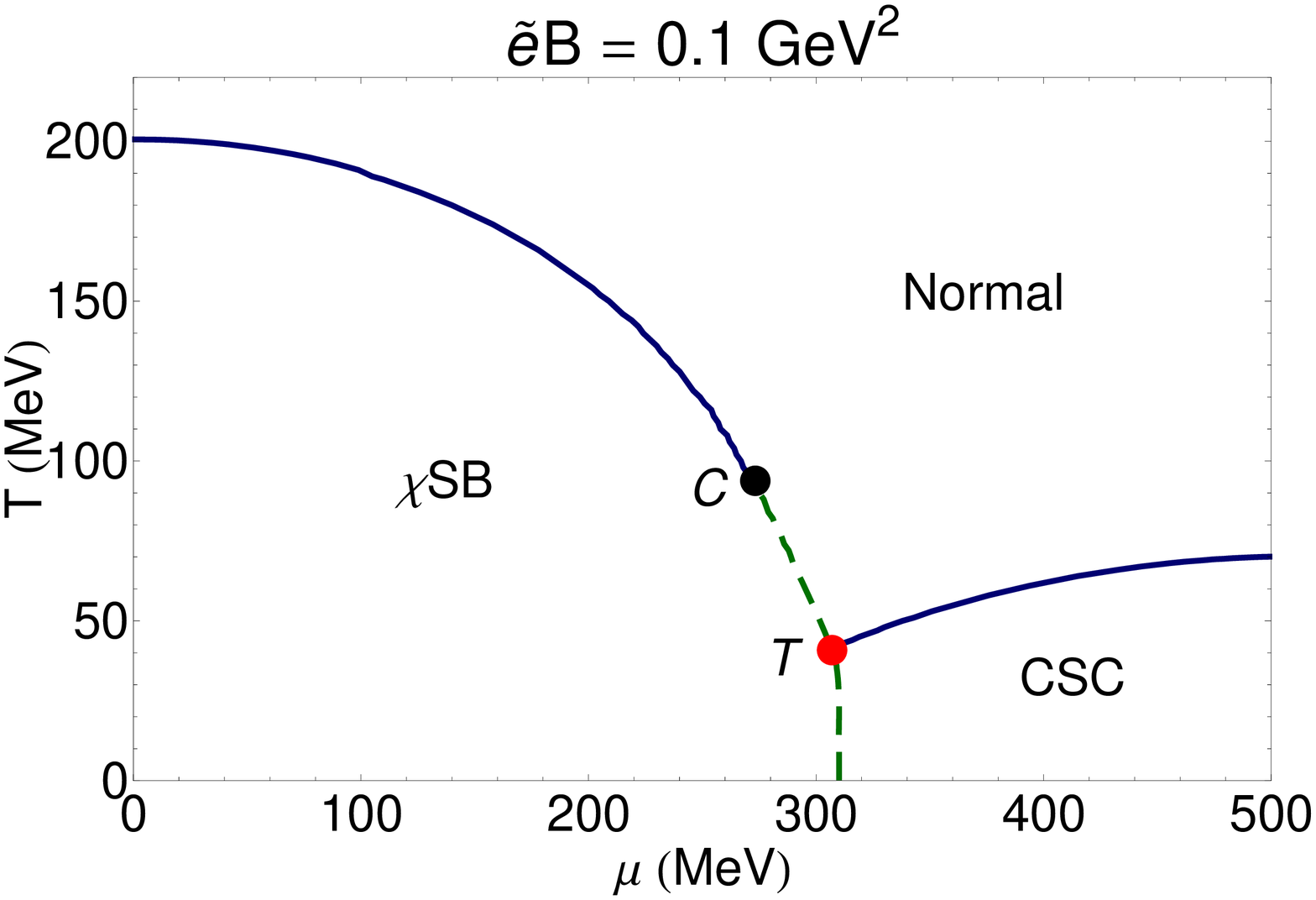}
\hspace{0.3cm}
\includegraphics[width=5cm,height=4cm]{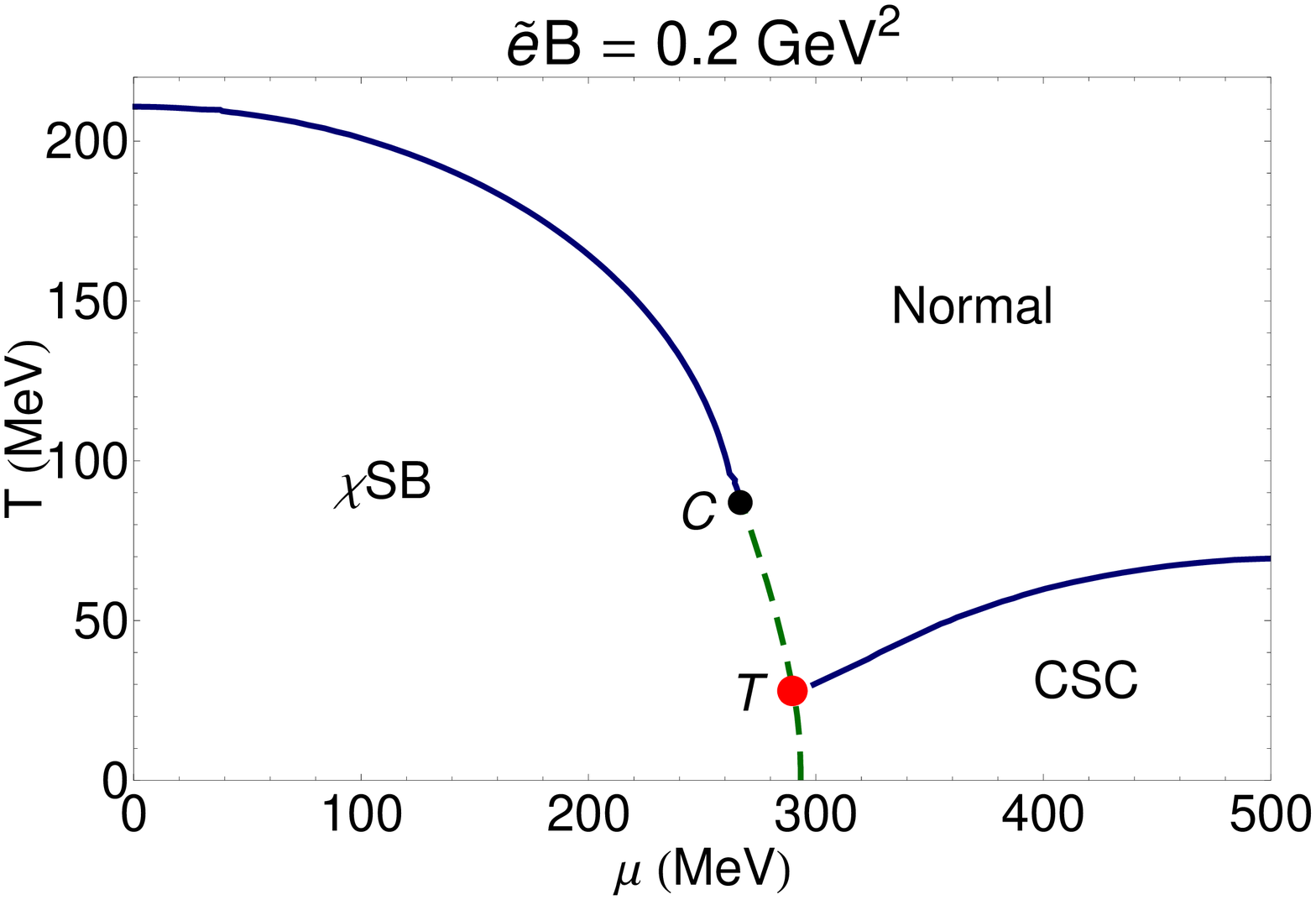}
\par\vspace{0.5cm}
\includegraphics[width=5cm,height=4cm]{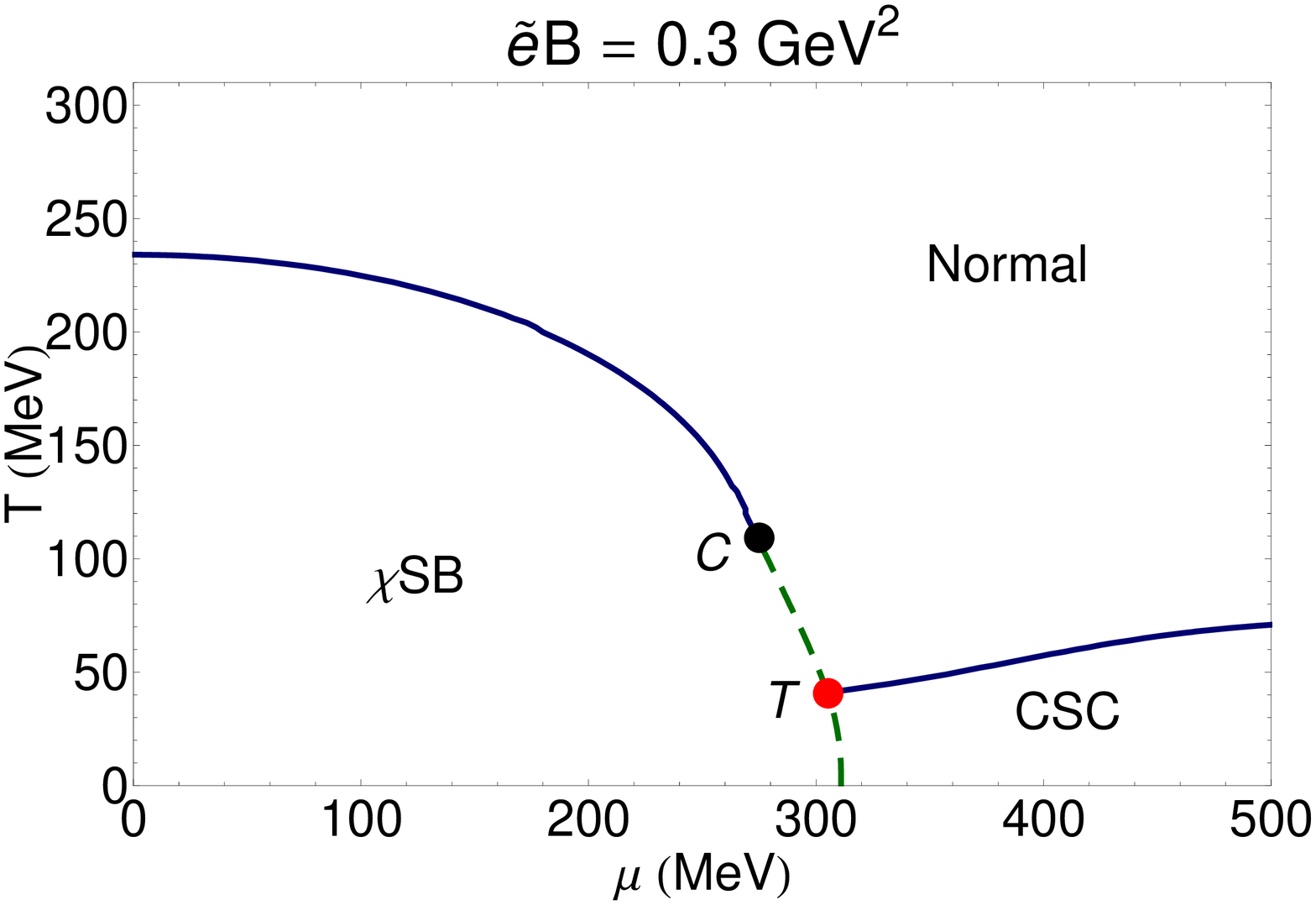}
\hspace{0.3cm}
\includegraphics[width=5cm,height=4cm]{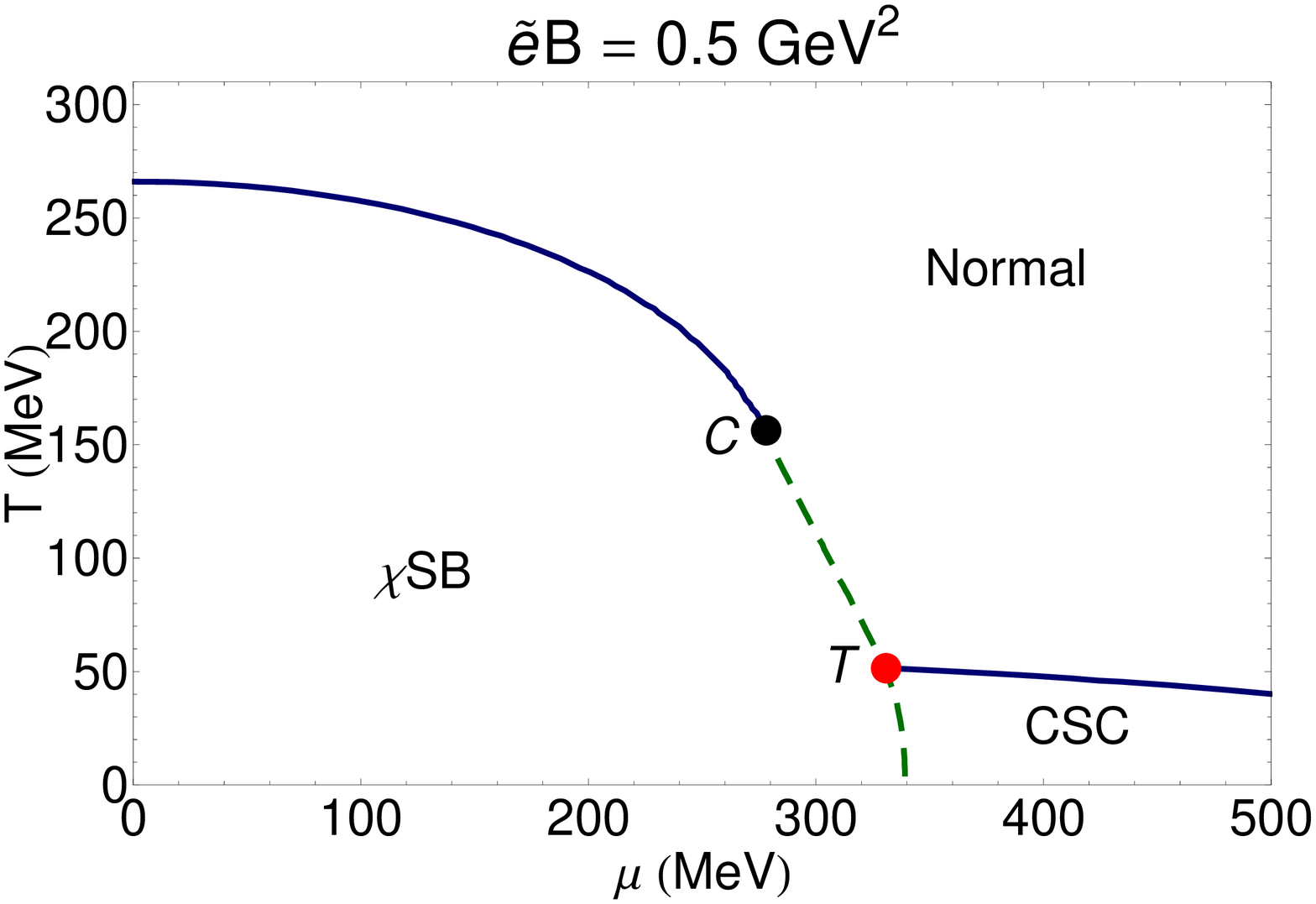}
\hspace{0.3cm}
\includegraphics[width=5cm,height=4cm]{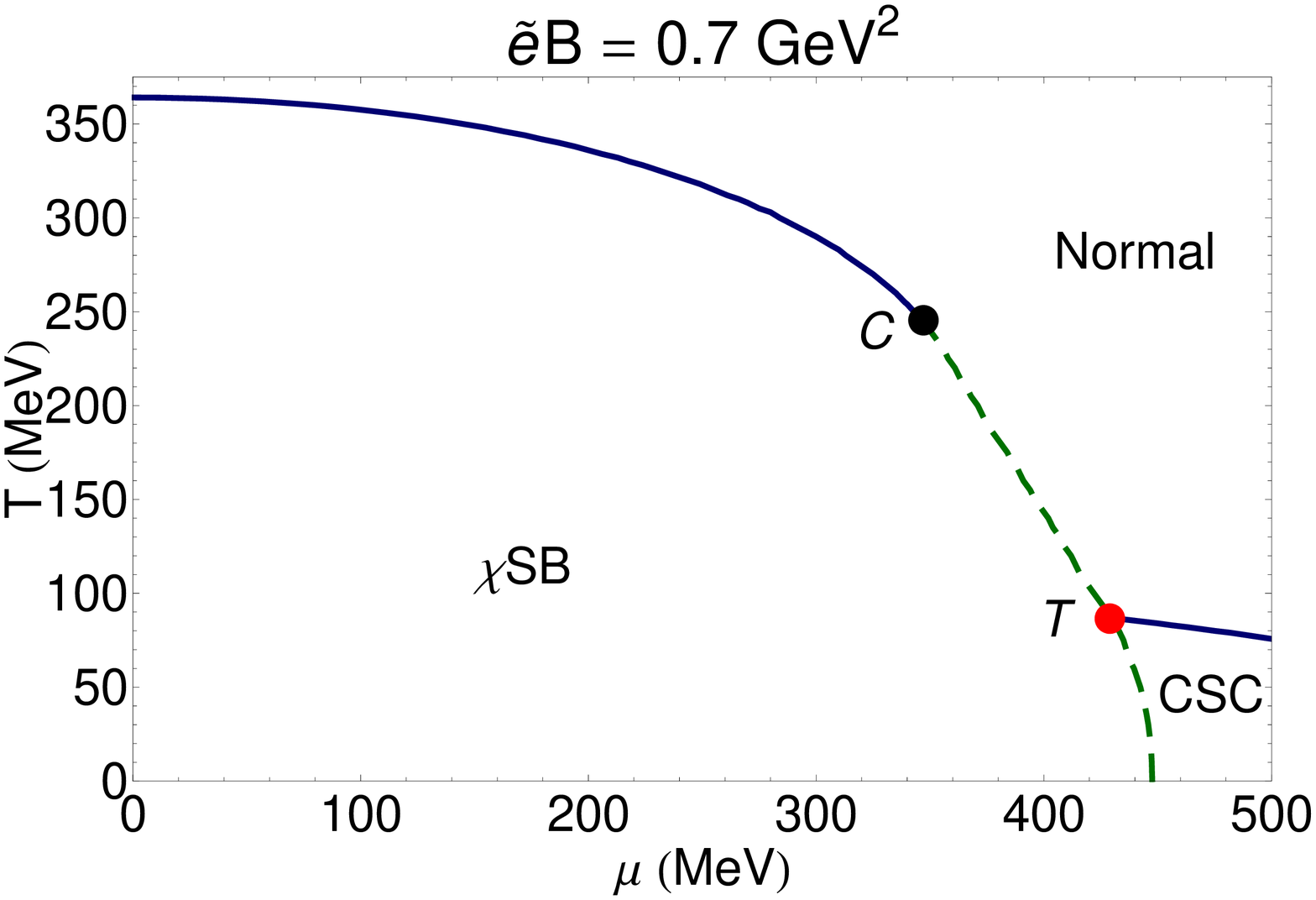}
 \caption{The $T-\mu$ phase diagram of
hot magnetized 2SC quark matter is presented for various
$\tilde{e}B$. Blue solid lines denote the second order phase
transitions and the green dashed lines the first order transitions.
The critical and tricritical points are denoted by $C$ and $T$,
respectively.}\label{figTmu}
\end{figure}

\noindent In Fig. \ref{figTmu}, the $T-\mu$ phase diagram of the
model is demonstrated for various fixed $\tilde{e}B$.
\par
As we have argued at the beginning of this section, three different
phases appear in this system; the $\chi$SB phase, the CSC phase and
the normal phase. To study the effects of constant magnetic fields
on the phase transition in the $T-\mu$ plane, let us compare the
plots from Fig. \ref{figTmu} with the plots from Figs.
\ref{fig1}-\ref{fig3NN}. As we have observed in Figs.
\ref{fig1}(a)-\ref{fig1}(c), at a fixed and low temperature
$T<T_{1}$, first order transitions occur between the $\chi$SB and
the CSC phase. By increasing the temperature to $T_{1}<T<T_{2}$
after a first order phase transition between the $\chi$SB to the
normal phase, the normal phase goes over into the CSC phase in a
second order phase transition [see 2(a)-2(b)]. At higher temperature
$T>T_{2}$, no CSC phase exists. Here, depending on the strength of
the external magnetic fields, only second or first order transitions
occur between the $\chi$SB and the normal phase [Figs.
\ref{fig3NN}(a)-\ref{fig3NN}(c)]. As it turns out, the values of
$T_{1}$ and $T_{2}$ depend on the external magnetic field, as it can
be seen also in the plots from Fig. \ref{figTmu}.
\begin{table}
\begin{tabular}{cccccc}
          \hline\hline
\multicolumn{6}{c}{\textbf{$T-\mu$ phase diagram (Fig. \ref{figTmu})}}\\
          \hline\hline
$\qquad\tilde{e}B\qquad$&&&& $\qquad(\mu_{cr},T_{cr})\qquad$&
$\qquad(\mu_{tr},T_{tr})\qquad$\\ \hline
$0$&&&&$(292,76)$&$(316,45)$\\
$0.01$&&&&$(280,84)$&$(309,43)$\\
$0.05$&&&&$(281,87)$&$(315,44)$\\
$0.10$&&&&$(273,94)$&$(307,41)$\\
$0.20$&&&&$(267,87)$&$(290,28)$\\
$0.30$&&&&$(275,109)$&$(305,41)$\\
$0.40$&&&&$(274,141)$&$(320,46)$\\
$0.50$&&&&$(278,156)$&$(331,52)$\\
$0.60$&&&&$(314,189)$&$(377,68)$\\
$0.70$&&&&$(347, 245)$&$(429,87)$ \\
$0.80$&&&&$(398,271)$&$(479,104)$\\
$0.90$&&&&$(444,304)$&$(566,111)$\\
$1.00$&&&&$(485,345)$&$(581,132)$\\
\hline\hline
\end{tabular}
\caption{Critical and tricritical points in $T-\mu$ phase diagrams
of Fig. \ref{figTmu}, denoted by $(\mu_{cr},T_{cr})$ and
$(\mu_{tr},T_{tr})$, respectively. Here, $\tilde{e}B$ is in
GeV$^{2}$, $T$ and $\mu$ are in MeV.}\label{table1}
\end{table}
\par
The phase transition from $\chi$SB to CSC phase and from CSC to the
normal phase are always of first and second order, respectively. The
order of phase transition between the $\chi$SB and the normal phase
depends, however, on the value of the external magnetic field
$\tilde{e}B$. To describe this effect, let us consider the blue
solid lines in Fig. \ref{figTmu}, that demonstrate the second order
phase transitions between the $\chi$SB and the normal phases.
Starting from high temperature and zero chemical potential, they all
end at the critical points denoted by $C$ (black bullets). They are
then followed by first order critical lines (green dashed lines)
between the $\chi$SB and the normal phase. The latter start at the
critical points $C$ and end up at the tricritical points $T$ (red
bullets), where three phases coexist. In Table \ref{table1}, the
values of the temperature and chemical potentials corresponding to
the critical points $(T_{cr},\mu_{cr})$ and the tricritical points
$(T_{tr},\mu_{tr})$ are presented.
\par
As we can see in Fig. \ref{figTmu}, by increasing the external
magnetic field, the black bullets are shifted more and more to
higher values of temperature and chemical potential (see also second
column in Table \ref{table1}), so that the distance between the
black and red bullets increases by increasing the strength of
external magnetic field. This is another effect of the external
magnetic field; increasing the magnetic field strength changes the
type of the $\chi$SB to normal phase transition from second to first
order. This can also be observed in Figs.
\ref{fig3NN}(a)-\ref{fig3NN}(c), where at high enough temperature,
the $\chi$SB phase transition to the normal phase is first of second
order and then, for larger magnetic fields, changes to a first order
phase transition.
\par
Increasing the magnetic field strength leads also to an increase in
the values of $(\mu_{tr},T_{tr})$ corresponding to the tricritical
point (see the third column of Table \ref{table1}). This confirms
the conclusion at the end of Sec. III.A.3, where it was stated that
the magnetic field above a certain threshold magnetic field enhances
the production of the diquark condensate $\Delta_{B}$, and the CSC
phase can therefore exists up to $T\simeq 100$ MeV. In what follows,
we will show that above certain threshold magnetic field,
$\tilde{e}B_{t}$, the analytical results arising in a LLL
approximation are comparable with the numerical results including
the contributions of all Landau levels. To do this, we will compare
the analytical expression for the second order critical lines of a
transition between the $\chi$SB and the normal phase with the
corresponding numerical data to determine the threshold magnetic
field for the LLL approximation. Similar comparison will be then
performed for the second order critical line of the transition
between the CSC and the normal phase.
\begin{figure}[hbt]
\includegraphics[width=7cm,height=5cm]{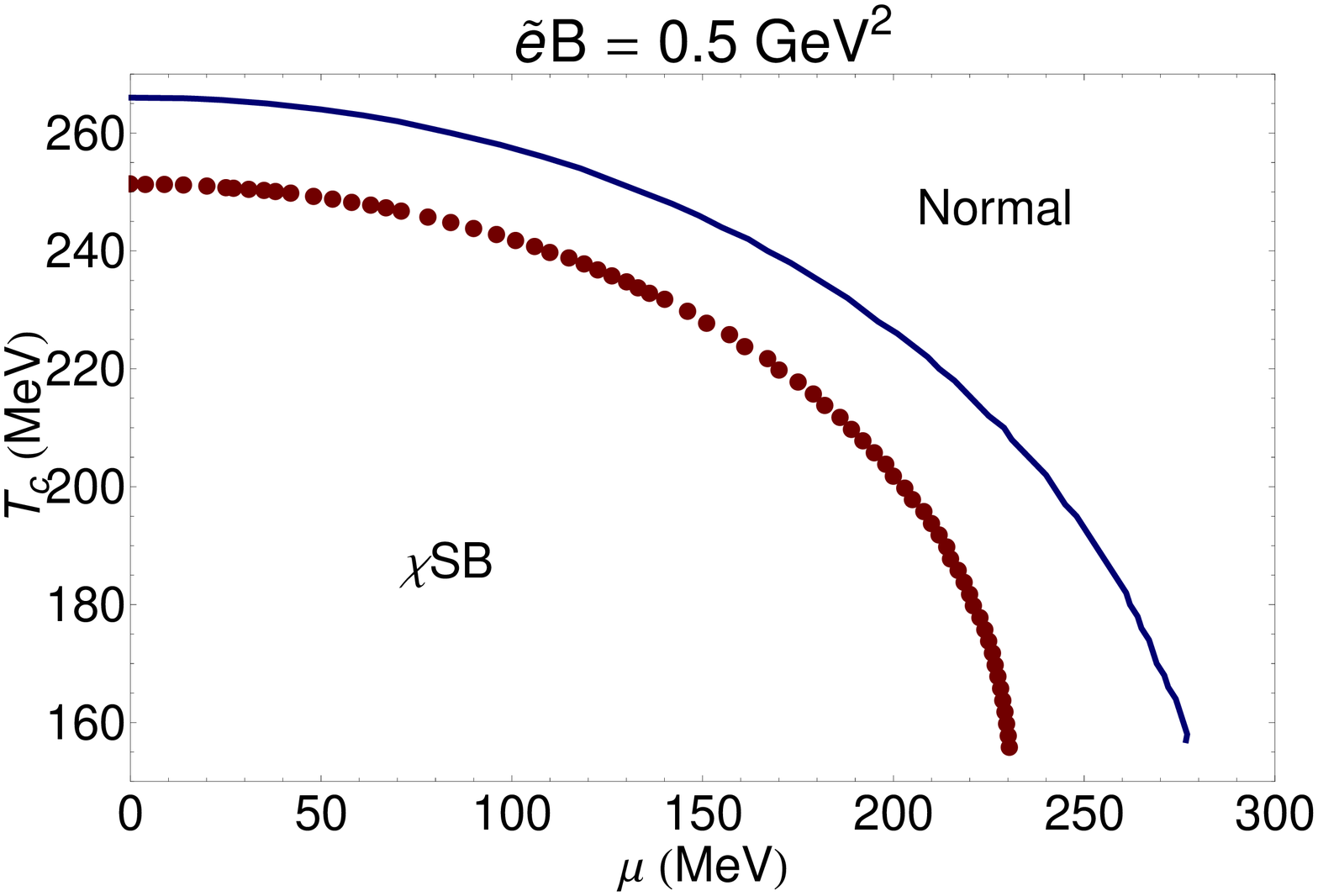}
\hspace{0.3cm}
\includegraphics[width=7cm,height=5cm]{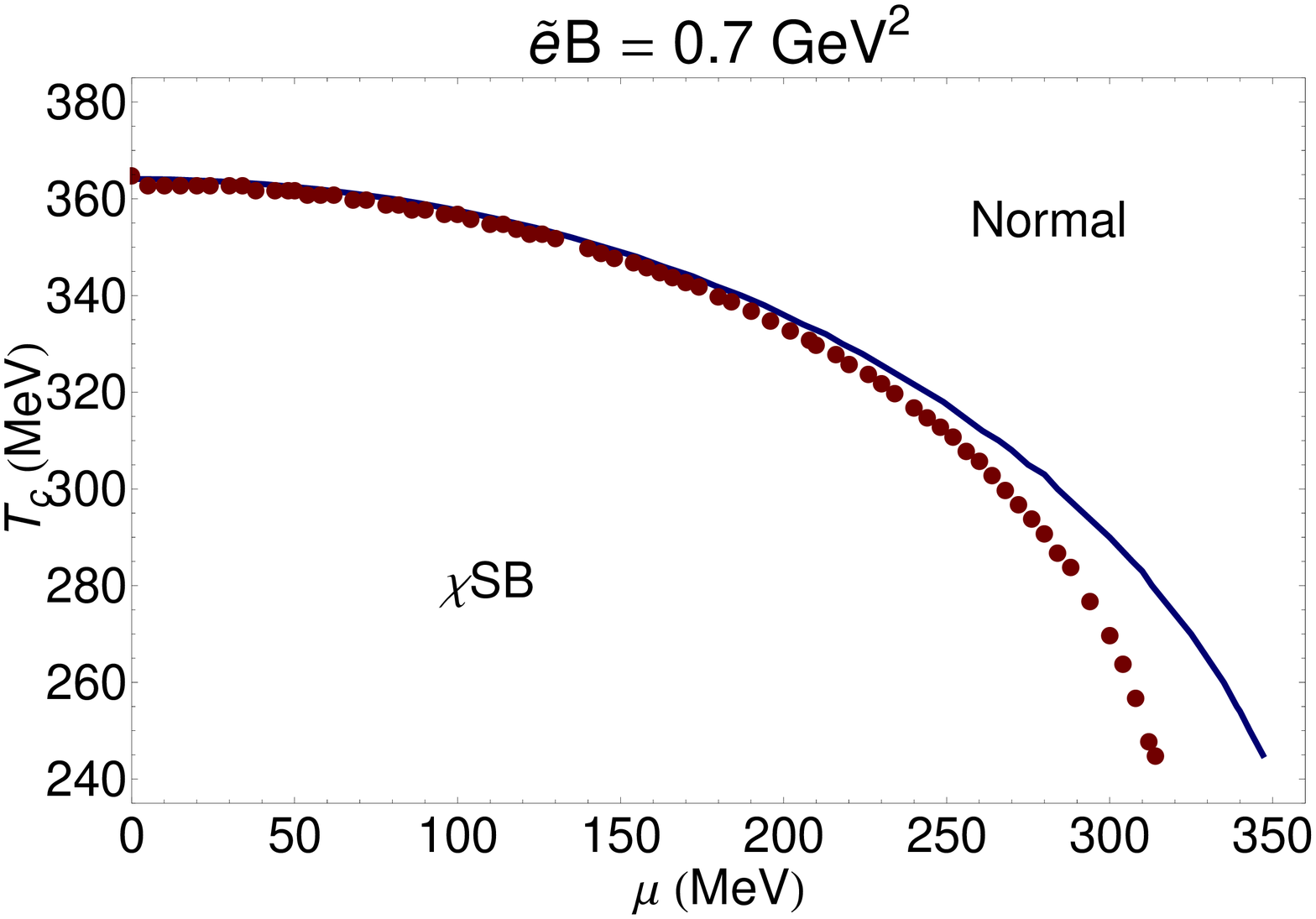}
 \caption{The second order critical lines of transition between the $\chi$SB and the normal phase. The red
 dots
 denote the analytical data that arise in a LLL approximation and the blue solid line the numerical
 data including the contributions of all Landau levels.
 For $\tilde{e}B=0.7$ GeV$^{2}$ stronger than $\tilde{e}B_{t}$, the analytical and numerical data exactly coincide.}\label{treTmu}
\end{figure}
\parspace
The second order critical surface of the transition between the
$\chi$SB and the normal phase is determined explicitly in App.
\ref{app1}. Here, we will present only the final results. Following
the method presented in \cite{sato1997}, in the phase space of the
thermodynamical parameters $(T,\mu,\tilde{e}B)$, the second order
critical surface is determined by solving \cite{sato1997,
inagaki2003}
\begin{eqnarray}\label{D11}
\lim\limits_{\sigma^{2}\to 0}\frac{\partial
\Omega_{\mbox{\tiny{eff}}}(\sigma,\Delta=0)}{\partial \sigma^{2}}=0.
\end{eqnarray}
Setting $n=0$ in $\Omega_{\mbox{\tiny{eff}}}$ from (\ref{Fb23}) and
plugging the resulting expression in (\ref{D11}), leads to the
second order critical surface in the phase space of these parameters
[see (\ref{K10}) in App. \ref{app1}]
\begin{eqnarray}\label{D12}
\frac{1}{4G_{S}}-\frac{1}{4\pi^{2}}\int_{0}^{\Lambda}dz~\bigg(2z+\frac{3\tilde{e}B}{z}\bigg)F[z;T,\mu]=0,
\end{eqnarray}
with
\begin{eqnarray}\label{D13}
F[z;T,\mu]\equiv \frac{\sinh(\beta z)}{\cosh(\beta z)+\cosh(\beta
\mu)}.
\end{eqnarray}
Expanding the integral in (\ref{D12}) in the orders of $\kappa\equiv
\frac{\mu}{\Lambda}$ and keeping only terms of the order
${\cal{O}}(\kappa^{3})$, we arrive after a straightforward
computation at [see also (\ref{K17}) in App. \ref{app1}]
\begin{eqnarray}\label{D14}
\mu^2(T,\tilde{e}B;\Lambda)\approx\frac{1}{\alpha}\bigg\{-\frac{\pi^2}{G_{S}}+4T^2\bigg[\lambda^2
-\mbox{Li}_2(-e^{-2\lambda})+2\lambda\ln\left(1+e^{-2\lambda}\right)-\frac{\pi^2}{12}\bigg]
+3\tilde{e}B\int_{0}^{\lambda}dz\frac{\tanh z}{z}\bigg\},\nonumber\\
\end{eqnarray}
where $\lambda\equiv \frac{\Lambda}{2T}$, and $\alpha$ is defined by
\begin{eqnarray}\label{D15}
\alpha(T,\tilde{e}B;\Lambda)\equiv\left(
\tanh\lambda+\lambda\tanh^2\lambda-\lambda\right)+\frac{3\tilde{e}B}{8T^2}\frac{\tanh^2\lambda}{\lambda}
+\frac{3\tilde{e}B}{8T^2}\int_{0}^{\lambda}dz\frac{\tanh^2z}{z^2}.
\end{eqnarray}
Moreover, Li$_{2}(z)$ in (\ref{D14}) is the dilogarithm function
defined by
\begin{eqnarray}\label{D19}
\mbox{Li}_{2}(z)\equiv-\int_{0}^{z}dz~\frac{\ln(1-z)}{z}.
\end{eqnarray}
To determine the second order critical line between the $\chi$SB and
the normal phase in the $T-\mu$ plane, we have to fix $\tilde{e}B$.
The analytical results arising from the LLL approximation are
demonstrated in Fig. \ref{treTmu} by red dots. Blue solid lines
denote numerical results for second order critical lines including
the contributions of all Landau levels. In Fig. \ref{treTmu}, the
analytical and the numerical results for $\tilde{e}B=0.5$ GeV$^{2}$,
approximately at the threshold, and for $\tilde{e}B=0.7$ GeV$^{2}$,
above the threshold magnetic field, are compared. The qualitative
behavior of two curves coincides above the threshold magnetic field
$\tilde{e}B_{t}\simeq 0.5$ GeV$^{2}$, where the system is in the
regime of LLL dominance.
\par
\begin{figure}[hbt]
\includegraphics[width=7cm,height=5cm]{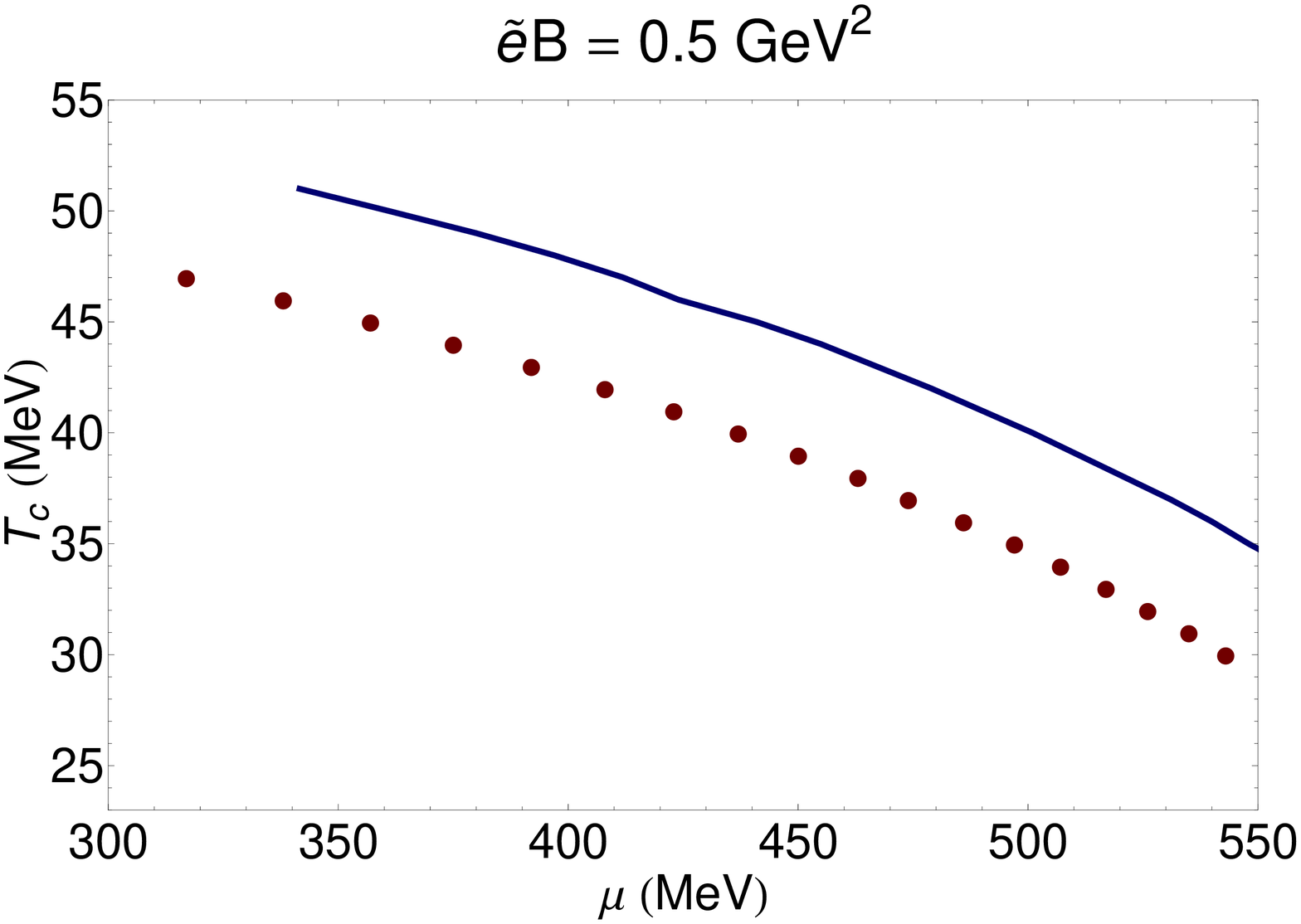}
\hspace{0.3cm}
\includegraphics[width=7cm,height=5cm]{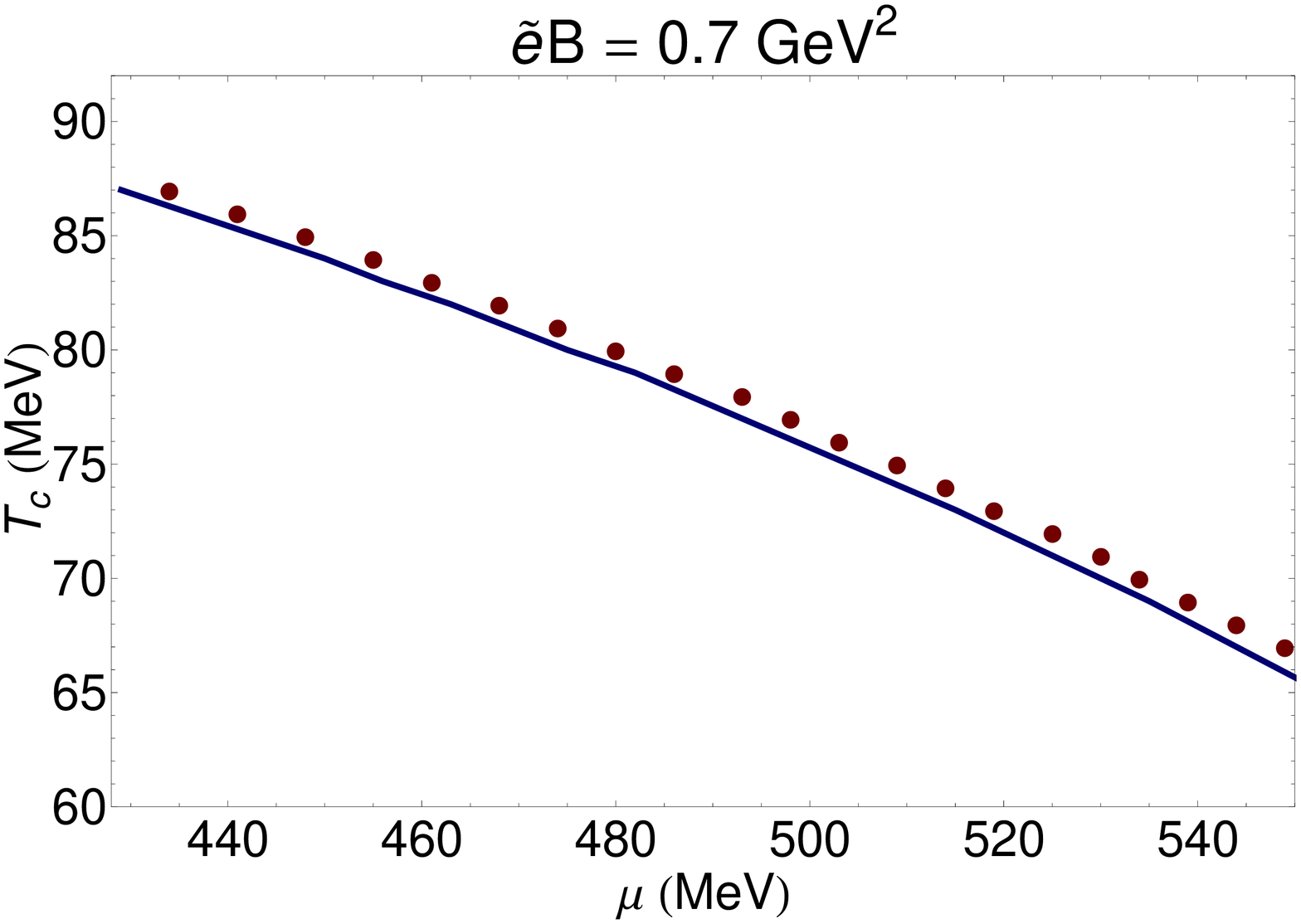}
 \caption{The second order critical lines of transition between the CSC and the normal phase. The red
 dots
 denote the analytical data that arise in a LLL approximation and the Blue solid line the numerical
 data including the contributions of all Landau levels. For $\tilde{e}B=0.7$ GeV$^{2}$ stronger than $\tilde{e}B_{t}$, the analytical and numerical data exactly coincide.}\label{comdiqTmu}
\end{figure}

Similarly, the second order critical surface between the CSC and the
normal phase is determined from \cite{sato1997,inagaki2003}
\begin{eqnarray}\label{D16}
\lim_{\Delta^{2}\rightarrow 0}\frac{\partial
\Omega_{\mbox{\tiny{eff}}}(\sigma=0,\Delta)}{\partial \Delta^{2}}=
0.
\end{eqnarray}
Setting $n=0$ in $\Omega_{\mbox{\tiny{eff}}}$ from (\ref{Fb23}) and
plugging the resulting expression in (\ref{D16}), we arrive at [see
(\ref{K21}) in App. \ref{app2}]
\begin{eqnarray}\label{D17}
\tilde{e}B^{-1}(T,\mu;\Lambda)=\frac{G_{D}}{\pi^2}\bigg[H\left(\frac{\Lambda+\mu}{2T}\right)+H\left(\frac{\Lambda-\mu}{2T}\right)
\bigg],
\end{eqnarray}
where $H(z)$ is
\begin{eqnarray}\label{D18}
H(z)\equiv
\sum\limits_{n=1}^{\infty}\frac{(-1)^{n-1}2^{2n}(2^{2n}-1)z^{2n-1}}{(2n-1)(2n)!}B_{n}.
\end{eqnarray}
Here, $B_{n}$ are the Bernoulli's numbers. Equation (\ref{D17})
leads to a relation between the phase space parameters
$(T,\mu,\tilde{e}B)$. In Fig. \ref{comdiqTmu}, we have fixed
$\tilde{e}B$ to be $\tilde{e}B=0.5,0.7$ GeV$^{2}$, and compared the
analytical data of the second order critical line (red dots),
arising from (\ref{D17})-(\ref{D18}) with the numerical data (blue
solid line).  For $\tilde{e}B>\tilde{e}B_{t}\simeq 0.5$ GeV$^{2}$,
the analytical and numerical data exactly coincide (see the plot of
$\tilde{e}B=0.7$ GeV$^{2}$ in Fig. \ref{comdiqTmu}). We conclude
therefore that for $\tilde{e}B>\tilde{e}B_{t}$ the higher Landau
level are decoupled and the properties of the phase transitions are
essentially affected by the contribution from the LLL.
\subsubsection{$T-\tilde{e}B$ phase diagram for various fixed $\mu$}
\begin{figure}[hbt]
\includegraphics[width=5cm,height=4cm]{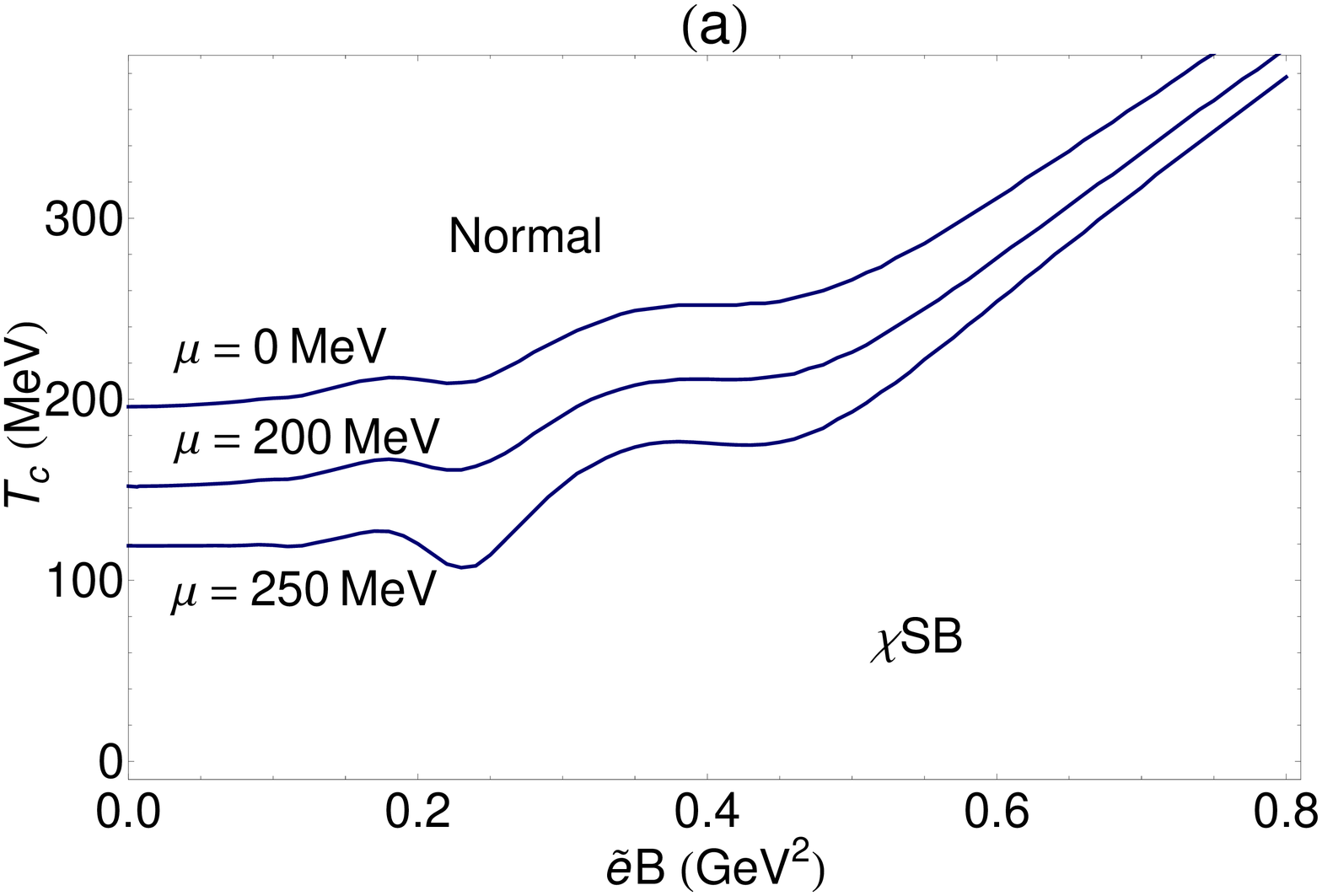}
\hspace{0.3cm}
\includegraphics[width=5cm,height=4cm]{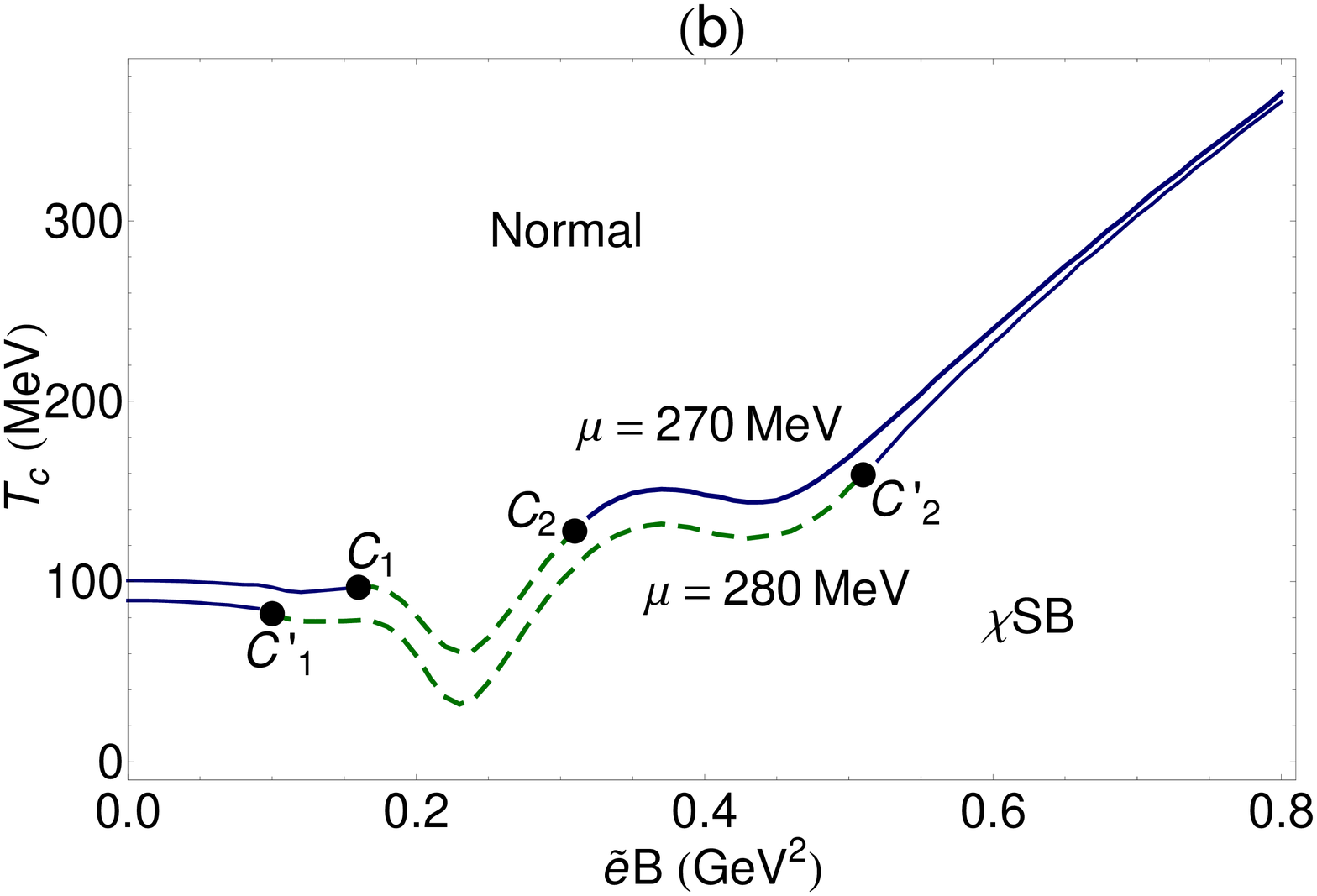}
\hspace{0.3cm}
\includegraphics[width=5cm,height=4cm]{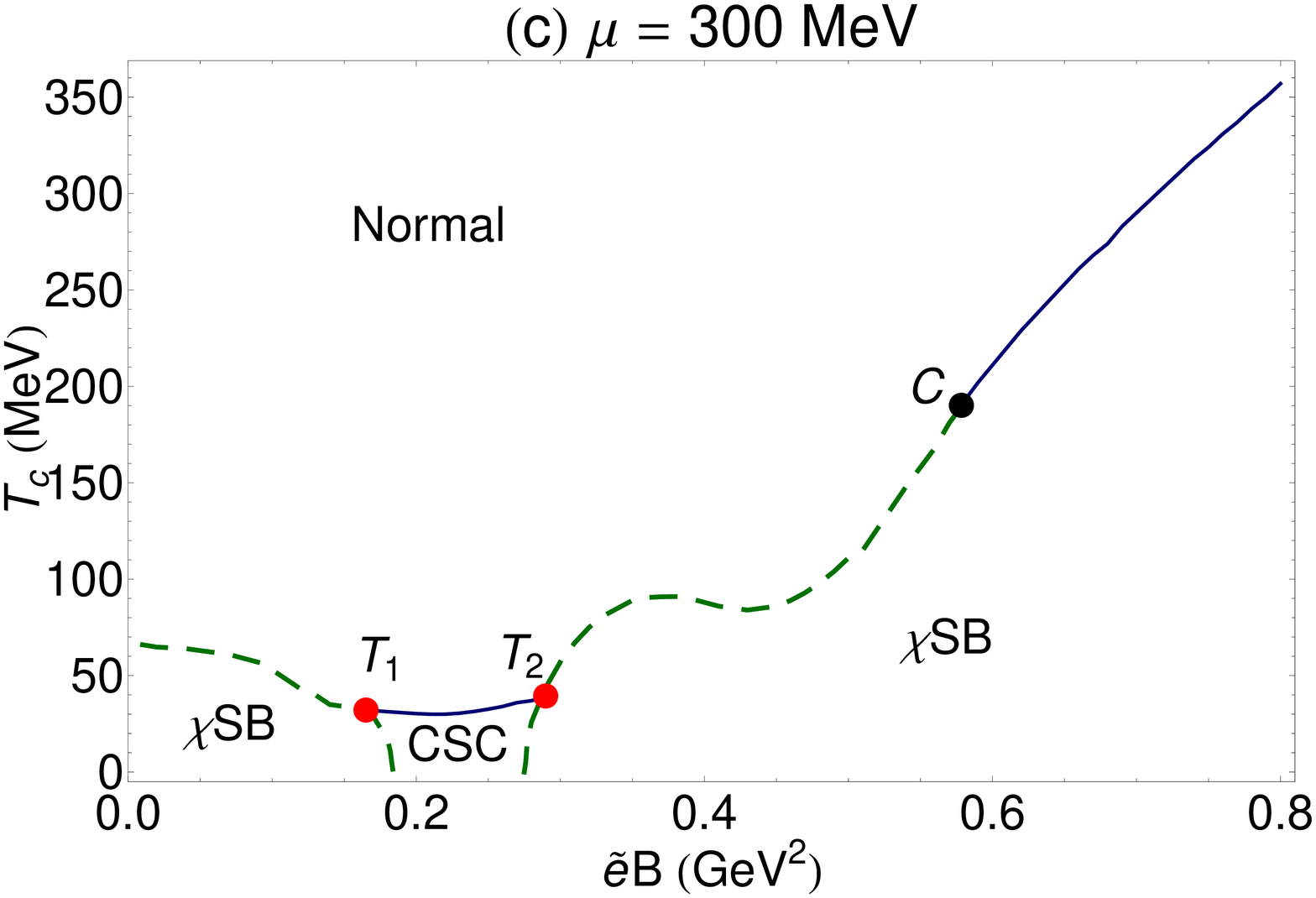}
\par\vspace{0.5cm}
\includegraphics[width=5cm,height=4cm]{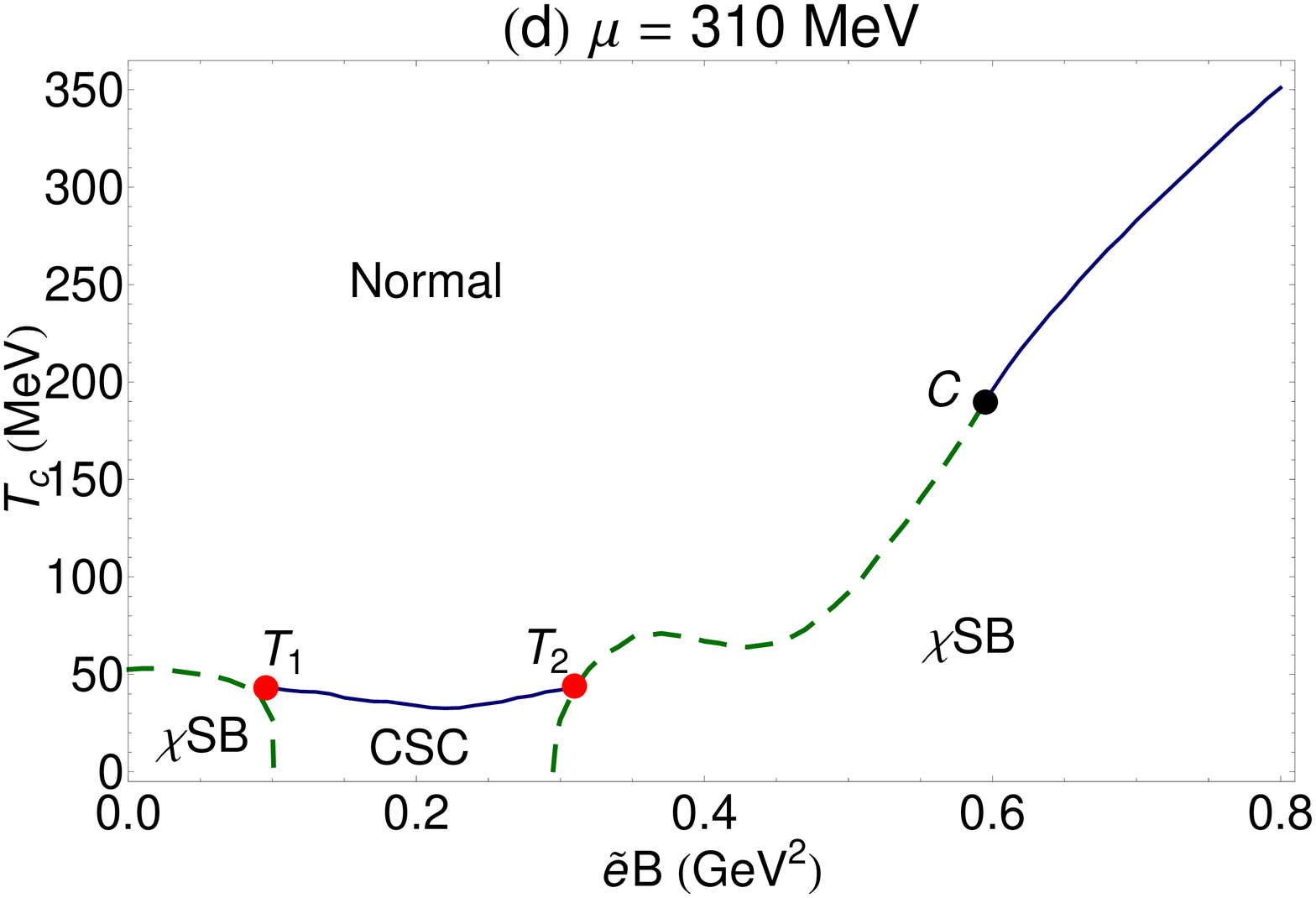}
\hspace{0.3cm}
\includegraphics[width=5cm,height=4cm]{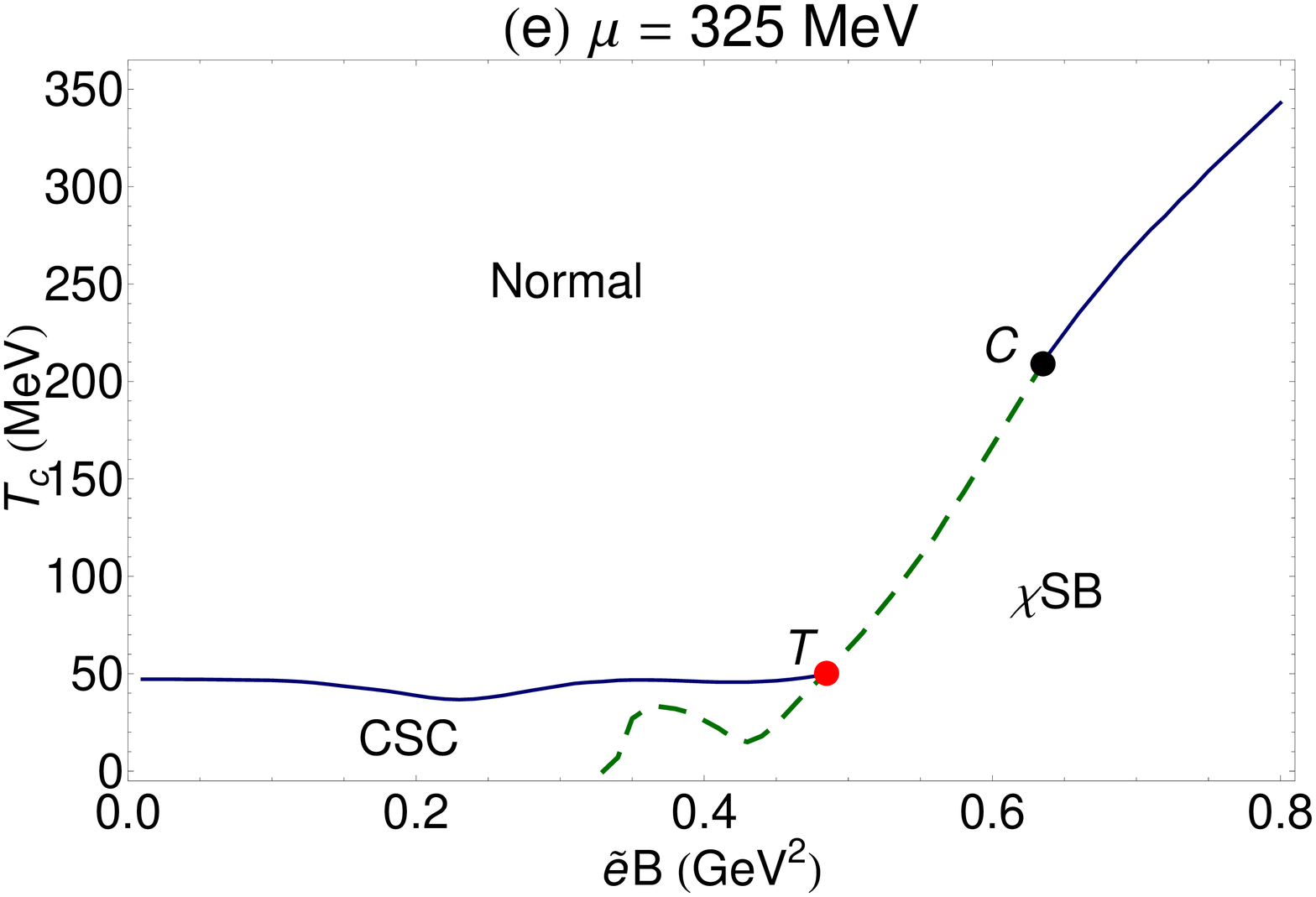}
\hspace{0.3cm}
\includegraphics[width=5cm,height=4cm]{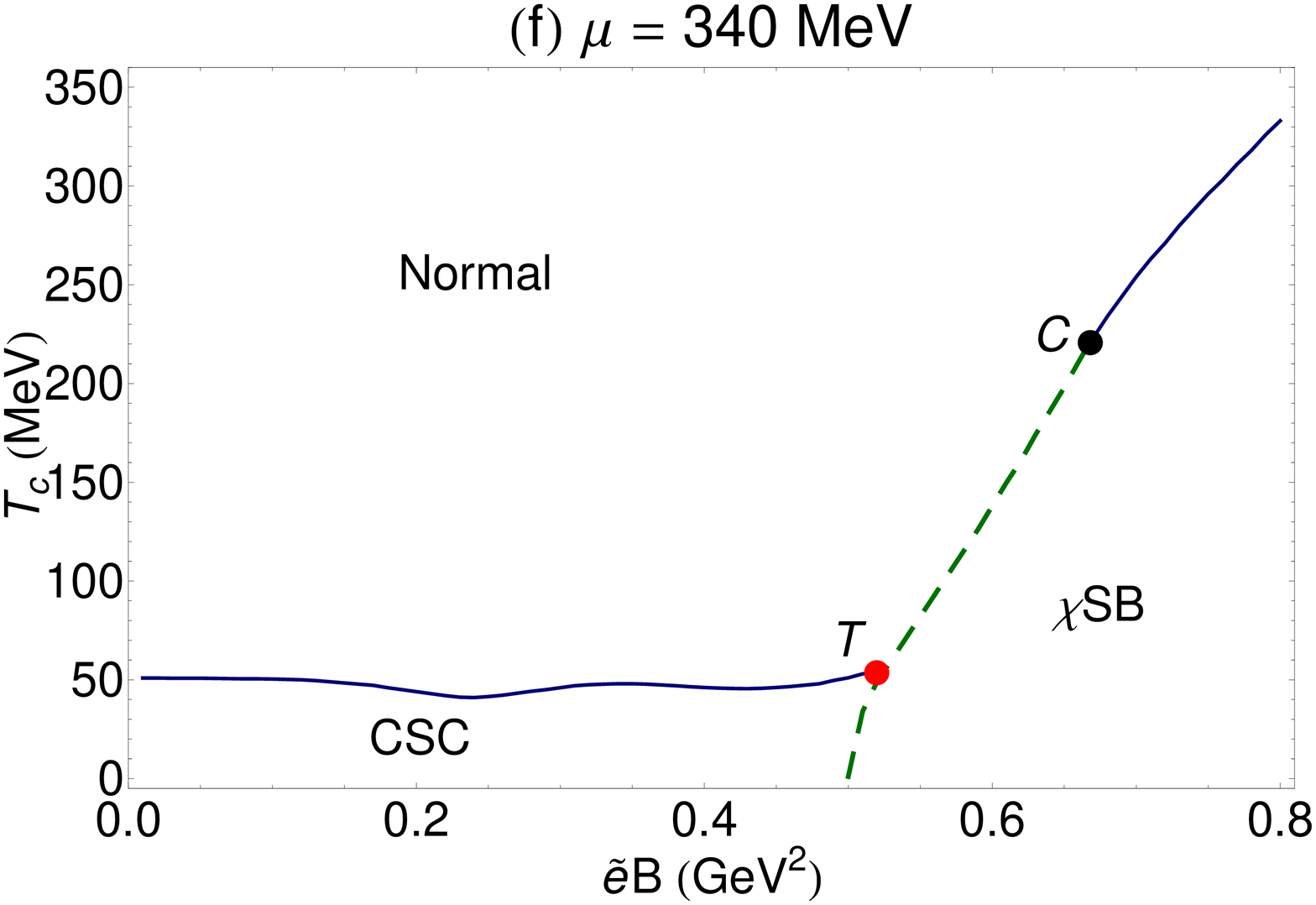}
\par\vspace{0.5cm}
\includegraphics[width=5cm,height=4cm]{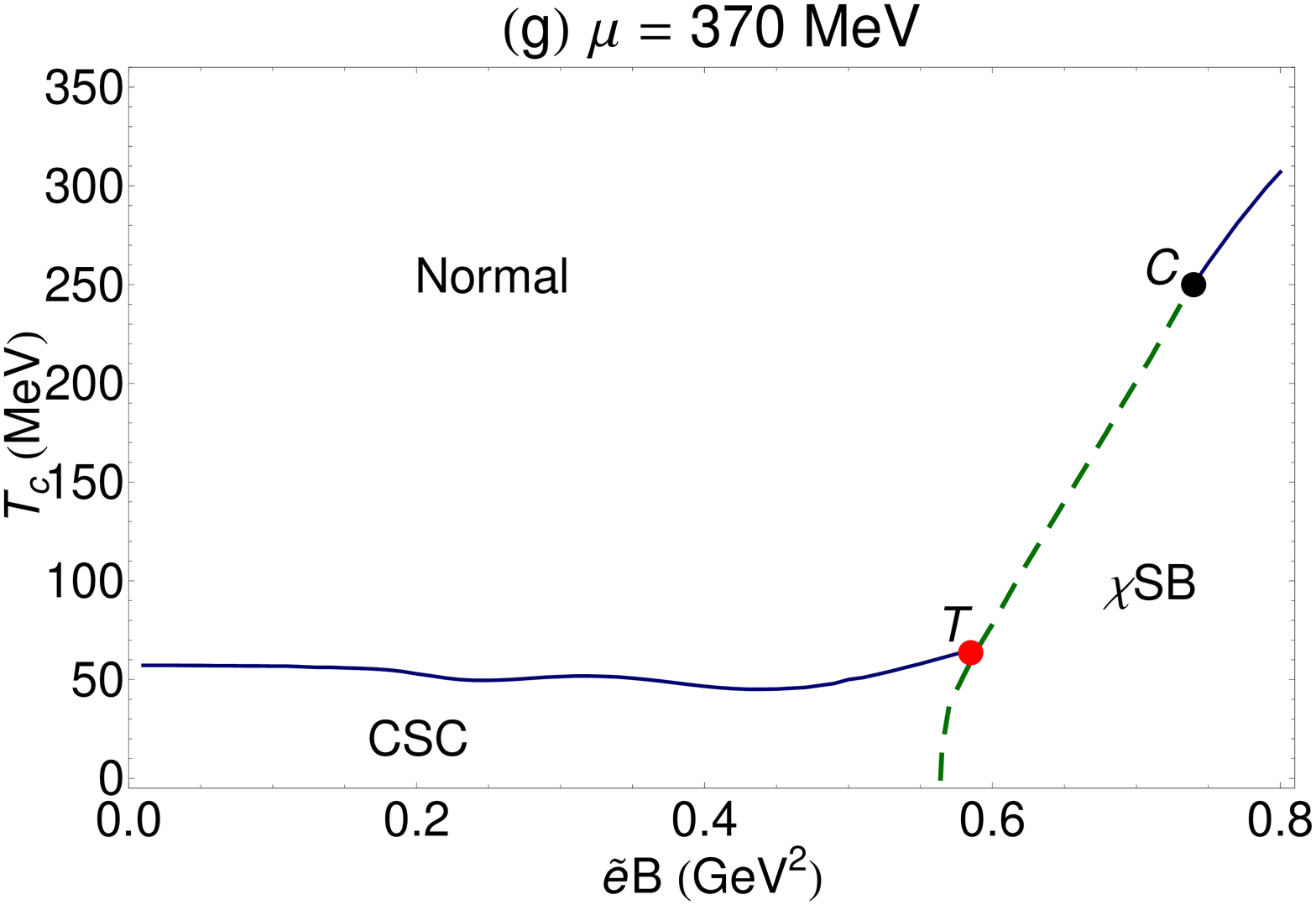}
\hspace{0.3cm}
\includegraphics[width=5cm,height=4cm]{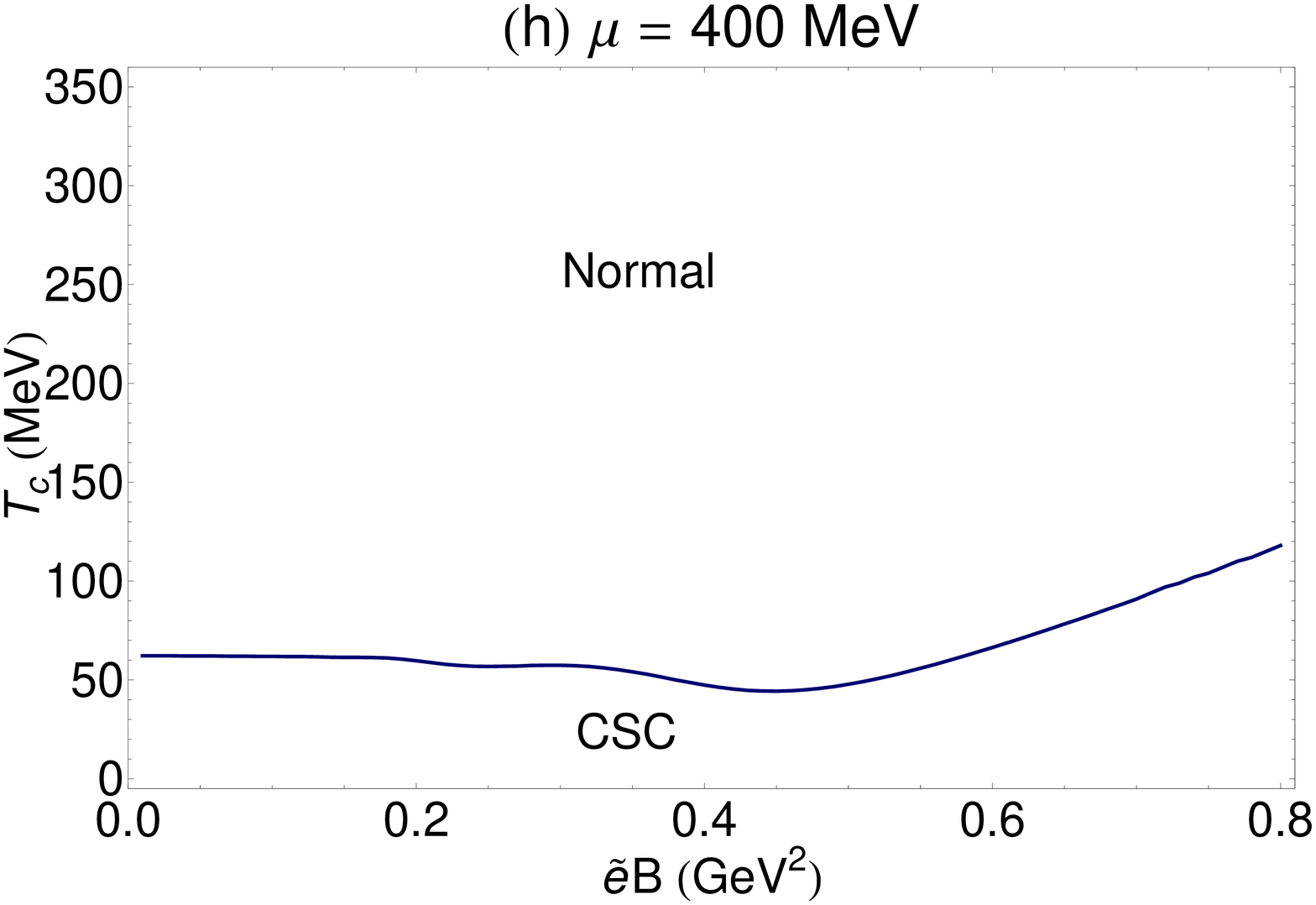}
\hspace{0.3cm}
\includegraphics[width=5cm,height=4cm]{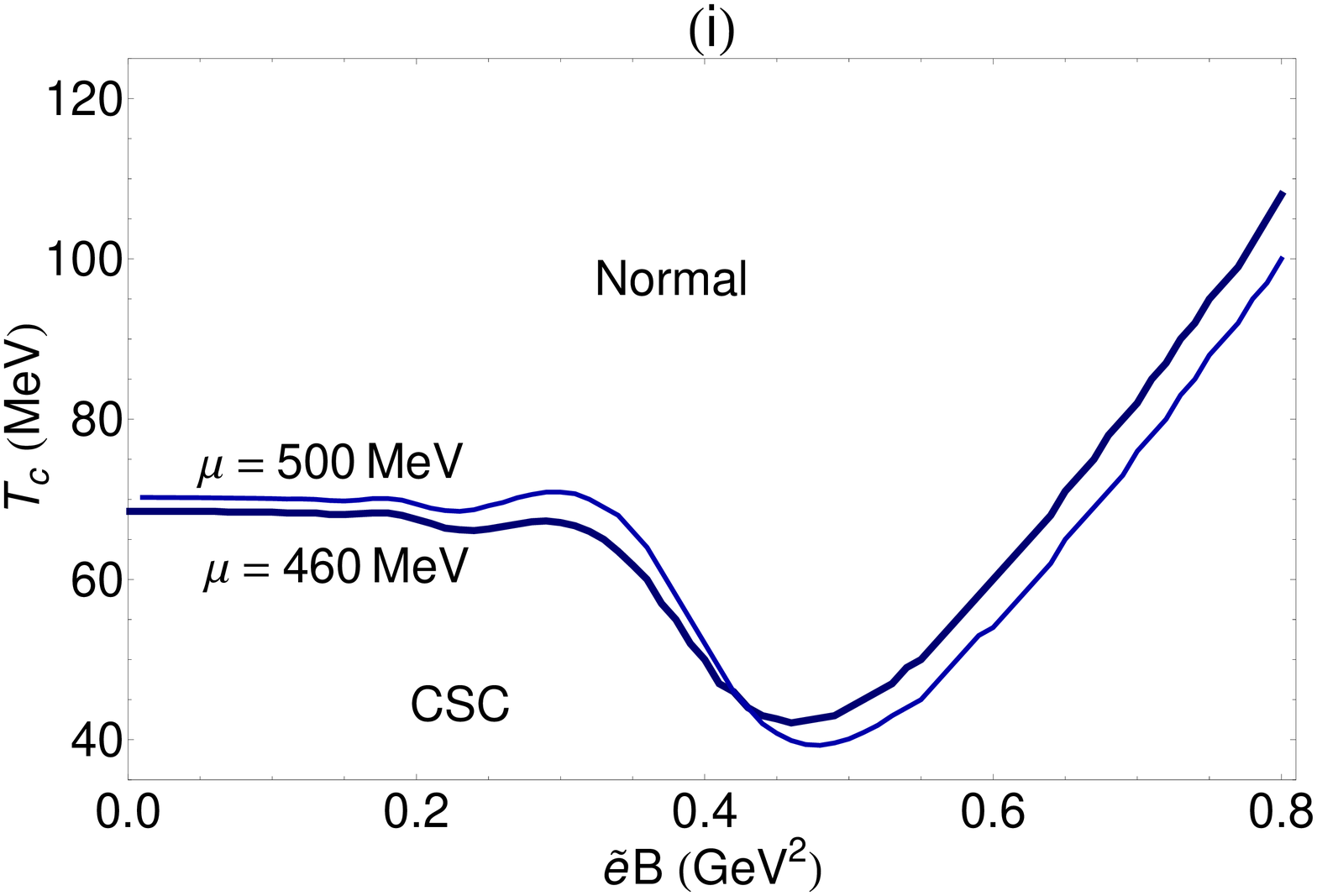}
 \caption{The $T-\tilde{e}B$ phase diagram of
hot magnetized 2SC quark matter is presented for various $\mu$. Blue
solid lines denote the second order phase transitions and the green
dashed lines the first order transitions. The critical points are
denoted by $C$ and the tricritical points by $T$.}\label{figTeB}
\end{figure}

\noindent In Fig. \ref{figTeB}, the $T-\tilde{e}B$ phase diagram of
hot 2SC quark matter is presented for various $\mu$. Blue solid
lines denote the second order phase transitions and the green dashed
lines the first order transitions. The critical points are denoted
by $C$ and the tricritical points by $T$. The exact numerical values
of the critical and tricritical points are presented in Table
\ref{table2}.
\par
\begin{table}[hbt]
\begin{tabular}{ccccccccccc}
          \hline\hline
\multicolumn{11}{c}{\textbf{$T-\tilde{e}B$ phase diagram (Fig. \ref{figTeB})}}\\
          \hline\hline
$\qquad\mu\qquad$&&&&
$\qquad(\tilde{e}B_{cr},T_{cr})_{1}\qquad$&&$\qquad(\tilde{e}B_{cr},T_{cr})_{2}\qquad$&&
$\qquad(\tilde{e}B_{tr},T_{tr})_{1}\qquad$&&$\qquad(\tilde{e}B_{tr},T_{tr})_{2}\qquad$\\
\hline
$\lesssim 250$&&&&--           &&--             &&--            &&--\\
$270$         &&&&$(0.16,97)$&&$(0.310,128)$ &&--            && --\\
$280$         &&&&$(0.10,83)$&& $(0.510,159)$&&--            &&--\\
$300$         &&&&--           &&$(0.578,190)$  &&$(0.165,32)$&&$(0.290,39)$\\
$310$         &&&&--           &&$(0.595,190)$  &&$(0.095,43)$&&$(0.310,44)$\\
$325$         &&&&--           &&$(0.635,209)$  &&--            &&$(0.485,50)$\\
$340$         &&&&--           &&$(0.668,221)$  &&--            &&$(0.520,54)$\\
$370$         &&&&--           &&$(0.740, 250)$ &&--            &&$(0.585,64)$\\
$\gtrsim 400$ &&&&--           &&--             &&--            && --\\
\hline\hline
\end{tabular}
\caption{Critical and tricritical points in $\tilde{e}B-T$ phase
diagrams of Fig. \ref{figTeB}, denoted by
$(\tilde{e}B_{cr},T_{cr})_{1,2}$ and
$(\tilde{e}B_{tr},T_{tr})_{1,2}$, respectively. Here, $\tilde{e}B$
is in GeV$^{2}$, $T$ and $\mu$ are in MeV. }\label{table2}
\end{table}
In Fig. \ref{figTeB}(a) and \ref{figTeB}(b), the critical lines are
plotted for $\mu=0,~200,~250$ MeV and $\mu=270,~280$ MeV,
respectively. For relatively small values of $\mu\lesssim 270$ MeV
the transition between the $\chi$SB and the normal phase is of
second order for the whole range of magnetic field
$\tilde{e}B\in[0,0.8]$ GeV$^{2}$. When we increase the chemical
potential to $\mu=270,280$ MeV in Fig. \ref{figTeB}(b), the type of
the phase transition changes from second to first order in the
regime $\tilde{e}B\in[0.1,0.5]$ GeV$^{2}$, below the threshold
magnetic field. By increasing the magnetic field and entering the
regime of LLL dominance, the first order phase transition goes over
into a second order transition [see in Table \ref{table2} for the
exact values of the critical points]. The same effect has been
previously observed in \cite{inagaki2003} (see Fig. 4(b) in
\cite{inagaki2003} and compare it with Fig. \ref{figTeB}(a) and
\ref{figTeB}(b) of the present paper). In Fig. \ref{figTeB}(c) the
chemical potential is increased to $\mu=300$ MeV. Here, a small
region of CSC phase appears in the regime $\tilde{e}B\in[0.1,0.3]$
GeV$^{2}$, between two regions of the $\chi$SB phase. The critical
points appearing in Fig. \ref{figTeB}(a) are shifted to higher
$(T,\tilde{e}B)$ [see also Table \ref{table2}]. The small CSC
``island'' enlarges by increasing the chemical potential to
$\mu=310$ MeV in Fig. \ref{figTeB}(d), and remove the $\chi$SB phase
appearing at $\tilde{e}B\lesssim 0.15$ GeV$^{2}$ totally at
$\mu=325$ MeV [Fig. \ref{figTeB}(e)]. This is expected from Figs.
\ref{fig1}(a)-\ref{fig1}(c), where the critical chemical potential
from the transition of $\chi$SB to CSC phase is approximately
$\mu\simeq 325$ MeV. The $\chi$SB phase appearing only for strong
magnetic field survives, but it is pushed away to the regime of LLL
dominance by increasing the chemical potential to $\mu=340$ and
$\mu=370$ MeV in Figs. \ref{figTeB}(f) and \ref{figTeB}(g). We
notice that the transition from $\chi$SB to the CSC phase is always
of first order, whereas the type of the phase transition from
$\chi$SB to the normal phase depends on the external magnetic field.
This confirm our conclusion from the previous section (See Fig.
\ref{figTmu} and the discussion in Sec. III.B.1). Increasing the
chemical potential to $\mu\geq 400$ MeV, the $\chi$SB region is
completely removed from the interval $\tilde{e}B\in [0,0.8]$
GeV$^{2}$, and a second order phase transition appears between the
CSC and the normal phase in Fig. \ref{figTeB}(h). In Fig.
\ref{figTeB}(i), the second order phase transition is plotted for
$\mu=460,~500$ MeV.\footnote{In Fig. \ref{figTeB}(i), we have
changed the scale of the plot from $T\in[0,360]$ MeV to $T\in
[35,125]$ MeV, in order to magnify the van Alfven-de Haas
oscillations that appear in the regime below the threshold magnetic
field $\tilde{e}B_{t}\simeq 0.5$ GeV$^{2}$.} Above the threshold
magnetic field, the critical temperature increases by increasing the
magnetic field. This may open the possibility to observe the 2SC
phase in future heavy ion experiments.
\par
Let us also emphasize that the plots in Fig. \ref{figTeB} are in
very good agreement with plots from previous sections. As an
example, let us consider Fig. \ref{mass-delta-eB}(b), the
$\tilde{e}B$ dependence of $\Delta_{B}$ at fixed $\mu=460$ MeV, and
compare it with the curve $\mu=460$ MeV in Fig. \ref{figTeB}(i). Let
us then focus on the range of $\tilde{e}B\in[0.4,0.6]$ GeV$^{2}$ in
both figures. As it turns out from Fig. \ref{mass-delta-eB}(b), at
low temperature $T=20$ MeV, $\Delta_{B}\simeq 70-80$ MeV in the
interval $\tilde{e}B\in[0.4,0.6]$ GeV$^{2}$, while at higher
temperature $T=60,70$ MeV, $\Delta_{B}$ vanishes and apparently a
normal phase appears in this regime of $\tilde{e}B$. The appearance
of a normal phase can be checked in Fig. \ref{figTeB}(i). For fixed
$T=20$ MeV, there is a CSC phase in the regime $\tilde{e}B\in
[0.4,0.6]$ GeV$^{2}$, while at higher temperature $T=60$ MeV and
$T=70$ MeV, a normal phase appears in the same regime of
$\tilde{e}B$. This phenomenon is because of van Alfven-de Haas
oscillations that occur in this regime of $\tilde{e}B\in [0.4,0.6]$
GeV$^{2}$ in Fig. \ref{mass-delta-eB} as well as in the phase
diagram of Fig. \ref{figTeB}(i).\footnote{Similar van Alfven-de Haas
oscillations occur in the ``gap vs. $eB/\mu^{2}$ graphs'' of a
magnetized CFL model studied in \cite{warringa2007} and
\cite{shovkovy2007}. In \cite{fayaz2010}, where the effect of
magnetic fields on the 2SC gap is studied at $T=0$, same
oscillations occur in the same regime of $\tilde{e}B$ for $\mu=460$
MeV (see Fig. \ref{fig2}a in \cite{fayaz2010}).} Moreover, the fact
that $\Delta_{B}$ increases for $\tilde{e}B$ stronger than a certain
threshold magnetic field $\tilde{e}B_{t}$ shows the important
interplay between the effect of temperature and external magnetic
field on the formation of bound states, which could be potentially
relevant in connection with the question of accessibility of 2SC
phase in experiments. Whereas below the threshold magnetic field, by
increasing the temperature, the diquark condensate is destroyed and
a normal phase is build up, the production of $\Delta_{B}$ is
enhanced by magnetic fields stronger than $\tilde{e}B_{t}$ in a
corresponding CSC phase. According to our previous studies in
\cite{fayaz2010}, we believe that above $\tilde{e}B_{t}$ the system
enters the LLL dominant regime, where the effect of magnetic
catalysis enhances the formation of mass gaps.\footnote{See the
explanation in Sec. I, for different mechanisms being responsible
for the production enhancement of meson and diquark mass gaps by
strong magnetic fields.}
\par
\begin{figure}[hbt]
\includegraphics[width=8cm,height=6cm]{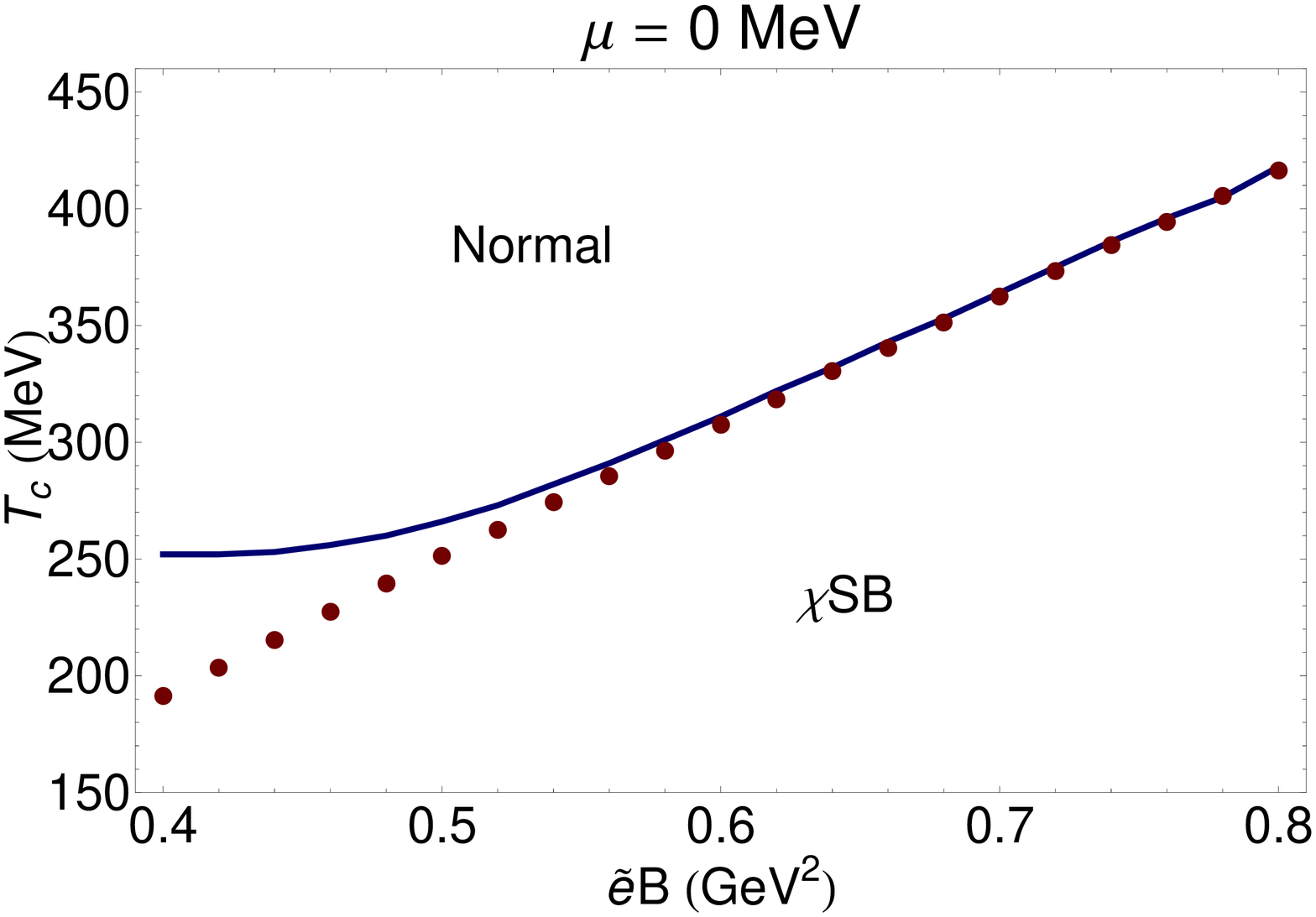}
 \caption{The second order critical lines of transition between the $\chi$SB and the normal phase. Red
 dots
 denote the analytical data that arise in a LLL approximation and the blue solid line denotes the numerical
 data including the contributions of all Landau levels. The threshold magnetic
 field is at $\tilde{e}B\simeq 0.5$ GeV$^{2}$. For $\tilde{e}B>\tilde{e}B_{t}$ the analytical and numerical data exactly coincides.}\label{comdiqTmu-2}
\end{figure}

To show how the numerical results including the contributions of all
Landau levels coincide with the analytical results consisting only
of the contribution of the LLL above the above mentioned threshold
magnetic field, we consider, as in the previous section, equation
(\ref{D12}), expressing the second order critical surface between
the $\chi$SB and the normal phase. Setting for instance $\mu=0$, we
get, as it is shown in App. \ref{app1}, the second order critical
line in the $T-\tilde{e}B$ phase [see (\ref{K12})]
\begin{eqnarray}\label{D20}
\tilde{e}B(T,\mu=0;\Lambda)=\frac{4\pi^{2}}{3H\left(\lambda\right)}\left\{
\frac{1}{4G_{S}}-\frac{\Lambda^{2}}{4\pi^{2}}+\frac{T^{2}}{12}+\frac{T^{2}}{\pi^{2}}\bigg[\mbox{Li}_{2}(-e^{-2\lambda})-2\lambda\ln(1+e^{-2\lambda})\bigg]\right\},
\end{eqnarray}
where $\lambda\equiv \Lambda/2T$. Moreover, the dilogarithm function
Li$_{2}(z)$ and $H(z)$ are defined in (\ref{D19}) and (\ref{D18}),
respectively. In Fig. \ref{comdiqTmu-2}, the analytical and
numerical data are compared. Red dots denote the analytical data and
the blue solid line the numerical data. Whereas at the threshold
magnetic field $\tilde{e}B_t\simeq 0.5$ GeV$^{2}$ the qualitative
behavior of both data are similar, for $\tilde{e}B>\tilde{e}B_{t}$
the analytical and numerical data exactly coincides.
\par
\subsubsection{$\mu-\tilde{e}B$ phase diagram for various fixed $T$}
\begin{figure}[hbt]
\includegraphics[width=5cm,height=4cm]{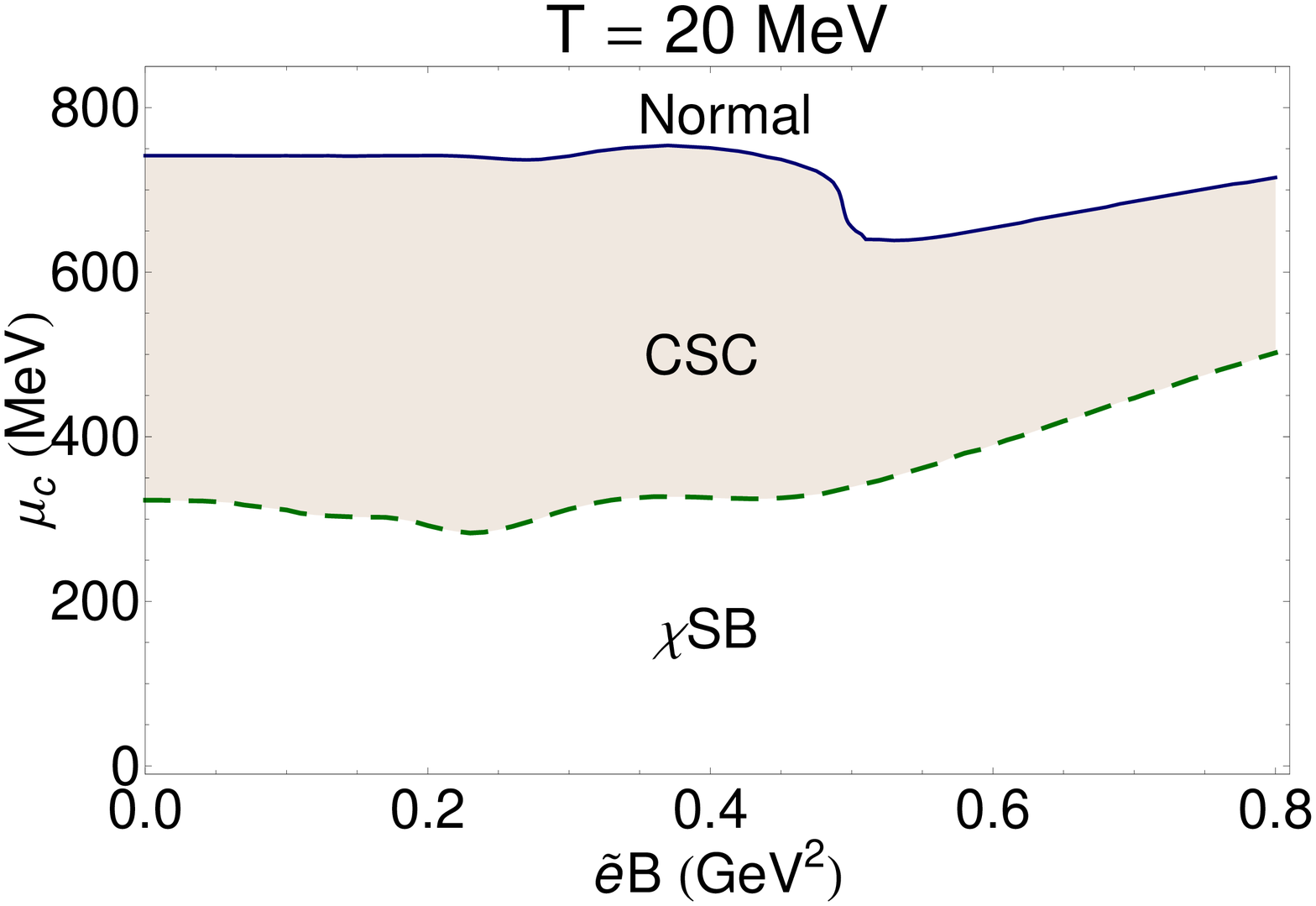}
\hspace{0.3cm}
\includegraphics[width=5cm,height=4cm]{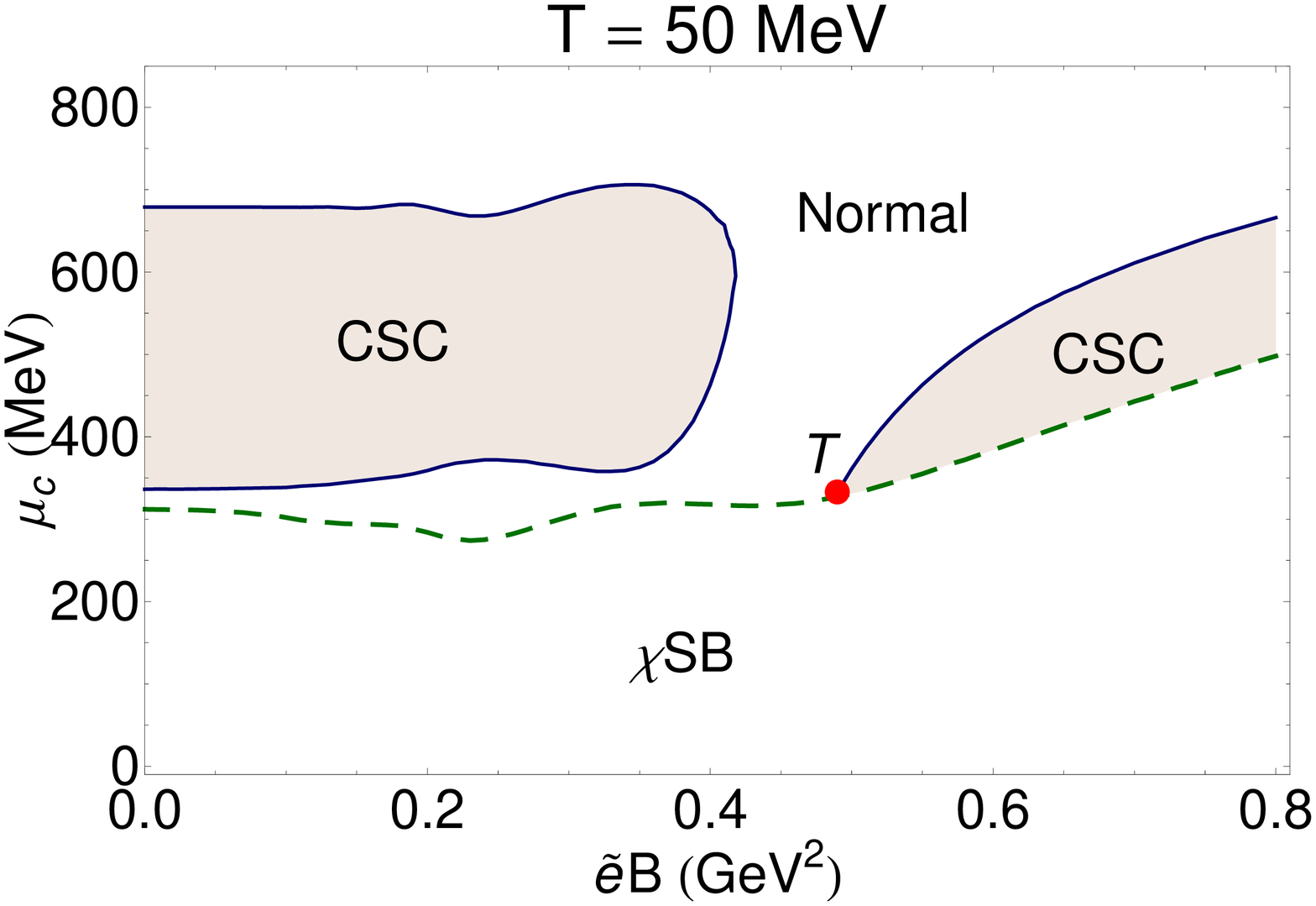}
\hspace{0.3cm}
\includegraphics[width=5cm,height=4cm]{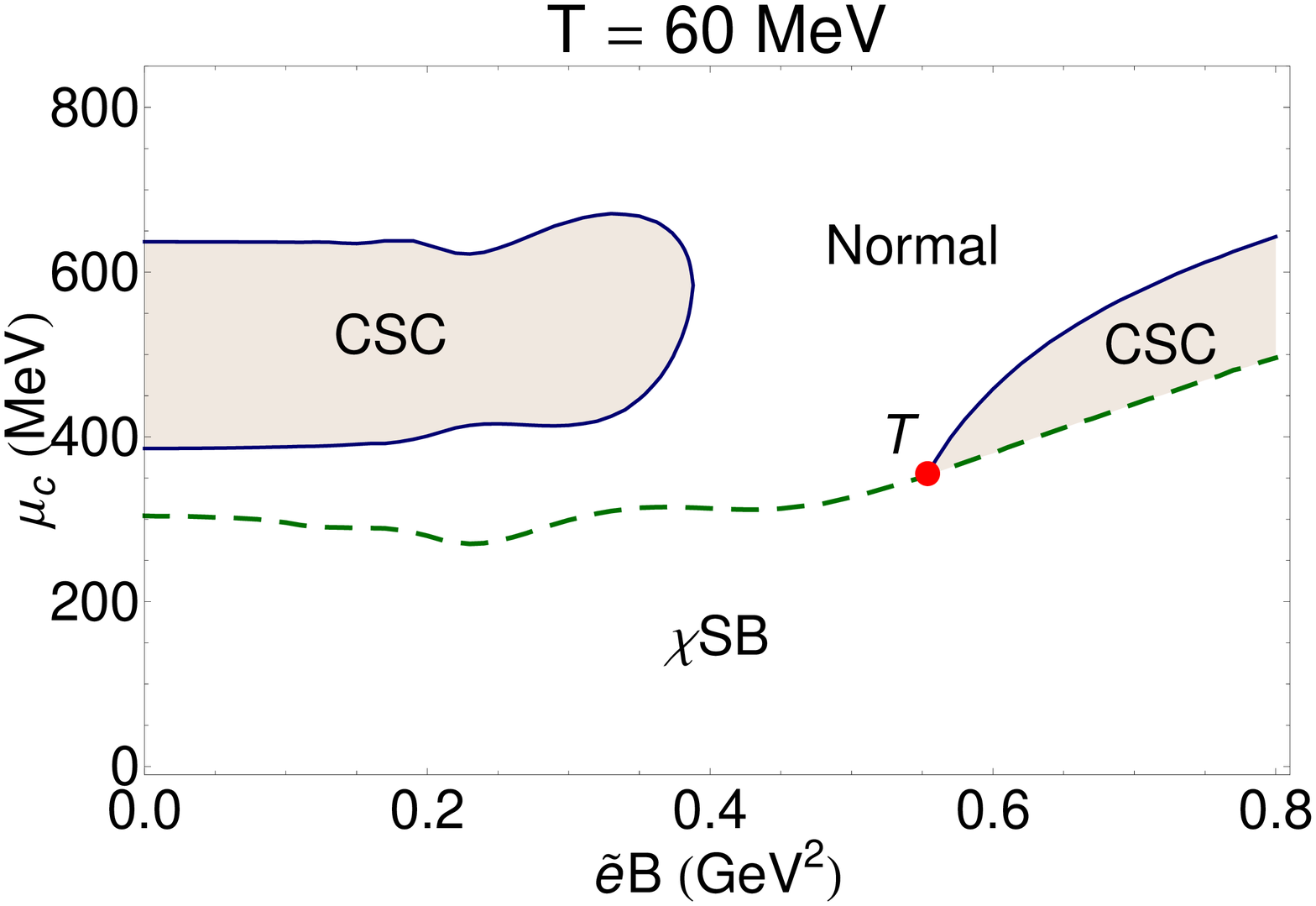}
\par\vspace{0.5cm}
\includegraphics[width=5cm,height=4cm]{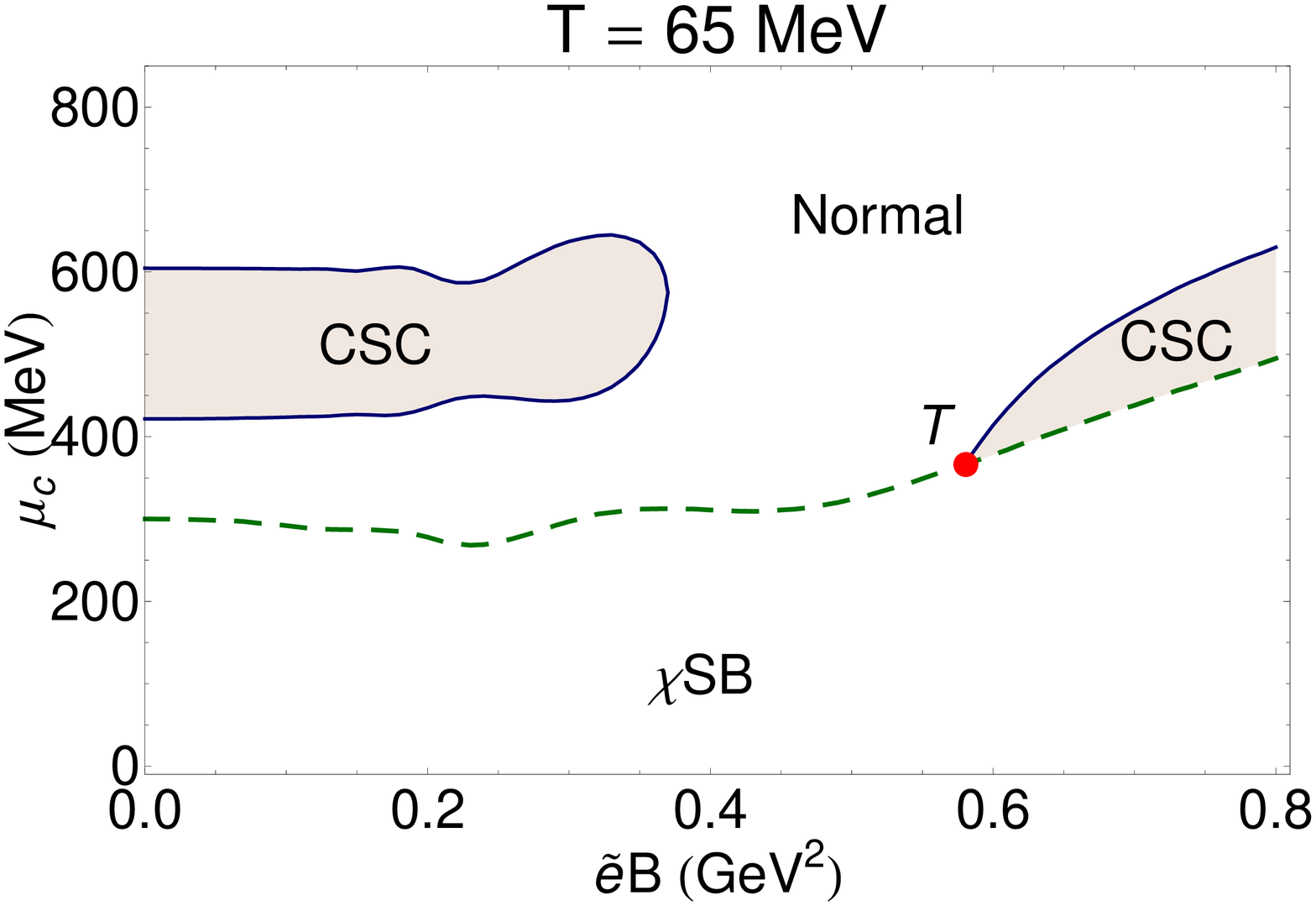}
\hspace{0.3cm}
\includegraphics[width=5cm,height=4cm]{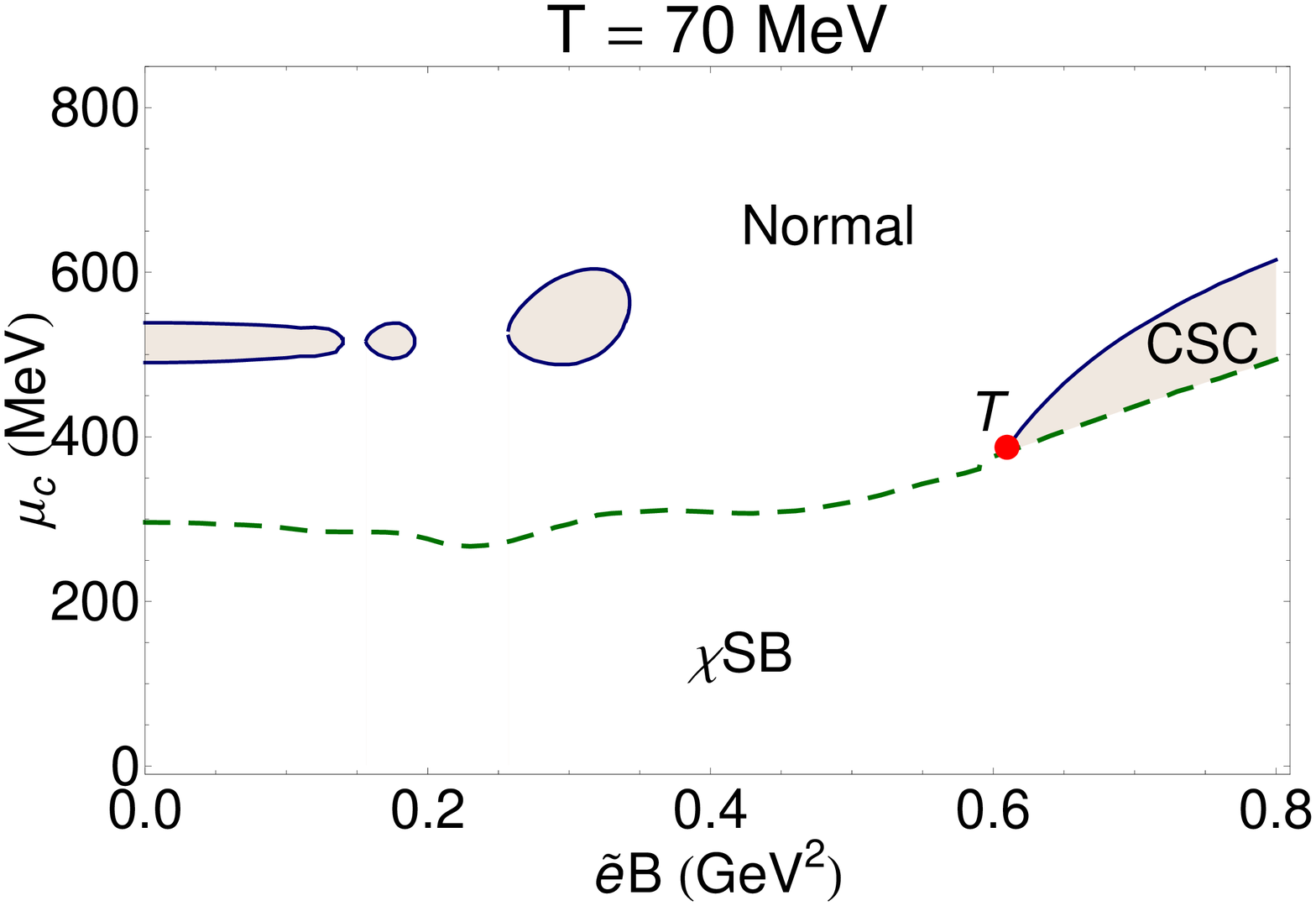}
\hspace{0.3cm}
\includegraphics[width=5cm,height=4cm]{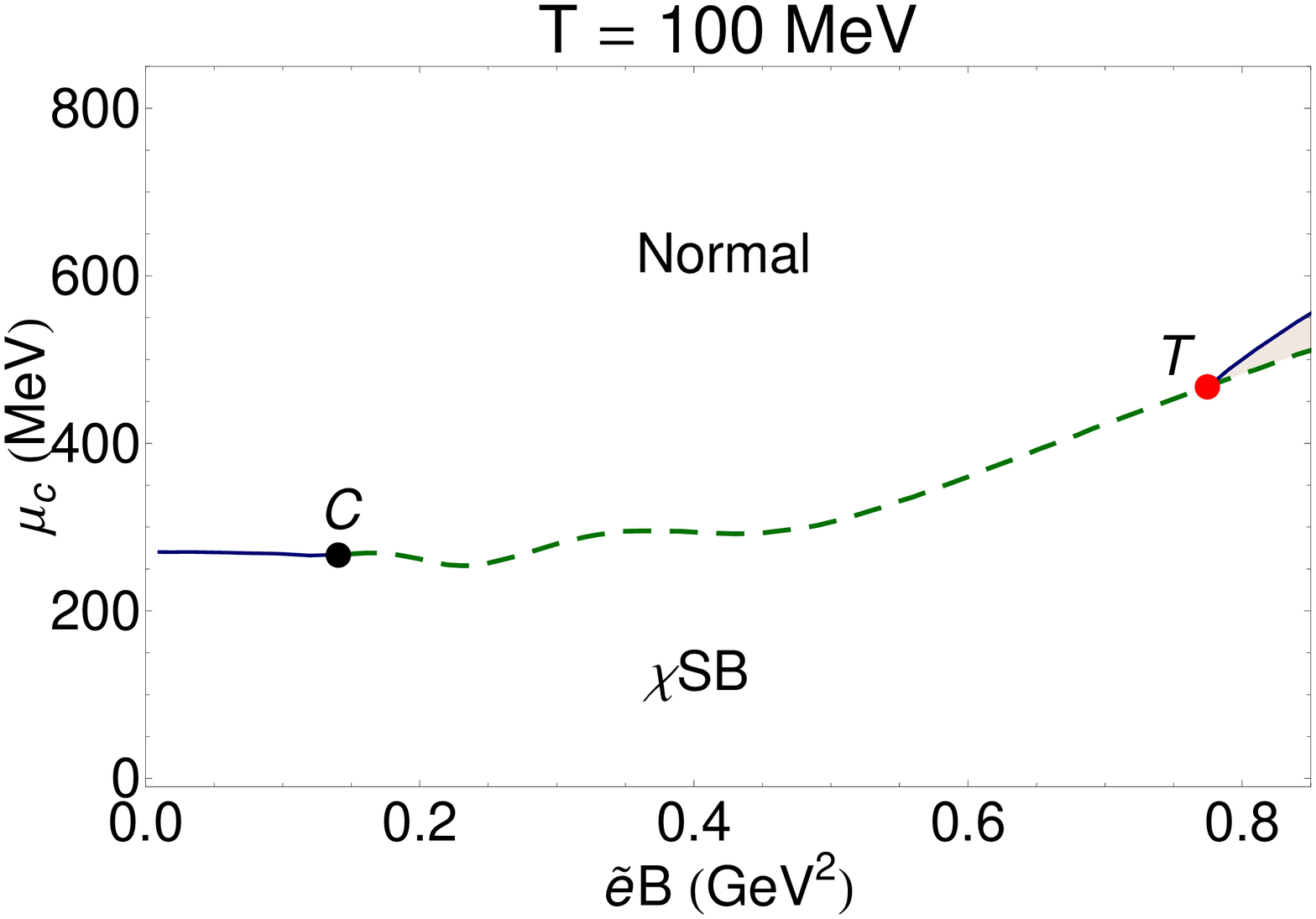}
\caption{The $\mu-\tilde{e}B$ phase diagram of hot magnetized 2SC
quark matter is presented for various $T$. Blue solid lines denote
second order phase transitions and the green dashed lines first
order transitions. The critical points are denoted by $C$ and the
tricritical points by $T$. The brown regions denote the CSC phases.
A small CSC phase appears at $T=100$ MeV on the right side of the
last plot. The tricritical point (red bullet) is shifted to the
regime $\tilde{e}B>0.8$ GeV$^{2}$ (see Table \ref{table3}). Black
bullet denotes the critical point.}\label{figmueBa}
\end{figure}

\begin{figure}[hbt]
\includegraphics[width=8cm,height=6cm]{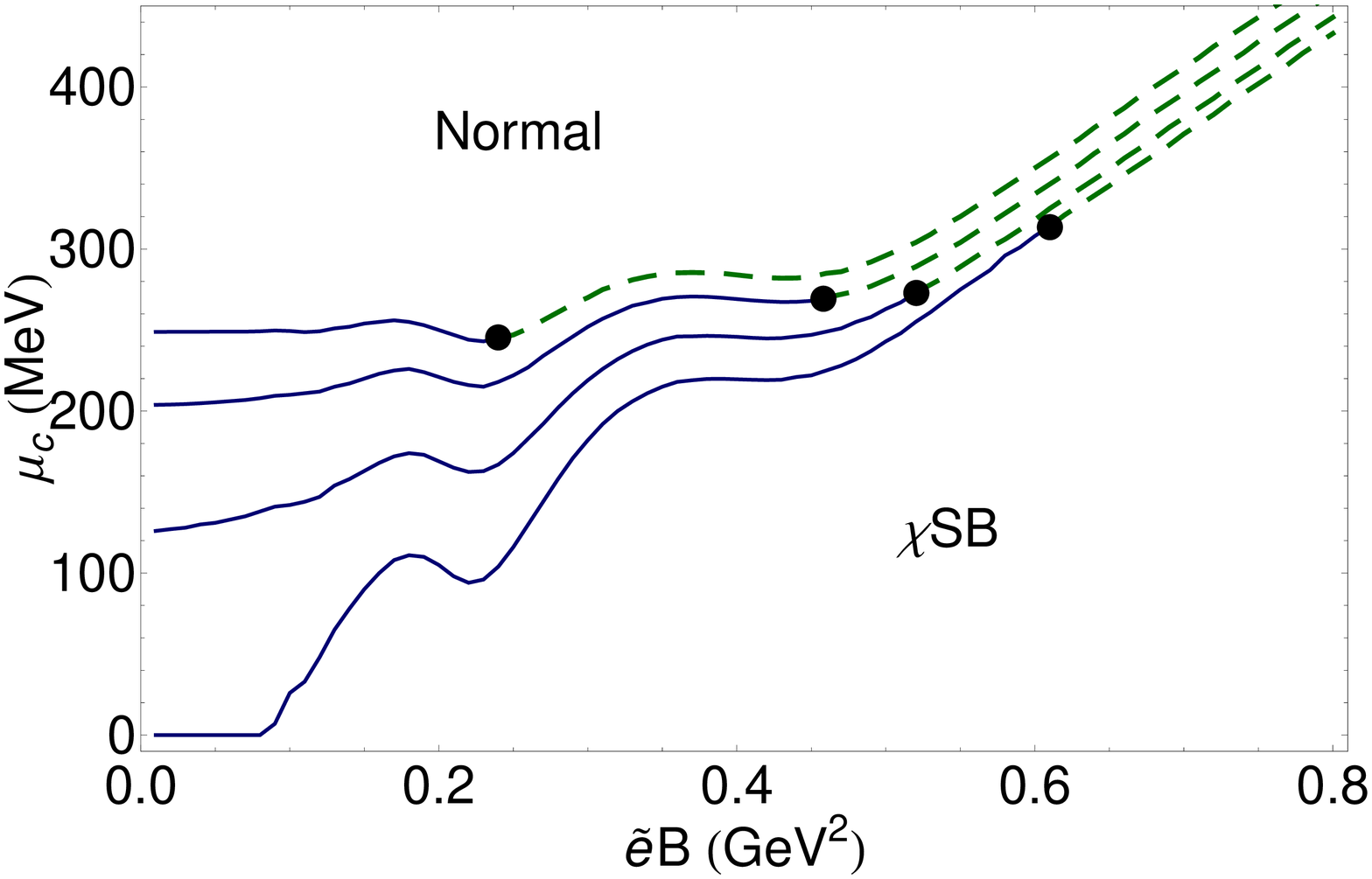}
\caption{$\mu-\tilde{e}B$ phase diagram for $T=120,150,180,200$ MeV
(from top to bottom).}\label{figmueBb}
\end{figure}

\noindent In Figs. \ref{figmueBa} and \ref{figmueBb}, the
$\mu-\tilde{e}B$ phase diagram of hot magnetized 2SC quark matter is
illustrated for various fixed temperature $T=20,\cdots 200$ MeV.
Green dashed lines denote the first order phase transitions and blue
solid lines the second order phase transitions. The critical points
are denoted by $C$ (black bullets) and the tricritical points by $T$
(red bullets).
\par
Let us consider the first plot in Fig. \ref{figmueBa}, the
$\mu-\tilde{e}B$ phase diagram at $T=20$ MeV. This plot is similar
to the phase diagram, which was found in \cite{fayaz2010} at $T=0$.
It includes a first order phase transition between the $\chi$SB and
the CSC phase. The latter goes over into the normal phase in a
second order phase transition.\footnote{As we have explained in
Footnote 2, in a three-flavor NJL model at low temperature and $\mu>
500$ MeV, the CSC phase goes over into a CFL color-superconducting
phase. In our model, where no CFL phase can be built, the second
order transition between the CSC and the normal phase at low
temperature and $\mu>500$ MeV is only assumed to exist. The same
assumption is also done in \cite{fayaz2010}.} This plot confirms our
findings in Figs. \ref{fig1}(a)-\ref{fig1}(c) and Fig. \ref{figTmu}.
At $T\simeq 50$ MeV, the CSC phase is broken into two separated
islands by the normal phase, and a tricritical point (red bullet)
appears at $(\tilde{e}B,\mu)=(0.490~\mbox{GeV}^2, 333~\mbox{MeV})$.
The normal phase between two CSC islands appears at
$\tilde{e}B\in[0.4,0.6]$ GeV$^{2}$. This confirms our results from
\ref{sigma-eB-mu}(b), where in the same regime of $\tilde{e}B$, the
CSC condensate $\Delta_{B}$ vanishes and a normal phase occurs. By
increasing the temperature both CSC islands shrink, so that at
$T=70$ MeV only three separated CSC islands are remained in the
regime below the threshold magnetic field $\tilde{e}B\leq 0.5$
GeV$^{2}$. They are then totally destroyed at $T\geq 100$ MeV and in
the relevant interval $\tilde{e}B\in[0,0.8]$ GeV$^{2}$. Moreover, by
increasing the temperature from $T\simeq 50$ MeV to $T\simeq 100$
MeV, the CSC island and the corresponding tricritical point that
exist in the regime above the threshold magnetic field are shifted
away to higher values of $\mu$ and $\tilde{e}B$ (for the exact
values of the tricritical points see Table \ref{table3}).
\par
\par
At $T\geq 100$ MeV, the $\mu-\tilde{e}B$ phase space consists only
of the $\chi$SB phase and the normal phase. The critical points that
appear at $T=100$ MeV at $(0.14~\mbox{GeV}^{2},267.3~\mbox{MeV})$ is
shifted to higher values of $\tilde{e}B$ and $\mu$ as it can be seen
in Fig. \ref{figmueBb} and Table \ref{table3}. High temperature
suppresses the production of the chiral condensate in the regime of
small magnetic fields (compare the curves in Fig. \ref{figmueBb} in
the regime $\tilde{e}B\in [0,0.4]$ GeV$^{2}$). This effect is
compensated by increasing the magnetic field to above the threshold
magnetic field.
\par
\par
\begin{table}[hbt]
\begin{tabular}{ccccccc}
          \hline\hline
\multicolumn{7}{c}{\textbf{$\mu-\tilde{e}B$ phase diagrams (Figs. \ref{figmueBa}, \ref{figmueBb})}}\\
          \hline\hline
$\qquad T\qquad$&&&& $\qquad(\tilde{e}B_{cr},\mu_{cr})\qquad$&&
$\qquad(\tilde{e}B_{tr},\mu_{tr})\qquad$\\ \hline
$\lesssim 20$&&&&--                                        &&-- \\
$50$         &&&&--                                        &&$(0.490, 333)$\\
$60$         &&&&--                                        &&$(0.554, 356)$\\
$65$         &&&&--                                        &&$(0.580, 365)$\\
$70$         &&&&--                                        &&$(0.610, 387)$\\
$100$        &&&&$(0.140,267)$                            && $(0.774, 467)$ \\
$120$        &&&&$(0.240,245)$                            && --\\
$150$        &&&&$(0.458,269)$                           && --\\
$180$        &&&&$(0.520,273)$                           && --\\
$200$        &&&&$(0.610,314)$                           && --\\
\hline\hline
\end{tabular}
\caption{Critical and tricritical points in $\tilde{e}B-\mu$ phase
diagrams of Figs. \ref{figmueBa} and \ref{figmueBb}, denoted by
$(\tilde{e}B_{cr},\mu_{cr})$ and $(\tilde{e}B_{tr},\mu_{tr})$,
respectively. Here, $\tilde{e}B$ is in GeV$^{2}$, $T$ and $\mu$ are
in MeV. }\label{table3}
\end{table}
\section{Concluding remarks}
\noindent In this paper, we have introduced a two-flavor
superconducting NJL type model including mesons and diquarks at
finite temperature $T$, chemical potential $\mu$, and in the
presence of a constant (rotated) magnetic field $\tilde{e}B$. We
were in particular interested on the effect of $(T,\mu,\tilde{e}B)$
on the formation of chiral and color symmetry breaking bound states,
and on the nature of phase transitions. One of the most important
effects of the magnetic field is its competition with $T$ and $\mu$
in the formation of bound states. This effect is demonstrated in
Figs. \ref{mass-delta-eB} and \ref{sigma-eB-mu}. As it is
illustrated in Fig. \ref{mass-delta-eB}, although it is expected
that increasing $T$ suppresses the formation of chiral and color
condensates, it turns out that this specific effect of $T$ is
minimized in the regime of strong magnetic field, for $\tilde{e}B$
larger than a certain threshold magnetic field $\tilde{e}B_{t}\simeq
0.5$ GeV$^{2}$. However, whereas the chiral condensate $\sigma_{B}$
increases by increasing $\tilde{e}B$ in the whole range of
$\tilde{e}B\in[0,0.8]$ GeV$^{2}$ [Fig. \ref{mass-delta-eB}(a)],
there are regions below $\tilde{e}B_{t}$, where the color condensate
$\Delta_{B}$ vanishes [Fig. \ref{mass-delta-eB}(b)]. Same phenomenon
is also observed in Fig. \ref{sigma-eB-mu}, where, at fixed
temperature and various chemical potentials, the dependence of
$\sigma_{B}$ and $\Delta_{B}$ on the magnetic field is plotted.
Here, as in the previous case, the effect of $\mu$ in suppressing
the production of the chiral and (2SC) color condensates is
compensated by $\tilde{e}B$ in the regime above the threshold
magnetic field.
\par
According to our results from \cite{fayaz2010}, we believe that in
the regime above $\tilde{e}B_{t}$, the dynamics of the system is
solely dominated by the lowest Landau level. In \cite{fayaz2010}, a
magnetized 2SC model at zero temperature is considered, and, the
numerical value of the threshold magnetic field is determined by
comparing the analytical $\tilde{e}B$ dependence of $\sigma_{B}$ and
$\Delta_{B}$, arising in a LLL approximation, with the corresponding
numerical data including the contribution of all Landau levels.
According to this comparison, the threshold magnetic field at $T=0$
turned out to be $\tilde{e}B_{t}\simeq 0.5$ GeV$^{2}$. In the
present paper, the same model is considered at finite temperature.
Comparing the analytical expression of the second order critical
line of the transitions from $\chi$SB to the normal phase, arising
from a LLL approximation, with the corresponding numerical data
including the contribution of all Landau levels in Fig.
\ref{comdiqTmu-2}, we arrive at the same threshold magnetic field as
in $T=0$ case. Note that in \cite{fayaz2010}, the color neutrality
condition was imposed on the magnetized 2SC quark matter. A
comparison between the presented results in this paper, where no
color neutrality is imposed, and the results from \cite{fayaz2010},
shows that color neutrality, being very small \cite{shovkovy2003},
has no significant effect on the value of the above mentioned
threshold magnetic field.\footnote{See Footnote 9.}
\par
The nature of the phase transition is also affected by the external
magnetic field. The most significant effect in this regard is
illustrated in Fig. \ref{figTmu}, where the $T-\mu$ phase space is
plotted for various $\tilde{e}B$. As it is demonstrated in this
figure, the phase space of the system includes three different
phases. We observe that the distance between the black and red
bullets, denoting the critical and the tricritical points, increases
by increasing the strength of the external magnetic field. This
implies that by increasing $\tilde{e}B$ and keeping $(T,\mu)$ fixed,
the second order transition from the $\chi$SB phase to the normal
quark matter changes into a first order transition.  This can also
be observed e.g. in Fig \ref{figmueBb}; on each isothermal critical
curve, a second order phase transition changes into a first order
one by increasing the strength of the magnetic field. Similar effect
occurs in the electroweak phase transition by implying an external
hypermagnetic field \cite{ayala2008} and in type I QED
superconductivity. Whereas the type of the phase transition from
$\chi$SB to the normal phase changes by increasing the strength of
$\tilde{e}B$, the latter has no effect on the nature of the phase
transition between the $\chi$SB and the CSC phase (first order) and
between the CSC and the normal phase (second order) [see the phase
diagrams in Figs. \ref{figTmu}, \ref{figTeB}, \ref{figmueBa}]. The
only crucial effect of $\tilde{e}B$ concerning the CSC phase, are
the observed CSC-Normal-CSC second order phase transition induced by
strong van Alfven-de Haas oscillations in the regime
$\tilde{e}B\in[0.4,0.6]$ GeV$^{2}$ [see e.g. Figs.
\ref{mass-delta-eB}(b) and \ref{figTeB}(i) at fixed $T=60$ MeV], and
an increase of the tricritical temperature and chemical potential by
increasing the magnetic field. The latter can be observed in Table
\ref{table1} for $\tilde{e}B>\tilde{e}B_{t}\simeq 0.5$ GeV$^{2}$.
This could be relevant in relation with the question addressing the
accessibility of 2SC phase in present or future heavy ion
experiments, which is recently posed in \cite{blaschke2010}. We have
extended the setup of the model used in \cite{blaschke2010} by
considering the effect of constant magnetic fields, which are
supposed to be created in the non-central heavy ion collisions
\cite{skokov2009} and are estimated to be of order $\tilde{e}B\simeq
0.03$ GeV$^{2}$ at RHIC energies, and $\tilde{e}B\simeq 0.3$
GeV$^{2}$ at LHC energies. These amounts of magnetic fields are,
according to our observation in this extended model, far below the
range of magnetic fields which could have significant effects on the
CSC phase transition by partly compensating the effects of
$(T,\mu)$, as described above. However, we believe, that more
realistic models are to be examined to find a satisfactory answer to
this interesting question.
\par
The model which is used in this paper can be extended in many ways,
e.g, by considering the 2SC-CFL phase including color neutrality.
Similar computation can also be performed within a color neutral
PNJL model. It is also intriguing to explore the effect of the axial
anomaly in the same context.
\section{Acknowledgments}
\noindent One of the authors (N.~S.) thanks the kind support and
warm hospitality of the physics department of Munich University of
Technology (TUM), where the final stage of this work is done.
\begin{appendix}
\section{Second order critical lines in the LLL approximation}
\setcounter{equation}{0} \noindent In this appendix, we will
determine the main equations leading to the second order critical
lines corresponding to the transition between the $\chi$SB and the
normal phase, as well as the transition between the $\chi$SB and the
CSC phase, in the limit of very strong magnetic field in the LLL
approximation.
\subsection{Transition between the $\chi$SB and the normal
phase}\label{app1}
\noindent In the phase space spanned by the intensive
thermodynamical parameters $(T,\mu,\tilde{e}B)$, the second order
critical surface between the $\chi$SB and the normal phase is
determined by \cite{sato1997,inagaki2003}
\begin{eqnarray}\label{K1}
\lim_{\sigma^{2}\rightarrow 0}\frac{\partial
\Omega_{\mbox{\tiny{eff}}}(\sigma,\Delta=0)}{\partial \sigma^{2}}=0.
\end{eqnarray}
Using $\Omega_{\mbox{\tiny{eff}}}$ from (\ref{Fb23}) and setting
$n=0$, to consider only the contribution of the lowest Landau level,
we get first
\begin{eqnarray}\label{K2}
&&\hspace{-0.2cm}\frac{1}{4G_{S}}-\frac{1}{2\sqrt{\pi}}\sum\limits_{\kappa=\pm{1}}\bigg\{\int\frac{d^{3}p}{(2\pi)^{3}}\int_{0}^{\infty}\frac{ds}{\sqrt{s}}\frac{p+\kappa\mu}{p}e^{-s(p+\kappa\mu)^2}
+3\tilde{e}B\int_{0}^{\infty}\frac{ds}{\sqrt{s}}\int_{0}^{\infty}\frac{dp_{3}}{4\pi^{2}}\frac{p_{3}+\kappa\mu}{p_{3}}e^{-s(p_{3}+\kappa\mu)^2}\bigg\}
\nonumber\\
&&\qquad\times
[1+2\sum\limits_{\ell=1}^{\infty}(-1)^{\ell}~e^{-\frac{\beta^{2}\ell^{2}}{4s}}]=0.
\end{eqnarray}
Denoting the temperature independent part of (\ref{K2}) with
$I_{T=0}$, and using
\begin{eqnarray}\label{K3}
\frac{1}{2\sqrt{\pi}}\sum\limits_{\kappa=\pm{1}}\int_{0}^{\infty}\frac{ds}{\sqrt{s}}\frac{x-\kappa\mu}{x}e^{-s(x-\kappa\mu)^{2}}=\frac{\theta(x-\mu)}{x},
\end{eqnarray}
from \cite{ebert2006}, we get
\begin{eqnarray}\label{K4}
I_{T=0}&=&\frac{1}{4G_{S}}-\int_{0}^{\infty}\frac{pdp}{2\pi^{2}}~\theta(p-\mu)
-3\tilde{e}B\int_{0}^{\infty}\frac{dp_{3}}{4\pi^{2}}\frac{\theta(p_{3}-\mu)}{p_{3}}\nonumber\\
&=&\frac{1}{4G_{S}}-\frac{(\Lambda^{2}-\mu^{2})}{4\pi^{2}}-\frac{3\tilde{e}B}{4\pi^{2}}\ln\left(\frac{\Lambda}{\mu}\right).
\end{eqnarray}
To evaluate the momentum integration, we have introduced the
ultraviolet cutoff $\Lambda$. Note that (\ref{K4}) is comparable
with our results from \cite{fayaz2010}. The temperature dependent
part of (\ref{K2}), which is denoted by $I_{T\neq 0}$, can be
evaluated using
\begin{eqnarray}\label{K5}
\sum\limits_{\kappa=\pm{1}}\int_{0}^{\infty}\frac{ds}{\sqrt{s}}\frac{z-\kappa\mu}{z}e^{-s(z-\kappa\mu)^{2}}e^{-\frac{\beta^{2}\ell^{2}}{4s}}=\frac{\sqrt{\pi}}{z}\left(e^{-\beta\ell
|z-\mu|}\mbox{sgn}(z-\mu)+e^{-\beta\ell(z+\mu)}\right),
\end{eqnarray}
and is given by
\begin{eqnarray}\label{K6}
I_{T\neq
0}&=&\sum\limits_{\ell=1}^{\infty}(-1)^{\ell}\bigg\{\int_{0}^{\mu}\frac{dz}{2\pi^{2}}e^{-\ell\beta\mu}\sinh(\ell\beta
z)\left(\frac{3\tilde{e}B}{z}+2z\right)-\int_{\mu}^{\Lambda}\frac{dz}{2\pi^{2}}e^{-\ell\beta
z}\cosh(\ell\beta
\mu)\left(\frac{3\tilde{e}B}{z}+2z\right)\bigg\},\nonumber\\
\end{eqnarray}
where $z$ is a generic integration variable replacing
$p=|{\mathbf{p}}|$ or $p_{3}$ in the one dimensional integrations of
(\ref{K2}). To perform the summation over $\ell$, we use
\begin{eqnarray}\label{K7}
\sum\limits_{\ell=1}(-1)^{\ell}\sinh\left(\ell\beta
z\right)e^{-\ell\beta\mu}&=&\frac{1}{2}F[z;T,\mu],\nonumber\\
\sum\limits_{\ell=1}^{\infty}(-1)^{\ell}\cosh\left(\ell\beta\mu\right)e^{-\ell\beta
z}&=&\frac{1}{2}\left(F[z;T,\mu]-1\right),
\end{eqnarray}
where
\begin{eqnarray}\label{K8}
F[z;T,\mu]\equiv \frac{\sinh(\beta z)}{\cosh(\beta z)+\cosh(\beta
\mu)}.
\end{eqnarray}
Plugging (\ref{K7}) in (\ref{K6}) and performing the integration
over $z$ as far as possible, the temperature dependent part
$I_{T\neq 0}$ is given by
\begin{eqnarray}\label{K9}
I_{T\neq 0}=-\frac{1}{4\pi^{2}}\int_{0}^{\Lambda}dz~
\left(2z+\frac{3\tilde{e}B}{z}\right)
F[z;T,\mu]+\frac{(\Lambda^2-\mu^2)}{4\pi^2}+\frac{3\tilde{e}B}{4\pi^{2}}\ln\left(\frac{\Lambda}{\mu}\right).
\end{eqnarray}
The second order critical surface of the transition between the
$\chi$SB and the normal phase as a function of the phase space
variables $(T,\mu,\tilde{e}B)$ is then given by adding (\ref{K4})
and (\ref{K9}) and reads
\begin{eqnarray}\label{K10}
\frac{1}{4G_{S}}-\frac{1}{4\pi^{2}}\int_{0}^{\Lambda}dz~\bigg(2z+\frac{3\tilde{e}B}{z}\bigg)F[z;T,\mu]=0,
\end{eqnarray}
where $F[z;T,\mu]$ is defined in (\ref{K8}). Note that (\ref{K10})
is only valid in the limit of strong magnetic field, where a LLL
approximation is justified. Choosing $\mu=0$, (\ref{K10}) reads
\begin{eqnarray}\label{K11}
\frac{1}{4G_{S}}=\frac{1}{4\pi^{2}}\int_{0}^{\Lambda}dz\bigg(2z+\frac{3\tilde{e}B}{z}\bigg)\tanh\frac{\beta
z}{2}.
\end{eqnarray}
This leads to the following relation between $T$ and $\tilde{e}B$
\begin{eqnarray}\label{K12}
\tilde{e}B(T,\mu=0;\Lambda)=\frac{4\pi^{2}}{3H\left(\lambda\right)}\left\{
\frac{1}{4G_{S}}-\frac{\Lambda^{2}}{4\pi^{2}}+\frac{T^{2}}{12}+\frac{T^{2}}{\pi^{2}}\bigg[\mbox{Li}_{2}(-e^{-2\lambda})-2\lambda\ln(1+e^{-2\lambda})\bigg]\right\},
\end{eqnarray}
where $\lambda\equiv \Lambda/2T$ and
\begin{eqnarray}\label{K13}
H(z)\equiv
\sum\limits_{n=1}^{\infty}\frac{(-1)^{n-1}2^{2n}(2^{2n}-1)z^{2n-1}}{(2n-1)(2n)!}B_{n},
\end{eqnarray}
and Li$_{2}(z)$ is the dilogarithm function defined by
\begin{eqnarray}\label{K14}
\mbox{Li}_{2}(z)\equiv-\int_{0}^{z}dy~\frac{\ln(1-y)}{y},
\end{eqnarray}
and $B_{n}$ are the Bernoulli's numbers. This result will be used in
Sec. III.B to determine the threshold magnetic field for the LLL
approximation at $\mu=0$. For arbitrary values of the chemical
potential, we expand (\ref{K10}) in the orders of
$\frac{\mu}{\Lambda}$ and keep terms up to
${\cal{O}}((\frac{\mu}{\Lambda})^{3})$, and arrive at the explicit
$(T,\mu)$ dependence of $\tilde{e}B$
\begin{eqnarray}\label{K15}
\tilde{e}B(T,\mu;\Lambda)&\approx&\frac{1}{3\gamma}\bigg\{-\frac{\pi^2}{G_{S}}+4T^2\bigg[\lambda^2
-\mbox{Li}_2(-e^{-2\lambda})+2\lambda\ln\left(1+e^{-2\lambda}\right)-\frac{\pi^2}{12}\bigg]\nonumber\\
&&\qquad-\mu^2\left(\tanh\lambda+\lambda\tanh^2\lambda-\lambda\right)\bigg\},
\end{eqnarray}
where $\gamma$ is defined by
\begin{eqnarray}\label{K16}
\gamma(T,\mu;\Lambda)\equiv\frac{\mu^2}{8T^2}\frac{\tanh^2\lambda}{\lambda}
-\int_{0}^{\lambda}dz\bigg(\frac{\tanh
z}{z}-\frac{\mu^2}{8T^2}\frac{\tanh^2 z}{z^2}\bigg).
\end{eqnarray}
Similarly, the $(T,\tilde{e}B)$ dependence of $\mu$ is given by
\begin{eqnarray}\label{K17}
\mu^2(T,\tilde{e}B;\Lambda)\approx\frac{1}{\alpha}\bigg\{-\frac{\pi^2}{G_{S}}+4T^2\bigg[\lambda^2
-\mbox{Li}_2(-e^{-2\lambda})+2\lambda\ln\left(1+e^{-2\lambda}\right)-\frac{\pi^2}{12}\bigg]
+3\tilde{e}B\int_{0}^{\lambda}dz\frac{\tanh z}{z}\bigg\},\nonumber\\
\end{eqnarray}
where $\alpha$ is defined by
\begin{eqnarray}\label{K18}
\alpha(T,\tilde{e}B;\Lambda)\equiv\left(
\tanh\lambda+\lambda\tanh^2\lambda-\lambda\right)+\frac{3\tilde{e}B}{8T^2}\frac{\tanh^2\lambda}{\lambda}
+\frac{3\tilde{e}B}{8T^2}\int_{0}^{\lambda}dz\frac{\tanh^2z}{z^2}.
\end{eqnarray}
Relations (\ref{K17}) will be evaluated numerically in Sec. III.B to
determine the threshold magnetic field for the LLL approximation.
\subsection{Transition between the CSC and the normal
phase}\label{app2}
\noindent For the transition between the CSC and the normal phase,
we use the equation \cite{sato1997, inagaki2003}
\begin{eqnarray}\label{K19}
\lim_{\Delta^{2}\rightarrow 0}\frac{\partial
\Omega_{\mbox{\tiny{eff}}}(\sigma=0,\Delta)}{\partial \Delta^{2}}=
0.
\end{eqnarray}
Setting $n=0$ in $\Omega_{\mbox{\tiny{eff}}}$ from (\ref{Fb23}) and
plugging the resulting expression in (\ref{K19}), we get
\begin{eqnarray}\label{K20}
\frac{1}{4G_{D}}-\frac{\tilde{e}B}{\sqrt{\pi}}\sum\limits_{\kappa=\pm{1}}\int_{0}^{\infty}\frac{ds}{\sqrt{s}}\int_{0}^{\infty}\frac{dp_{3}}{4\pi^{2}}e^{-s(p_3+\kappa\mu)^2}[1+2\sum\limits_{\ell=1}^{\infty}
(-1)^{\ell}~e^{-\frac{\beta^{2}\ell^{2}}{4s}}]=0.
\end{eqnarray}
Evaluating the $T$-dependent and independent parts of the expression
in the r.h.s. of (\ref{K20}) using (\ref{K3}) and (\ref{K5}),
respectively, we arrive after some straightforward manipulations at
the relation between the phase space parameters
$(T,\mu,\tilde{e}B)$,
\begin{eqnarray}\label{K21}
\tilde{e}B^{-1}(T,\mu;\Lambda)&=&\frac{G_{D}}{\pi^2}\int_{0}^{\frac{\Lambda+\mu}{2T}}dz
\frac{\tanh z}{z}+\int_{0}^{\frac{\Lambda-\mu}{2T}}dz\frac{\tanh z}{z}\nonumber\\
&=&\frac{G_{D}}{\pi^2}\bigg[H\left(\frac{\Lambda+\mu}{2T}\right)+H\left(\frac{\Lambda-\mu}{2T}\right)
\bigg],
\end{eqnarray}
where $H(z)$ is defined in (\ref{K13}). Fixing one of these
parameters in (\ref{K21}), the second order critical lines for the
transition from CSC to the normal phase arises in the phase space of
two other parameters.
\end{appendix}

\end{document}